\documentclass[a4paper,11pt]{article}
\pdfoutput=1
\usepackage[utf8]{inputenc}
\usepackage{jheppub_mod}
\usepackage{comment}
\usepackage{bm}
\usepackage[dvipsnames]{xcolor}
\usepackage{tikz, pgfplots}
\usepackage{caption,subcaption}

\usepackage{tocloft}
\newcommand{\I}{\mathrm{i}}
\newcommand{\D}{\mathrm{d}}

\renewcommand{\O}{\mathcal{O}}
\newcommand{\x}{\bs{x}}
\renewcommand{\k}{a}

\DeclareMathOperator{\im}{Im}
\DeclareMathOperator{\re}{Re}

\DeclareMathOperator{\Li}{Li}

\newcommand{\PLp}{\mathcal{D}^{(+)}}  
\newcommand{\PLm}{\mathcal{D}^{(-)}} 
\newcommand{\act}{\mathfrak{a}}
\newcommand{\bct}{\mathfrak{b}}
\newcommand{\jsrc}{\varphi_{(0)}}

\newcommand{\<}{\langle}
\renewcommand{\>}{\rangle}
\newcommand{\lla}{\langle \! \langle}
\newcommand{\rra}{\rangle \! \rangle}

\newcommand{\bs}[1]{\boldsymbol{#1}}
\newcommand{\reg}[1]{\hat{#1}}
\newcommand{\dreg}{\hat{d}}
\newcommand{\Dreg}{\hat{\Delta}}

\newcommand{\nn}{\nonumber}

\newcommand{\ep}{\epsilon}

\renewcommand{\j}{\varphi}

\newcommand{\ino}{i}
\newcommand{\ireg}{i^{\text{reg}}}

\newcommand{\ifin}{i^{\text{fin}}}
\newcommand{\iren}{i^{\text{ren}}}
\newcommand{\ict}{i^{\text{ct}}}

\newcommand{\dsno}{ds}
\newcommand{\dsreg}{ds^{\text{reg}}}

\newcommand{\dsfin}{ds^{\text{fin}}}
\newcommand{\dsren}{ds^{\text{ren}}}
\newcommand{\dsct}{ds^{\text{ct}}}

\newcommand{\s}[2]{\sigma_{(#1) #2}}
\newcommand{\m}[1]{l_{#1 -}}
\newcommand{\p}[1]{l_{#1 +}}

\newcommand{\vphi}{\varphi}
\renewcommand{\[}{\begin{equation}}
\renewcommand{\]}{\end{equation}}

\begin{document}

\title{Renormalisation of IR divergences and holography\\[0.5ex] in de Sitter}

\author[a]{Adam Bzowski,}
\affiliation[a]{
Institute of Physics, 
University of Silesia,
75 Pu\l ku Piechoty 1, 
41-500 Chorz\'{o}w, 
Poland
}

\author[b]{Paul McFadden}
\affiliation[b]{School of Mathematics, 
	Statistics \& Physics, Newcastle University, 
	Newcastle NE1 7RU, U.K.} 

\author[c]{and Kostas Skenderis.}
\affiliation[c]{STAG Research Center \& Mathematical Sciences, 
	University of Southampton,
	Highfield, \\
	Southampton 
	SO17 1BJ, U.K.}

\emailAdd{adam.bzowski@us.edu.pl}
\emailAdd{paul.l.mcfadden@newcastle.ac.uk}
\emailAdd{k.skenderis@soton.ac.uk}

\abstract{
We formulate a renormalisation procedure for IR divergences of tree-level in-in late-time de Sitter (dS) correlators. These divergences are due to the infinite volume of spacetime and are analogous to the 
divergences that appear in AdS  dealt with by holographic renormalisation. Regulating the theory using dimensional regularisation, we show that one can remove all infinities by adding local counterterms at the future boundary of dS in the Schwinger-Keldysh path integral. The counterterms amount to renormalising the late-time bulk field. We frame the discussion in terms of bulk scalar fields in dS${}_{d+1}$, using the computation of tree-level correlators involving massless and conformal scalars for illustration. The relation to AdS via analytic continuation is discussed, and we show that different versions of the analytic continuation appearing in the literature are equivalent to each other. In AdS, one needs to add counterterms that are related to conformal anomalies, and  also to renormalise the source part of the bulk field. The analytic continuation to dS projects out the traditional AdS counterterms, and links the renormalisation of the sources to the renormalisation of the late-time bulk field. We use these results to establish holographic formulae that relate tree-level dS${}_{d+1}$ in-in correlators to CFT correlators at up to four points, and we provide two proofs: one using the connection between the dS wavefunction and the partition function of the dual CFT, and a second  by direct evaluation of the in-in correlators using the Schwinger-Keldysh formalism. The renormalisation of the bulk IR divergences is mapped by these formulae to UV renormalisation of the dual CFT via local counterterms, providing structural support for a possible duality. We also recast the regulated holographic formulae in terms of the AdS amplitudes of shadow fields, but  show that this relation breaks down when renormalisation is required.

}

\maketitle

\section{Introduction and summary of results}

Holography for cosmology, and in particular, for de Sitter, has been an active topic of research since the early days of AdS/CFT.
A correspondence between de Sitter and CFT was proposed in \cite{Strominger:2001pn, Strominger:2001gp}; see also \cite{Hull:1998vg, Witten:2001kn} for earlier work. Despite the many results that have been obtained over the years, the status of this correspondence has remained controversial. One reason for this is that we do not have a concrete realisation of this correspondence in string theory akin to that 
for the AdS/CFT correspondence \cite{Maldacena:1997re}, and there are no examples where both sides of the duality are defined independently (apart from the example in \cite{Anninos:2011ui}, which involves however 
an exotic theory of gravity in the bulk).
Another reason is that, through the years, different-looking 
formulations have been introduced and it has not always been clear what is the relation between these and whether they are consistent with one another.  Nevertheless, as we will review below,  there is  significant supporting evidence for a correspondence -- including some evidence for the existence of a duality -- as well as agreement between different approaches. One of the purposes of this paper is to bring out the similarities and differences between these different approaches.

The asymptotic symmetry of de Sitter is the same as the Euclidean conformal group in one dimension less. It follows that observables defined at future infinity will satisfy the same kinematical constraints as that of a CFT. However, this on its own, while it may be useful, does not imply the existence of a duality between de Sitter and a local CFT. 
A similar question was raised in the early days of the AdS/CFT correspondence: was the agreement between bulk and boundary computations due to the high amount of (super)symmetry realised by the original AdS/CFT examples, or is there  support for the duality that does not rely on a high amount of symmetry? A trademark of a local QFT is that its UV divergences are local, and any dual formulation should have this property. This implies that IR divergences in AdS (namely, divergences due to the infinite volume of spacetime) should be local, as they are linked to the UV divergences of the CFT via the UV/IR correspondence. It turns out that this is indeed the case for AdS gravity \cite{deHaro:2000vlm, Papadimitriou:2004ap}. This is a non-trivial property as it does not hold automatically for any theory of gravity. For example, this is not the case for gravity with zero cosmological constant near spatial infinity \cite{deHaro:2000wj}. In this paper, we will address this question in the context of de Sitter.

Through the years different approaches to dS/CFT have been pursued. In \cite{Maldacena:2002vr}, Maldacena proposed that the putative duality relates the partition function of the dual CFT, $Z_{CFT}$, to the wavefunction of the universe, $\Psi_{dS}=Z_{CFT}$, upon a specific analytic continuation. The analysis took place in the regime where gravity is perturbative, and the paper also discussed the relation between AdS  and dS computations due to analytic continuation. This approach was further discussed and extended in, for example, \cite{Larsen:2002et, Larsen:2003pf, Seery:2006tq, 
Harlow:2011ke, Maldacena:2011nz,  
Mata:2012bx,  Ghosh:2014kba, 
Anninos:2014lwa, Kundu:2014gxa}. 

In \cite{McFadden:2009fg} we initiated an ``agnostic'' approach, where instead of postulating the existence and/or the form of a possible duality, we aimed to investigate whether any relation between standard observables in cosmology and QFT correlators in one dimension less  is possible
for generic accelerating cosmologies. 
The starting point was a one-to-one correspondence between FLRW cosmologies and domain-wall spacetimes \cite{Skenderis:2006jq}, which may be viewed as generalisation of the analytic continuation of dS to AdS, but now applied to general FLRW metrics \cite{Skenderis:2006fb}. A special case of this correspondence is that between accelerating FLRW metrics, either asymptotically de Sitter or asymptotically power-law, and 
holographic RG flows. For such cosmologies, we computed cosmological 2- and 3-point functions at tree-level
using the standard in-in formalism, both for scalar and tensor modes, and for arbitrary potential  \cite{McFadden:2010na, McFadden:2010vh, McFadden:2011kk,  Bzowski:2011ab}.
These computations boil down to  
solving specific differential equations satisfying certain initial conditions. As mentioned, these cosmologies are in   correspondence with spacetimes describing holographic RG flows, and for the latter, one can use standard holographic methods to compute the 2- and 3-point functions of the holographic energy-momentum tensor. These latter computations also boil down to solving specific differential equations satisfying a regularity condition in the interior. As it  turns out, the cosmological in-in correlations can then be re-expressed in terms of holographic correlators in an essentially unique way, for general potential, upon using a specific analytic continuation. The analytic continuation was  such that the differential equations and initial conditions that arose in the cosmology computation exactly mapped to those needed for the holographic computations, thus guaranteeing agreement for any model.  

The resulting holographic formulae were similar to those appearing in the wavefunction of the universe approach, except that the analytic continuation appeared to be different. In  \cite{Maldacena:2002vr} one analytically continued the AdS radius $L_{AdS}$ to the de Sitter radius $L_{dS}$, $L_{AdS} = i L_{dS}$, and the AdS radial coordinate $z$ to the dS conformal time $\tau$, $z=-i \tau$. In \cite{McFadden:2009fg}, we worked in momentum space along the boundary directions and  analytically continued the magnitude of momentum, $q_{AdS}=i q_{dS}$,  together with the Planck scale, $\ell_P^{(AdS)} = -i \ell_P^{(dS)}$. In large-$N$ $SU(N)$ theories, this is equivalent to $N^2 \to -N^2$. In this paper we show that these two analytic continuations are equivalent, extending an argument given in \cite{Garriga:2014fda}.\footnote{More precisely, the signs  that appear in the analytic continuation used in \cite{McFadden:2009fg} are such that one finds agreement with \cite{Maldacena:2002vr} based on the identification $\Psi^*_{dS} = Z_{CFT}$, which leads to the same results for observables as $\Psi_{dS} = Z_{CFT}$. Here, and elsewhere in this paper, we have adjusted the signs in the continuation used in \cite{McFadden:2009fg} so that there is an agreement with $\Psi_{dS} = Z_{CFT}$, see also footnote \ref{ft:an_cont}.} 

There are two natural scales that appear in the bulk theory: the Planck scale and the scale associated with the cosmological constant, the (A)dS radius. One may choose to work in Planck units, effectively setting $\ell_P=1$, both in AdS and dS. Then, relating AdS to dS yields the continuation discussed in  \cite{Maldacena:2002vr}. This  is natural from the bulk perspective, namely, when all computations are done in the bulk -- for example, when relating in-in dS correlations to AdS amplitudes computed via Witten diagrams. In these units, the dual CFT lives in a space with metric $ds^2 = L_{AdS}^2 \D \x^2$, and one would need to include (and analytically continue) these factors of $L_{AdS}$ in all CFT computations if one wishes to express the dS results in terms of CFT correlators. Alternatively, one may work in (A)dS units, effectively setting $L_{(A)dS}=1$, and this leads to the continuation used in  \cite{McFadden:2009fg}. Now the boundary geometry is $ds^2 = \D \x^2$, and as such, this setup is more natural from the perspective of the dual CFT. Moreover, this continuation refers only to CFT variables and parameters  and is therefore available when the bulk theory is not weakly coupled. We show explicitly that one can go from the one analytic continuation to the other simply by performing a Weyl rescaling that adds or removes the factor of $L_{(A)dS}^2$ from the boundary metric.
Thus, taking this into account, the approaches in  \cite{Maldacena:2002vr} and  \cite{McFadden:2009fg} are in fact equivalent.
 
The analytic continuation in these works is done in an {\it ad hoc} fashion, and an important question is whether it can be understood from a more fundamental perspective and whether it makes sense non-perturbatively in $1/N^2$. A quantum cosmology perspective has been discussed in \cite{Hertog:2011ky, Hartle:2012qb, Conti:2015ruo, Hertog:2015nia}. For other more recent perspectives, see \cite{Araujo-Regado:2022gvw, Antonini:2022ptt}. The analytic continuation has also been embedded in supergravity  \cite{Bergshoeff:2007cg, Skenderis:2007sm}. In these  works, the AdS and dS supergravities appear as different real slices of a single theory with complex action, and there is a  connection with $II^*$ string theories and $M^*$ theories \cite{Hull:1998vg}, see also \cite{Dijkgraaf:2016lym, Blumenhagen:2020xpq}.  
Recall that in standard QFT, analytic continuation in the momenta relates different correlators (time-ordered products, retarded/advanced correlators, {\it etc.}), each having different physical meaning. 
The results described above suggest that  meaningful observables may also be obtained by analytically continuing the parameters that appear in (a class of) QFTs.

The holographic formulae derived in \cite{Maldacena:2002vr} and  \cite{McFadden:2009fg}  are valid at tree-level in the bulk, or at leading order as $N \to \infty$ from the perspective of the dual QFT.  However, the dual QFT need not be strongly coupled. A weakly coupled dual QFT would correspond to a stringy non-geometric bulk, described by a strongly coupled worldsheet theory in string theory. While we do not currently have any such models where both sides of the duality are tractable, one can nevertheless work out the predictions of such models using holography to see if their phenomenology is interesting. Models where the dual QFT is perturbative were introduced in  \cite{McFadden:2009fg}, and their predictions for spectra and bi-spectra were computed in \cite{McFadden:2010na, McFadden:2010vh,McFadden:2011kk, Bzowski:2011ab, Coriano:2012hd, Kawai:2014vxa}. 
These models have been custom fitted to the WMAP \cite{Dias:2011in, Easther:2011wh}  and Planck data \cite{Afshordi:2016dvb, Afshordi:2017ihr}, showing they provide an excellent fit to cosmological data and thus an alternative to conventional inflation. These models also provide a new perspective on the resolution of the classic puzzles of hot Big Bang cosmology (the horizon, flatness and relic problems) that historically motivated the introduction of cosmological inflation \cite{Nastase:2019rsn, Nastase:2020uon}. As time evolution is mapped to inverse RG flow, the resolution of the Big Bang singularity  is mapped to the question of the IR finiteness of the dual QFT; a question which is non-trivial but tractable for the QFTs featured  in the non-geometric models  \cite{Jackiw:1980kv, Appelquist:1981vg, Cossu:2020yeg}.

Yet a different approach to the relation between late-time de Sitter correlators and conformal field theory is to use the fact that de Sitter isometries are the same as conformal transformations in one dimension less, and thus the late-time de Sitter correlators should satisfy the conformal Ward identities.  Indeed, in \cite{Bzowski:2012ih,McFadden:2013ria}, we showed that the inflationary predictions for slow-roll inflation, to second order in slow roll,  are exactly reproduced by conformal perturbation theory, and related work has appeared in \cite{Antoniadis:2011ib, Maldacena:2011nz, Creminelli:2011mw, 
Kehagias:2012pd, Kehagias:2012td,Schalm:2012pi, Mata:2012bx, Garriga:2013rpa, Garriga:2014fda}. 
The topic received a new impetus following \cite{Arkani-Hamed:2015bza} and \cite{Arkani-Hamed:2018kmz} under the umbrella of cosmological collider physics and the cosmological bootstrap. In the cosmological bootstrap, one aims to obtain the late-time de Sitter correlators by solving the conformal Ward identities, using as input  expected physical properties, such as analyticity, that the correlators should have. Recent works in this direction include \cite{Arkani-Hamed:2017fdk,Benincasa:2018ssx,   Sleight:2019mgd, Sleight:2019hfp, Baumann:2019oyu, Baumann:2020dch, Pajer:2020wnj,Sleight:2020obc, Goodhew:2020hob, Cespedes:2020xqq, Pajer:2020wxk, Jazayeri:2021fvk, Melville:2021lst, Baumann:2021fxj,  Meltzer:2021zin, Hogervorst:2021uvp, DiPietro:2021sjt, Sleight:2021plv, 
 Pimentel:2022fsc,  Bonifacio:2022vwa, Salcedo:2022aal, Wang:2022eop, Albayrak:2023jzl, Armstrong:2023phb, Albayrak:2023hie, Arkani-Hamed:2023bsv, Arkani-Hamed:2023kig}.  

In this paper, we aim to use light scalar fields in de Sitter (with mass $0<m^2 L_{dS}^2< d^2/4$) to further develop the correspondence between cosmological in-in correlators and CFT correlation functions, and to  compare and contrast different approaches that have appeared in the literature. We present holographic formulae for correlation functions of light scalars at up to four points.   These formula express the de Sitter correlators as analytic continuations of the 2-, 3- and 4-point functions of the CFT dual to the corresponding AdS spacetime.   We supply two proofs of these holographic formulae.  The first is based on analytically continuing the AdS partition function to the wavefunction of the universe, and holds to all orders in perturbation theory.  The second is a tree-level analysis showing how the Schwinger-Keldysh diagrams for dS correlators can be re-expressed in terms of AdS Witten diagrams. 

The relation between dS and CFT is easiest to describe using the wavefunction coefficients. Let $\vphi_{(0)}(\x)$ be the field 
parametrising the late-time behavior of the bulk field $\varphi(\tau, \x)$ (see \eqref{dSasympt}), and $\Psi_{dS}[\vphi_{(0)}]$ the dS wavefunction. Expanding perturbatively in powers of $\vphi_{(0)}$, we define the wavefunction coefficients $\psi_n$ by
\[
\Psi_{dS}[\vphi_{(0)}]  = \exp\Big(\sum_{n=2}^\infty \frac{(-1)^n}{n!}\int [\D \bs{q}_n] \psi_n(\bs{q}_1,\ldots, \bs{q}_n) \vphi_{(0)}(-\bs{q}_1)\ldots\vphi_{(0)}(-\bs{q}_n)\Big). \label{intro:psiexp}
\]
A holographic formula then relates these coefficients to CFT correlation functions,
\begin{equation}\label{intro:corrtopsi3}
\psi_n(\bs{q}_1,\ldots,\bs{q}_n) = (-i)^{d} \lla \O(\bs{q}_1)\ldots \O(\bs{q}_n)\rra \Big|_{N^2 \rightarrow -N^2, \  q_i \to i q_i}
\end{equation}
where the correlator on the right-hand side is a Euclidean CFT correlator in flat space, $ds^2 = d \x^2$. The double-bracket notation means that we have extracted the momentum-conserving delta function, see \eqref{eq:stripped}. We assume (as usual) that the CFT admits a 't Hooft large $N$ limit with $N^2 \to \infty$ the leading contribution, 
as in $SU(N)$ theories with matter in the adjoint of $SU(N)$. Essentially the substitution $N^2 \to -N^2$ reverses the sign of the coupling that suppresses non-planar loops. (For $SO(N)$ theories, the analogous continuation is $N \to -N$.) In the bulk, this corresponds to reversing the sign of the Planck constant (working in (A)dS units), or more precisely continuing $\ell_P^{(AdS)} = -i\ell_P^{(dS)}$. The substitution $q_i \to i q_i$ analytically continues the magnitude of the momenta. The relation \eqref{intro:corrtopsi3} holds to all orders in bulk perturbation theory, and we establish this by showing that the respective Feynman rules map to each other under the analytic continuation. (The correlator on the right-hand side is computed via an AdS computation using the rules of the AdS/CFT correspondence.) The results on the non-geometric models described above suggest that this relation may also hold in the stringy regime, where the bulk is described by a strongly coupled sigma model and the boundary theory is at weak coupling.

Starting from this relation, one may then compute the in-in correlators by functionally integrating over $\vphi_{(0)}(\x)$ with measure provided by the square of the wavefunction,
\begin{equation} \label{intro:eq: wavefn}
\<\vphi_{(0)}(\bs{x}_1)\ldots \vphi_{(0)}(\bs{x}_n)\>=\int\mathcal{D}\vphi_{(0)}\, \vphi_{(0)}(\bs{x}_1)\ldots \vphi_{(0)}(\bs{x}_n)
|\Psi_{dS}[\vphi_{(0)}]|^2.
\end{equation}
Performing this integral at tree-level leads to the following holographic formulae:
\begin{align}\label{intro:2ptholformula}
\lla\vphi_{(0)}(\bs{q})\vphi_{(0)}(-\bs{q})\rra &=- \frac{1}{2}\frac{1}{\re [ (-i)^d \lla\O(\bs{q})\O(-\bs{q})\rra]},\\[2ex]
\lla\vphi_{(0)}(\bs{q}_1)\vphi_{(0)}(\bs{q}_2)\vphi_{(0)}(\bs{q}_3)\rra &=\frac{1}{4}\frac{\re [(-i)^d\lla \O(\bs{q}_1)\O(\bs{q}_2)\O(\bs{q}_3)\rra]}{\prod_{i=1}^3\re [(-i)^d \lla\O(\bs{q}_i)\O(-\bs{q}_i)\rra]}, 
\end{align}
\begin{align}
\lla\vphi_{(0)}(\bs{q}_1)\vphi_{(0)}(\bs{q}_2)\vphi_{(0)}(\bs{q}_3)\vphi_{(0)}(\bs{q}_4)\rra  &
=\frac{1}{8}\Big[\frac{\re [(-i)^d \lla\O(\bs{q}_1)\O(\bs{q}_2)\O(\bs{q}_3)\O(\bs{q}_4)\rra]}{\prod_{i=1}^4\re [(-i)^d \lla\O(\bs{q}_i)\O(-\bs{q}_i)\rra]}\nn
\\[2ex]&  \hspace{-6.0cm}
-\Big(\frac{\re [(-i)^d\lla\O(\bs{q}_1)\O(\bs{q}_2)\O(\bs{q}_{12})\rra] \re [(-i)^d\lla\O(-\bs{q}_{12})\O(\bs{q}_3)\O(\bs{q}_4)\rra] }{\re [(-i)^d \lla\O(\bs{q}_{12})\O(-\bs{q}_{12})\rra] \prod_{i=1}^4\re [(-i)^d \lla\O(\bs{q}_i)\O(-\bs{q}_i)\rra]}
+(2\leftrightarrow 3) + (2\leftrightarrow 4)\Big)\Big]
\label{intro:4ptholformula}
\end{align}
where the real parts of all correlators on the right-hand sides are taken after the continuation described above.  
Note that while \eqref{intro:corrtopsi3} holds to all orders in bulk perturbation theory, \eqref{intro:2ptholformula}-\eqref{intro:4ptholformula} are only valid at tree-level. The reason is that they have been derived by performing the functional 
integral over $\vphi_{(0)}(\x)$ only at tree-level. It would be straightforward but tedious to extend the formulae to higher order.
Note also that the two terms on the right-hand side of \eqref{intro:4ptholformula} do not represent bulk contact and exchange diagrams.
For example, the 4-point function $\lla \O(\bs{q}_1)\O(\bs{q}_2)\O(\bs{q}_3)\O(\bs{q}_4) \rra$ at bulk tree-level is itself a sum of contact and exchange diagrams. We argued above that the results for the non-geometric models suggest that \eqref{intro:corrtopsi3} may also hold in the regime where the dual QFT is perturbative (and the bulk is stringy). The holographic formulae \eqref{intro:2ptholformula}-\eqref{intro:4ptholformula} would then hold at string tree-level order.

To make contact with recent literature, we also present a second derivation of the above holographic formulae by using now the in-in formalism.  
We consider a bulk Lagrangian that involves general cubic and quartic vertices.  For such a theory, we computed the tree-level in-in dS late-time amplitudes and compared the results with the corresponding tree-level AdS amplitudes. At tree-level one can readily track the signs due to the analytic continuation $\ell_P^{(AdS)} = -i\ell_P^{(dS)}$ and include them explicitly in the holographic formula. The results up to 4-point functions are given by
\begin{align}
\dsren_{[\Delta \Delta]}(q)  &= - \frac{1}{2} \frac{1}{\im \iren_{[\Delta \Delta]}(\I q; \mu, \bct)}. \label{intro:Amp2} \\
 \dsren_{[\Delta_1 \Delta_2 \Delta_3]}(q_i; \mu, \act_i) &= - \frac{1}{4} \frac{\im \iren_{[\Delta_1 \Delta_2 \Delta_3]}(\I q_i)}{\prod_{j=1}^3 \im \iren_{[\Delta_j \Delta_j]}(\I q_j)}, \label{intro:Amp3} \\
 \dsren_{[\Delta_1 \Delta_2 \Delta_3 \Delta_4]}(q_i; \mu, \act_i) &= - \frac{1}{8} \frac{\im  \iren_{[\Delta_1 \Delta_2 \Delta_3 \Delta_4]}(\I q_i; \mu, \bct_k(\act_i))}{\prod_{j=1}^4 \im \iren_{[\Delta_j \Delta_j]}(\I q_j)}, \\
 \dsren_{[\Delta_1 \Delta_2; \Delta_3 \Delta_4 x \Delta_x]}(q_i, s; \mu, \act_i) &= \frac{1}{8} \prod_{j=1}^4 \frac{1}{\im \iren_{[\Delta_j \Delta_j]}(\I q_j)} \left[ \im \iren_{[\Delta_1 \Delta_2; \Delta_3 \Delta_4 x \Delta_x]}(\I q_i, \I s; \mu, \bct_k(\act_i)) \right.\nn\\
& 
\hspace{-1.5cm}
\left. - \, \frac{\im \iren_{[\Delta_1 \Delta_2 \Delta_x]}(\I q_1, \I q_2, \I s; \mu, \bct_k(\act_i)) \im \iren_{[\Delta_x \Delta_3 \Delta_4]}(\I s, \I q_3, \I q_4; \mu, \bct_k(\act_i))}{\im \iren_{[\Delta_x \Delta_x]}(\I s)} \right]. \label{intro:Amp4x}
\end{align}
The left-hand side denotes dS tree-level in-in late-time amplitudes with $\ell_P^{(dS)}=L_{dS}=1$.
We use conformal dimensions to denote the external states and exchange particles. For example, 
$\dsren_{[\Delta_1 \Delta_2 \Delta_3]}(q_i; \mu, \act_i)$ denotes the renormalised scalar 3-point in-in correlator associated scalars of mass $m_i^2=- \Delta_i(\Delta_i -d)L_{dS}^{-2}$. The dimensions $\Delta_i$ are the canonical dimension specified by AdS/CFT (and not the shadow dimensions $\bar{\Delta}_i=d-\Delta_i$). The correlator depends on the magnitude $q_i$ of the momenta $\bs{q}_i$ (conjugate to the insertion points $\x_i$). The superscript ``ren'' indicates that these are renormalised correlators,  where renormalisation  refers to renormalisation of IR divergences due to the infinite volume of spacetime, which we discuss in greater detail below. (Here, $\mu$  is the associated renormalisation scale and $\act_i$ the constants parametrising the scheme dependence).  For 4-point functions, we have contact diagrams $\dsren_{[\Delta_1 \Delta_2 \Delta_3 \Delta_4]}(q_i; \mu, \act_i) $, and exchange diagrams $\dsren_{[\Delta_1 \Delta_2; \Delta_3 \Delta_4 x \Delta_x]}(q_i, s; \mu, \act_i) $, where $\Delta_x$ is the dimension associated with the exchange particle 
and $s$ is the magnitude of the momentum carried by it. On the right-hand side, we have the corresponding renormalised AdS amplitudes, $i^{\rm ren}$, with $\ell_P^{(AdS)}=L_{AdS}=1$. Without loss of generality we can choose the renormalisation scale in AdS and dS to be the same, but in general there is a non-trivial map that relates the scheme-dependent constants $\bct_k$ that appear in the AdS computation to those of the dS computation, $\act_i$. For concreteness, we focused on the case of two bulk scalars: one massless and another with mass $m_{dS}^2 = (d^2-1)/( 4 L_{dS}^2)$ describing  a conformal scalar in de Sitter. These correspond to a marginal operator with $\Delta=d$ and an operator of dimension $\Delta=(d+1)/2$.
When final results are quoted we typically 
restrict to $d=3$.

Converting the AdS amplitudes to CFT correlators in \eqref{intro:Amp2}-\eqref{intro:Amp4x} yields again \eqref{intro:2ptholformula}-\eqref{intro:4ptholformula} (more precisely, the generalisation of these formulae involving multiple bulk fields of the appropriate dimensions on the left-hand side, and the corresponding CFT operators on the right-hand side). Note that for general cubic and quartic couplings, the CFT 4-point functions involves a sum of contact and exchange diagrams and these combine to yield \eqref{intro:4ptholformula}. As discussed in \cite{Bzowski:2022rlz}, each Witten diagram is a CFT correlator for some holographic CFT specified by a bulk action, and in this case each of \eqref{intro:Amp2}-\eqref{intro:Amp4x} becomes one of \eqref{intro:2ptholformula}-\eqref{intro:4ptholformula}.
It is a peculiarity of the tree-level diagrams that each diagram on its own has a well-defined map under the analytic continuation. In general, one would expect that the analytic continuation only maps observables of the one theory to the other: in-in correlators to CFT correlators. From this perspective,  \eqref{intro:2ptholformula}-\eqref{intro:4ptholformula} are more fundamental than \eqref{intro:Amp2}-\eqref{intro:Amp4x}.  

As mentioned, we did the dS computations in two ways: first by using  
\eqref{intro:eq: wavefn}, and second, by using the in-in or Schwinger-Keldysh formalism. In the Schwinger-Keldysh formalism one performs the path integral over a path in the complex time plane. For in-in correlators this is a closed path that starts and ends at the same time. In dS this is a real path that starts at future infinity ($\tau=0$ in the coordinates used in \eqref{eq:dSmetric}), moves backward in time to $\tau=-\infty$, and then reverses and moves forward in time till it reaches future infinity again. Operators may be inserted at any point along the time contour and the correlators may be computed using the Schwinger-Keldysh path integral,
\begin{equation} \label{intro:Z_SK}
Z[J_{+}, J_{-}] = \hspace{-3cm}\int\limits_{\hspace{3.1cm}\vphi_+(0,\x)=\vphi_-(0,\x) \sim \varphi_{(0)}(\x)} \hspace{-3cm} 
\mathcal{D} \j_{+}\ \mathcal{D} \j_{-} \  \exp\Big(\I S_{+}[\j_{+}] - \I S_{-}[\j_{-}] + \I \int \D^{d+1} x \sqrt{-g} \left( J_{+} \j_{+} - J_{-} \j_{-} \right)\Big),
\end{equation}
where the subscript $+$ indicates fields in the forward path and $-$ in the backward path, and $J_{\pm}$ are sources that generate insertions in the two paths. The fields on the forward and backward contour have the same limit as we go to future infinity, and it is the correlators of the late-time field $\varphi_{(0)}$ that we are interested in computing. 
One may directly link \eqref{intro:eq: wavefn} with \eqref{intro:Z_SK} by identifying $\Psi_{dS}$ with the forward path and $\Psi_{dS}^*$ with the backward path.

The dS correlators generically exhibit singularities as $\tau \to 0^-$, as was noticed, for example, in  \cite{Falk:1992sf, Zaldarriaga:2003my, Seery:2008qj, Creminelli:2011mw, Sleight:2019hfp}. In previous literature the singularities were regulated by putting a late-time cut-off $\tau_0$, but they were not renormalised. Here, we show how to renormalise such infinities: it suffices to introduce a counterterm at $\tau=0$ in the Schwinger-Keldysh path integral
of the following schematic form,
\begin{equation} \label{intro:dS_ct}
S_{\rm ct}^{dS}[\varphi_{(0)}^l; (J^l_+-J_-^l) ; \mu, \act_j; \epsilon] = \int_{\tau=0} d^d x \sum_k (J^k_+-J_-^k) f^k\left(\varphi_{(0)}^l; \mu, \act_j; \epsilon \right),
\end{equation}
where $\varphi_{(0)}^l$ denotes collectively all (late-time) bulk fields (labelled by $l$), $\epsilon$ denotes the regulator (which could be a late-time cutoff $\tau_0$, or the regulator parameter of dimensional regularisation, which is the regulator we use in this paper), $\mu$ the renormalisation scale and $\act^j$ is a set of constants that parametrise the scheme dependence.  For each bulk field we have sources $J_{\pm}^l$, and the corresponding $f^l$ are {\it local functions} of the late-time fields $\varphi_{(0)}^l$ which essentially describe how the late-time fields are renormalised.
More precisely, at $\tau=0$ the source terms in \eqref{intro:Z_SK} become $(J_+^l - J^l_-) \varphi_{(0)}^l$ and combine with \eqref{intro:dS_ct} to yield
\begin{equation} \label{eq: ren_sou}
 \int_{\tau=0}  d^d x \sum_k (J^k_+-J_-^k) \varphi_{R (0)}^k  
\end{equation}
with 
\begin{equation} \label{eq: field_ren}
\varphi_{R (0)}^k = \varphi_{(0)}^k + f^k\left(\varphi_{(0)}^l; \mu, \act_j; \epsilon \right).    
\end{equation}
The function $f^k$ is a polynomial of degree $n-1$ in the $\varphi_{(0)}^l$ and their derivatives, and the possible terms are restricted by dimensional analysis: $\varphi^k_{(0)}$ has dimension $d-\Delta_k$, so non-trivial contributions are possible only when there are operators whose dimensions are such that one can construct non-linear terms, using $\varphi_{(0)}^l$  and their derivatives, whose dimensions are precisely $d-\Delta_k$. It turns out that bulk correlators exhibit IR singularities exactly when counterterms of the form \eqref{intro:dS_ct} exist.

To illustrate the above for the cases studied in this paper,
consider the case $d=3$ with a massless scalar $\varphi^{[0]}$ and conformal scalar $\varphi^{[1]}$. The numbers in brackets are equal to $d-\Delta_i$, where $\Delta_i$ is the conformal dimension of dual operator, which is related to the de Sitter mass-squared as $m_i^2=- \Delta_i(\Delta_i -d)L_{dS}^{-2}$. This means that 
\begin{equation} \label{intro:f}
f^{[0]} = a^{[0]}_1 (\varphi^{[0]}_{(0)})^2 + a^{[0]}_2 (\varphi^{[0]}_{(0)})^3 + \cdots, \qquad 
f^{[1]} = a^{[1]}_1 \varphi^{[0]}_{(0)} \varphi^{[1]}_{(0)}+ a^{[1]}_2 (\varphi^{[0]}_{(0)})^2 \varphi^{[1]}_{(0)} + \cdots,   \end{equation}
where the constants $a^{[i]}_j$ are adjusted so that all infinities are cancelled. Most of the 3- and 4-points dS amplitudes involving these fields are actually divergent, see tables \ref{fig:deg3} and \ref{fig:deg4}. We have shown that all such infinities can be cancelled using the counterterms in \eqref{intro:dS_ct} with the $f$'s in \eqref{intro:f} for suitable choice of the constants $a^{[i]}_j$. For example, for the case of a single massless scalar field with the (regulated) action \eqref{SdSall3} the constants are given by \eqref{eq: a1}-\eqref{eq: 3333x3}. After adding the counterterm \eqref{intro:dS_ct}, one may remove the regulator resulting in finite renormalised correlators, which now depend on the renormalisation scale $\mu$ and the constants $\act_i$ parametrising the scheme dependence. To fix this scheme dependence one would need to impose normalisation conditions. 

We have just established that the tree-level dS IR divergences are local and can be dealt with by renormalising the dS late-time fields $\varphi_{(0)}$, as in  \eqref{eq: field_ren}, \eqref{intro:f}. 
The dS IR infinities are the analogue of the IR infinities in AdS that are dealt with by holographic renormalisation \cite{deHaro:2000vlm}. We would like to emphasise however that \eqref{intro:dS_ct} is not the analytic continuation of the conventional AdS counterterms derived in \cite{deHaro:2000vlm}.
As shown in \cite{Skenderis:2002wp}, and also discussed here for the cases of interest, the asymptotic solutions are mapped to each other by the analytic continuation as is the on-shell value of the action. However, the contributions of the conventional AdS counterterms drop out after analytic continuation to dS. There are a number of ways to see this. 
From the perspective of the Schwinger-Keldysh path integral \eqref{intro:Z_SK}, the conventional counterterms amount to $S_\pm(\varphi_\pm) \to S_\pm(\varphi_\pm) + S_{\rm ct}[\varphi_{(0)}]$, where we use the fact that $\varphi_+=\varphi_-\sim \varphi_{(0)}$ at $\tau=0$, and the counterterm action $S_{\rm ct}[\varphi_{(0)}]$ drops out since \eqref{intro:Z_SK} contains the difference $S_+ - S_-$. A second way to see the same is to note that conventional counterterms contribute ultra-local terms in CFT correlators ({\it i.e.} terms where all insertions are coincident).
Such terms are analytic in all squared momenta when written in momentum space, and such contributions are projected out when taking the imaginary parts in \eqref{intro:Amp2}-\eqref{intro:Amp4x}.
This means in particular that the conformal anomalies are projected out, a point which was also recently made in \cite{Chakraborty:2023los, Chakraborty:2023yed}. 

The infinities that we found in dS in-in amplitudes are also present in CFT correlators and AdS amplitudes \cite{Bzowski:2015pba}. In CFT,  the counterterm corresponding to \eqref{intro:dS_ct}, renormalises the coupling between the sources, $\varphi_{(0)}^k$, and the CFT operator, $\O_k$, 
$S_{\rm CFT}[\varphi^k_{(0)}, \O_k] = 
\int d^d x \sum_k \varphi^k_{(0)} \O_k$ to
\begin{equation} \label{eq: CFT_ren_sou}
S_{\rm CFT}[\varphi^k_{R(0)}, \O_k] = \int d^d x \sum_k \varphi^k_{R (0)} \O_k\, ,   
\end{equation}
with $\varphi^k_{R(0)}$ as in (\ref{eq: field_ren}). While $\beta$-functions vanish at the critical point, their derivatives (with respect to other couplings) do not in general vanish, and these counterterms encode this CFT data (while counterterms associated with anomalies encode the anomaly coefficients). In AdS/CFT, the coupling between the source and operator is implicit, and as such one does not directly add a new counterterm, but rather, one specifies how the source depends on the cut-off ({\it i.e.}, (\ref{eq: field_ren})) before removing the cut-off \cite{Bzowski:2015pba}.
The analytic structure of these divergences is different from the ones associated with anomalies: these are semi-local, meaning that only a subset of operator insertions are coincident (while, as mentioned above, divergences associated with anomalies are ultra-local). In momentum space, these divergences are analytic in some but not all of the squared momenta. This implies that they survive when taking the imaginary part. Indeed, we have shown that the holographic formulae \eqref{intro:Amp2}-\eqref{intro:Amp4x}  hold at the renormalised level, with the left-  and the right-hand sides computed independently. This is a non-trivial agreement and it involves a non-trivial map between the scheme-dependent terms. 

Comparing \eqref{eq: ren_sou} and \eqref{eq: CFT_ren_sou}, the CFT operators $\O_k$ are essentially identified with the (difference of the) Schwinger-Keldysh sources $J^k_+-J^k_-$. The coupling \eqref{eq: ren_sou} and the integration over the $\varphi_{(0)}^k$ then implements a Legendre transform. In CFT, and for operators of generic conformal dimension, this has the effect of exchanging operators of dimension $\Delta$ with operators of shadow dimension, $\bar{\Delta}=d-\Delta$. This relation always holds in a dimensionally regulated theory, but it  breaks down when renormalisation is needed \cite{Bzowski:2015pba} -- we will discuss what happens in such cases below.
We thus expect that the {\it regulated} holographic formulae can be expressed in terms of CFT correlation functions of operators of shadow dimensions. 
Indeed, starting from \eqref{intro:Amp2}-\eqref{intro:Amp4x}, explicitly taking  the imaginary parts and converting to 
correlators of operators of shadow dimension we obtain,
\begin{align}
\label{intro:2ptshadowdS0}
ds^{\rm reg}_{[\Delta\Delta]}&=\frac{1}{(2\bar{\beta})^2\mathcal{C}_{[\bar{\Delta}\bar{\Delta}]}}\,\ireg_{[\bar{\Delta}\bar{\Delta}]},\\[2ex]
\label{intro:3ptshadowdS0}
ds^{\rm reg}_{[\Delta_1 \Delta_2\Delta_3]} &=\prod_{j=1}^3\Big(\frac{1}{2\bar{\beta}_j \mathcal{C}_{[\bar{\Delta}_j\bar{\Delta}_j]}}\Big)\,\mathcal{C}_{[\bar{\Delta}_1\bar{\Delta}_2\bar{\Delta}_3]}\ireg_{[\bar{\Delta}_1\bar{\Delta}_2\bar{\Delta}_3]},\\[2ex]
\label{intro:4ptcontactshadow0}
ds^{\rm reg}_{[\Delta_1\Delta_2\Delta_3\Delta_4]} &=\prod_{j=1}^4\Big(\frac{1}{2\bar{\beta}_j \mathcal{C}_{[\bar{\Delta}_j\bar{\Delta}_j]}}\Big)\,\mathcal{C}_{[\bar{\Delta}_1\bar{\Delta}_2\bar{\Delta}_3\bar{\Delta}_4]}\ireg_{[\bar{\Delta}_1\bar{\Delta}_2\bar{\Delta}_3\bar{\Delta}_4]},\\[2ex]
\label{intro:shadowdSexch0}
 \dsno^{\rm reg}_{[\Delta_1 \Delta_2; \Delta_3 \Delta_4 x \Delta_x]} &=\prod_{j=1}^4\Big(\frac{1}{2\bar{\beta}_j \mathcal{C}_{[\bar{\Delta}_j\bar{\Delta}_j]}}\Big)\,\Big[
\frac{\mathcal{C}_{[\bar{\Delta}_1\bar{\Delta}_2\bar{\Delta}_x]}\mathcal{C}_{[\bar{\Delta}_3\bar{\Delta}_4\bar{\Delta}_x]}}{\mathcal{C}_{[\bar{\Delta}_x\bar{\Delta}_x]}}\ireg_{[\bar{\Delta}_1 \bar{\Delta}_2; \bar{\Delta}_3 \bar{\Delta}_4 x \bar{\Delta}_x]} 
\nn\\&\qquad\qquad\qquad\qquad \qquad  +
\frac{\mathcal{C}_{[\bar{\Delta}_1\bar{\Delta}_2\Delta_x]}\mathcal{C}_{[\bar{\Delta}_3\bar{\Delta}_4\Delta_x]}}{\mathcal{C}_{[\Delta_x\Delta_x]}}\ireg_{[\bar{\Delta}_1 \bar{\Delta}_2; \bar{\Delta}_3 \bar{\Delta}_4 x \Delta_x]} \Big], 
\end{align}
where 
\[\label{intro: Cdef}
\bar{\beta}_j = \bar{\Delta}_j-\frac{d}{2}  = -\beta_j,\qquad
\mathcal{C}_{[\bar{\Delta}_1,\ldots,\,\bar{\Delta}_n]} = 2\sin\Big[\frac{\pi}{2}\Big(d-\sum_{j=1}^n\bar{\Delta}_j\Big)\Big].
\]
Note that all analytic continuations have been performed. 
This reproduces results in \cite{Sleight:2020obc, DiPietro:2021sjt, Sleight:2021plv}.

One may have anticipated this result based on the dS Ward identities and  general considerations. Indeed, as the group of de Sitter isometries is the same as the Euclidean conformal group in one dimension less, the dS Ward identities become the conformal Ward identities at late times, acting on fields of the shadow dimension, $\bar{\Delta}_i = d -\Delta_i$, exactly as expected given the dimension of $\varphi_{(0)}^i$ is $d-\Delta_i$. The conformal Ward identities have a unique solution through to 3-point functions, and the AdS amplitudes are a solution of the conformal Ward identities, so for \eqref{intro:2ptshadowdS0}-\eqref{intro:3ptshadowdS0} the only 
issue is to explain the coefficients. For the 4-point function, we use in addition the fact that we are considering tree-level correlators, and the corresponding analytic structure in momentum space. This then implies \eqref{intro:4ptcontactshadow0}, up to a constant. For the exchange diagram, noting that the sum over bulk dS Schwinger-Keldysh propagators is invariant under $\beta_x \to -\beta_x$, which amounts to $\Delta_x \to \bar{\Delta}_x$, the right-hand side of \eqref{intro:shadowdSexch0} must be symmetric under $\Delta_x \leftrightarrow \bar{\Delta}_x$ and thus involve both the AdS exchange diagram with $\Delta_x$, and its shadow, $\bar{\Delta}_x$. 

The constants may be fixed by requiring that the right-hand sides have the same singularity structure as the left-hand sides. As we already discussed, there are no possible anomalies in dS. Anomalies appear in  AdS $n$-point functions, however,  whenever \cite{Petkou:1999fv,Bzowski:2013sza,Bzowski:2019kwd}, 
\[\label{intro: anomalies}
(n-2)\frac{d}{2} - \sum_{j=1}^n \beta_j = -2k \quad  
\Leftrightarrow \quad d-\sum_{j=1}^n\bar{\Delta}_j = 2k, \qquad k=0,1,2,\ldots \, .
\]
This means that the AdS amplitudes of operators of shadow dimension will have zeros and singularities when \eqref{intro: anomalies} hold, which the dS amplitudes do not have. To cancel those the coefficients that relate the dS amplitudes to AdS amplitudes of operators of shadow dimension must have compensating zeros and poles. This is indeed manifestly the structure of the coefficients in \eqref{intro:2ptshadowdS0}-\eqref{intro:4ptcontactshadow0}.\footnote{The factors of $1/(2 \bar{\beta}_j)$ are due to different normalisation of AdS amplitudes relative to dS amplitudes, see section \ref{sec:shadowdS}.} This is also the case in \eqref{intro:shadowdSexch0}, even though this is not manifestly so, see appendix \ref{sec:shadowsingcompatibility}.

Having established these relations, it is tempting to contemplate a direct link between dS and a CFT with operators of shadow dimension. This has indeed been suggested in \cite{DiPietro:2021sjt}, building on \cite{Sleight:2020obc}, where it was also proposed that this CFT may be holographically constructed via an AdS action with appropriate couplings to account for the coefficients that relate the dS and shadow AdS amplitudes in \eqref{intro:2ptshadowdS0}-\eqref{intro:shadowdSexch0}. Such a duality however is not possible as the relations \eqref{intro:2ptshadowdS0}-\eqref{intro:shadowdSexch0} cannot be renormalised. In contrast, as we have already mentioned, \eqref{intro:Amp2}-\eqref{intro:Amp4x} hold at the level of renormalised correlators. To see why there is no renormalised version of the shadow formulae, it suffices to discuss a counterexample.
Consider the case of $ds_{[322]}$. This dS amplitude has IR singularities and can be renormalised as discussed earlier, yielding
\begin{equation}
\dsren_{[322]}  = \frac{1}{4 q_1^3 q_2 q_3} \left\{ - q_1 + (q_2 + q_3) \left[ \log \left( \frac{q_t}{\mu} \right) + \act_{[322]}^{(1)} - 1 \right] \right\}.
\end{equation}
The corresponding shadow AdS amplitude has $(\bar{\Delta}_1,\bar{\Delta}_2,\bar{\Delta}_3)=(0,1,1)$ and reads
\begin{equation}
    \iren_{[011]} =c_{[011]} \frac{(q_2+q_3)}{q_1^3 q_2 q_3},
\end{equation}
where we have allowed for arbitrary 3-point constant $c_{[011]}$, and the  coefficient of proportionality on the right-hand side of \eqref{intro:3ptshadowdS0} is equal to $
\mathcal{C}_{[011]}/(2 \bar{\beta}_1 \mathcal{C}_{[00]} (2 \bar{\beta}_2 \mathcal{C}_{[11]})^2)=1/12$.
We conclude that \eqref{intro:3ptshadowdS0} cannot hold at the level of renormalised correlators. In fact, the right-hand side of \eqref{intro:3ptshadowdS0} can only account for the scheme-dependent part of the renormalised dS correlator. This is typical: when renormalisation is needed, the shadow formulae compute only the scheme-dependent part of the renormalised correlator. Note that the disagreement is not just on the details: the left-hand side transforms anomalously under dS isometries while the right-hand side transforms covariantly. While we focused here on one counterexample, the same conclusion can be reached for any correlators that requires renormalisation. The underlying reason is that the shadow CFT does not have suitable local counterterms built from sources and operators of the shadow dimensions to remove the infinities one encounters on the dS side. In contrast, the original CFT has precisely the counterterms needed to remove these infinities. 

This is a long paper and to accommodate readers with different interests the different sections have been written so as to allow them to be read independently of each other. 
In section \ref{sec: AdS_to_dS}, we describe the relation between the AdS partition function and the dS wavefunction. The section starts with two subsections outlining the AdS and dS computations, presented in parallel. This serves both to make the paper self-contained and also to emphasise the similarity of the computations. To facilitate the discussion of analytic continuation, we explicitly include  the Planck and (A)dS lengths in all formulae. Then, we show how the seemingly different analytic continuations that have appeared in the literature are related to each other, and  we obtain the precise relation between the coefficients of the dS wavefunction and CFT correlators, to all orders in the bulk perturbation theory. This section finishes with the tree-level holographic formulae that relate dS in-in  and CFT correlators. 

Section \ref{sec:SK_approach} is devoted to the Schwinger-Keldysh formalism. We present the computation of the tree-level dS in-in  correlators using the Schwinger-Keldysh path integral, and derive their relation with the corresponding AdS amplitudes confirming the holographic formulae derived in the previous section. This section is self-contained and does not assume prior knowledge of the Schwinger-Keldysh approach.  

In section \ref{sec:regandren}, we discuss the regularisation and renormalisation of tree-level dS divergences. We start by reviewing how holographic renormalisation works in AdS and outline the similarities and differences with the case of de Sitter. We use dimensional regularisation to regulate the IR divergences. The discussion is  illustrated throughout for the cases of massless and conformal scalars. We tabulate and explain the degree of divergences of AdS and dS diagrams: generally all such diagrams are IR divergent; typically AdS ones are more divergent than dS ones, but there are counterexamples. We discuss in complete detail the renormalisation of correlators of a single massless scalar in dS and its compatibility with the renormalisation of corresponding AdS amplitudes, and we present the corresponding results for all such correlators involving both massless and conformal scalars. We briefly discuss weight-shift operators and bulk derivative couplings.
Amplitudes constructed using derivative couplings are less singular than the corresponding amplitudes without derivative interactions, but not all of them are finite.  

In section \ref{sec: shadow}, we discuss a recent approach that aims to connect de Sitter amplitudes to AdS amplitudes involving operators of the shadow dimension \cite{Sleight:2020obc, DiPietro:2021sjt, Sleight:2021plv}. We start by showing that the dS Ward identities acting on the late-time correlators take the form of conformal Ward identities in one dimension less, acting on operators of the shadow dimension. We then show by explicit computation that one may rewrite the regularised tree-level holographic formulae in terms of AdS amplitudes involving shadow fields, and explain that the resulting structure is essentially dictated by conformal symmetry and the singularity structure of the correlators. We show however that the connection between dS correlators and shadow AdS amplitudes breaks down when renormalisation is needed.

The paper also contains four appendices. In appendix \ref{sec:notation} we summarise our notation, and in appendix \ref{1pt_app} we summarise the holographic derivation of the 1-point functions in the presence of sources. The AdS discussion is well known, and here we also present the analogous discussion for dS. Appendix \ref{app:sings} contains a derivation of the singularities of AdS and dS contact and exchange diagrams for general values of the operator and spacetime dimensions. Finally, appendix \ref{app: shadow} lists the shadow relations up to 4-point functions and briefly discusses their derivation via a Legendre transform.

\section{Relating the AdS partition function to the dS wavefunction} \label{sec: AdS_to_dS}

In this section we review the  AdS partition function and the dS wavefunction, showing how their perturbative expansions are naturally related by analytic continuation.

\subsection{Perturbative expansion of the Euclidean AdS partition function}

We consider the Euclidean bulk scalar field action 
\begin{align}\label{EAdSaction}
S_{AdS} &= (\ell_{P}^{(AdS)})^{1-d}\int\D^{d+1} x\sqrt{g}\,\Big(\frac{1}{2}(\partial\vphi)^2+\frac{1}{2}m_{AdS}^2\vphi^2 +(\ell_P^{(AdS)})^{-2}  V_{int}(\vphi)\Big),
\end{align}
where factors of the Planck length $\ell_P^{(AdS)}$ have been introduced so that $\vphi$ and the interaction potential $V_{int}(\vphi)$ (containing terms of cubic order and higher) are both dimensionless.  This convention is convenient in supergravity where $\vphi$ parametrises a dimensionless metric component in a Kaluza-Klein reduction, and will turn out to simplify our later analytic continuation to dS.  In order to understand the form of this continuation when working in {\it either}
Planck units ($\ell_P^{(AdS)}=\ell_P^{(dS)}=1$) {\it or} (A)dS units ($L_{AdS}=L_{dS}=1$) we will keep both these quantities explicit throughout.  As we will see, the different analytic continuation prescriptions proposed in the holographic cosmology literature  simply reflect   
 this choice of units.
 As one of these continuations involves the Planck length, we have chosen to explicitly differentiate this constant in the AdS set-up ($\ell_P^{(AdS)}$) from its counterpart in dS ($\ell_P^{(dS)}$).
Finally, while for simplicity we focus initially on a single scalar field, our results will later generalise to multiple interacting scalars.

On a $(d+1)$-dimensional Euclidean AdS (EAdS) background
\[ \label{AdSmetric}
\D s_{AdS}^2 = \frac{L_{AdS}^2}{z^2}(\D z^2+\D \x^2),
\]
where the radial coordinate $0<z<\infty$, 
the action reduces to
\begin{align}\label{AdSaction1}
S_{AdS} &=
\frac{1}{2}\Big(\frac{L_{AdS}}{\ell_P^{(AdS)}}\Big)^{d-1}\int_0^{\infty}\frac{\D z}{z^{{d+1}}}\, \int\D^{d} \bs{x}\,\Big(
z^2(\partial_{z}\vphi)^2+z^2 (\partial_i\vphi)^2 \nn\\&\qquad \qquad\qquad\qquad\qquad \qquad\qquad\quad+m_{AdS}^2L_{AdS}^2 \vphi^2+\Big(\frac{L_{AdS}}{\ell_P^{(AdS)}}\Big)^2 2V_{int}(\vphi)\Big).
\end{align}
With mass
\[\label{AdSmass}
m_{AdS}^2 = \Delta(\Delta-d)L_{AdS}^{-2},
\]
the asymptotic behaviour of the scalar field is 
\begin{align}\label{AdSasympt}
\vphi(z,\x) = z^{d-\Delta}\vphi_{(0)}(\x) +\ldots +z^\Delta \vphi_{(\Delta)}(\x)+\ldots, \qquad z\rightarrow 0^+
\end{align} 
where the dimensions $[\vphi_{(0)}]=d-\Delta$ and $[\vphi_{(\Delta)}]=\Delta$ correspond respectively to those of  the source and vev of a dual scalar operator $\O$ of dimension $\Delta$. 

The EAdS partition function is given by the path integral
\[
Z_{AdS}[\vphi_{(0)}] = \int\mathcal{D}\vphi\, e^{-S_{AdS}}\, ,
\]
subject to the boundary conditions
\[ \label{def:source}
\lim_{z\rightarrow\infty}\vphi(z,\x)\ \ {\rm regular}, \qquad \lim_{z\rightarrow 0} z^{\Delta-d}\vphi(z,\x) = \vphi_{(0)}(\x),
\]
and may be expanded into boundary CFT correlators as
\[\label{ZAdSdef}
Z_{AdS}[\vphi_{(0)}] 
=\exp\Big(\sum_{n=2}^\infty \frac{(-1)^n}{n!}\int [\D\bs{q}_n]
\lla \O(\bs{q}_1)\ldots \O(\bs{q}_n)\rra\, \vphi_{(0)}(-\bs{q}_1)\ldots\vphi_{(0)}(-\bs{q}_n)\Big).
\]
Here, $\vphi_{(0)}$ acts as the source for a dual scalar operator $\O$ of dimension $\Delta$ in a $d$-dimensional Euclidean CFT and we have Fourier transformed along the boundary directions so that $\bs{q}_i$ is the momentum conjugate to the insertion $\x_i$.
The measure $[\D\bs{q}_n]$ is
\[
[\D\bs{q}_n] =\Big(\prod_{i=1}^n \frac{\D^d \bs{q}_i}{(2\pi)^d}\Big)\,(2\pi)^d\delta\big(\sum_{j=1}^n\bs{q}_j\big)\, ,
\label{measure}
\]
and the double brackets represent the stripped correlators
\[ \label{eq:stripped}
\<O(\bs{q}_1)\ldots \O(\bs{q}_n)\> = \lla\O(\bs{q}_1)\ldots \O(\bs{q}_n)\rra \,(2\pi)^d\delta\big(\sum_{j=1}^n\bs{q}_j\big).
\]
In the saddle point approximation, the 1-point function in the presence of sources is \cite{deHaro:2000vlm}
\[\label{1ptfn}
\<\O\>_s = -(2\Delta-d) \Big(\frac{L_{AdS}}{\ell_P^{(AdS)}}\Big)^{d-1}\vphi_{(\Delta)}.
\]
The derivation of this formula is reviewed in appendix \ref{1pt_app}.
Correlation functions in the CFT then follow by repeated functional differentiation with respect to the source,
\[\label{fndiff}
\<\O(\bs{q}_1)\ldots \O(\bs{q}_n)\> =\prod_{i=2}^n \Big(-\frac{\delta}{\delta\vphi_{(0)}(-\bs{q}_i)}\Big)\<\O(\bs{q}_1)\>_s \Big|_{\vphi_{(0)}\rightarrow 0}.
\]
To determine $\vphi_{(\Delta)}$ as a function of $\vphi_{(0)}$ we must solve the bulk equation of motion
\begin{align}
\label{AdSeom}
(-\Box+m_{AdS}^2)\vphi&=
L_{AdS}^{-2}\Big(-z^2\partial_z^2 + (d-1)z\partial_z + z^2 q^2+\Delta(\Delta-d)\Big)\vphi \\
&= -(\ell_P^{(AdS)})^{-2}\partial_\vphi V_{int}. \nonumber
\end{align}
This can be accomplished perturbatively through an expansion in Witten diagrams.
For concreteness, let us consider the interaction 
\[
V_{int}(\vphi) = \frac{1}{k!}\lambda_k\vphi^k.
\] 
Expanding the bulk scalar in powers of the coupling
\[\label{fieldexpinlambda}
\vphi(\bs{q},z) = \sum_{j=0}^\infty (\lambda_k)^j \vphi_{\{j\}}(\bs{q},z),
\]
the solution of the bulk equations of motion is then given by 
\begin{align}\label{AdSinteqn}
\vphi(\bs{q}_1,z) &= \vphi_{\{0\}}(\bs{q}_1,z) + \lambda_k \vphi_{\{1\}}(\bs{q}_1,z) + \cdots ,\nn \\[1ex]
\vphi_{\{0\}}(\bs{q}_1,z) &= \mathcal{K}^{AdS}_\Delta(q_1,z)\vphi_{(0)}(-\bs{q}_1) \, ,\\[1ex]
\vphi_{\{1\}}(\bs{q}_1,z)& =\frac{-1}{(k-1)!} \Big(\frac{L_{AdS}}{\ell_P^{(AdS)}}\Big)^{d+1}
\int_0^\infty\frac{\D z'}{z'^{d+1}} \mathcal{G}^{AdS}_\Delta(q_1;z,z')\nn\\&\qquad\qquad\qquad  \times \Big(
\prod_{i=2}^k\int\frac{\D^d\bs{q}_i}{(2\pi)^d} 
\vphi_{\{0\}}(-\bs{q}_i,z')\Big)   (2\pi)^d\delta(\sum_{j=1}^k\bs{q}_j).\nn
\end{align}
Here, the bulk-to-boundary propagator $\mathcal{K}^{AdS}_\Delta(q,z)$ satisfies
\[\label{AdSKeom}
(-\Box+m_{AdS}^2)\mathcal{K}^{AdS}_\Delta(q,z) = 0,
\]
with boundary conditions
\[
\mathcal{K}^{AdS}_\Delta(q,z)\rightarrow \begin{cases} z^{d-\Delta} \,\, &\mathrm{as}\,\, z\rightarrow 0\\ 0\,\,&\mathrm{as}\,\,z\rightarrow \infty\end{cases}
\]
so that 
\[\label{AdSK}
\mathcal{K}^{AdS}_\Delta(q,z) = \frac{z^{d/2}q^\beta }{2^{\beta-1}\Gamma(\beta)}K_\beta(q z), \qquad \beta=\Delta-d/2, 
\]
where $q=+\sqrt{\bs{q}^2}$ and $K_\beta$, $I_\beta$ denote  modified Bessel functions.  
The bulk-to-bulk propagator $\mathcal{G}^{AdS}_\Delta(q;z,z')$  satisfies
\[\label{AdSGeom}
(-\Box+m_{AdS}^2)\mathcal{G}^{AdS}_\Delta(q;z,z') = (\ell_P^{(AdS)})^{d-1}\frac{1}{\sqrt{g}}\delta(z-z'),
\]
with boundary conditions
\[\label{GAdSbdy}
\mathcal{G}^{AdS}_\Delta(q;z,z')\rightarrow\begin{cases} \dfrac{z^\Delta}{2\beta}\,\Big(\dfrac{L_{AdS}}{\ell_P^{(AdS)}}\Big)^{1-d}\mathcal{K}^{AdS}_\Delta(q,z')\, \,&\mathrm{as}\,\, z\rightarrow 0\\[1ex]\quad 0 \,\,&\mathrm{as}\,\,z\rightarrow\infty\end{cases}
\]
such that
\[\label{AdSG}
\mathcal{G}^{AdS}_\Delta(q;z,z') = 
\Big(\frac{L_{AdS}}{\ell_P^{(AdS)}}\Big)^{1-d} (z z')^{d/2}\Big[ K_\beta(pz)I_\beta(p z')\Theta(z-z')
+(z\leftrightarrow z')\Big].
\]

Witten diagrams are then constructed from bulk-to-bulk propagators and bulk-to-boundary propagators with their sources $\vphi_{(0)}(\bs{q})$ stripped off.  Vertices correspond to radial integrations with the factor
\[\label{AdSvertex}
\mathcal{V}_k^{AdS} = -\frac{\lambda_k}{(k-1)!} \Big(\frac{L_{AdS}}{\ell_P^{(AdS)}}\Big)^{d+1}\int_0^\infty\frac{\D z'}{z'^{d+1}}
\]
and momentum conservation enforced.
Summing up all $n$-point diagrams to a given order in $\lambda_k$, the correlator is then found by multiplying by $(-1)^n$ to account for the signs in \eqref{fndiff}.
Notice also that the factor of $(2\beta)^{-1}(L_{AdS}/\ell_P^{(AdS)})^{1-d}$ in $\mathcal{G}_\Delta^{AdS}(q;z,z')$ as $z\rightarrow 0$ cancels with that in \eqref{1ptfn}.

The various factors of $L_{AdS}$ and $\ell_P^{(AdS)}$ entering the propagators can be accounted for as follows.
The AdS radius enters the bulk-to-bulk propagator \eqref{AdSG} through \eqref{AdSGeom}, which reads
\[\label{AdSGreens}
\Big(-z^2\partial_z^2 + (d-1)z\partial_z + z^2 q^2+\Delta(\Delta-d)\Big)\mathcal{G}_\Delta^{AdS}(q;z,z') =\Big(\frac{L_{AdS}}{\ell_P^{(AdS)}}\Big)^{1-d}z^{d+1}\delta(z-z').
\]
The factor of $(\ell_{P}^{(AdS)})^{d-1}$ follows from the normalisation of the bulk action (the bulk-to-bulk propagator being the inverse of the operator in the quadratic part of the action). Note however that this factor
has no effect on the correlators since the vertices carry cancelling factors arising from the integral equation \eqref{AdSinteqn}.
By contrast, the bulk-to-boundary propagator \eqref{AdSK} obeys  the  homogeneous equation  \eqref{AdSKeom} from which all factors of $L_{AdS}$  cancel.   The boundary condition  $\mathcal{K}^{AdS}_{\Delta}(q,z)\rightarrow z^{d-\Delta}$ as $z\rightarrow 0$ then ensures the asymptotic behaviour matches that for a source in \eqref{AdSasympt} and prohibits any factors of $L_{AdS}$ or $\ell_{P}^{(AdS)}$.
A final point that will be relevant later when we consider analytically continuing to de Sitter spacetime is that  the boundary metric on which the dual CFT lives is
\[\label{bdymetric}
\D s^2 = L_{AdS}^2\, \D \x^2  ,
\]
as follows from (\ref{AdSmetric}).

\subsection{Perturbative expansion of the  dS wavefunction}

We now review the corresponding perturbative expansion for the de Sitter wavefunction, which is closely modelled on that for AdS above.  The relation between the dS wavefunction and the actual observables of interest --  the correlation functions on late-time slices --  will be examined subsequently in section \ref{dScorrsec}.

We start with the Lorentzian bulk action 
\begin{align}\label{dSaction}
S_{dS} &=- (\ell_{P}^{(dS)})^{1-d}\int\D^{d+1} x\sqrt{-g}\,\Big(\frac{1}{2}(\partial\vphi)^2+\frac{1}{2}m_{dS}^2\vphi^2 +(\ell_P^{(dS)})^{-2}  V_{int}(\vphi)\Big),
\end{align}
evaluated on the fixed $(d+1)$-dimensional de Sitter background
\[ \label{eq:dSmetric}
\D s_{dS}^2 = \frac{L_{dS}^2}{\tau^2}(-\D \tau^2+\D \x^2),
\]
where the conformal time $\tau$ takes values in the range $-\infty<\tau<0$.  The overall minus sign in  \eqref{dSaction} relative to \eqref{EAdSaction}, along with the additional sign in the metric, ensure that the action is  the kinetic energy  minus the energy in field gradients and the potential as required in Lorentzian signature:
\begin{align}\label{dSaction2}
S_{dS} &=
\frac{1}{2}\Big(\frac{L_{dS}}{\ell_P^{(dS)}}\Big)^{d-1}\int_{-\infty}^0\frac{\D \tau}{(-\tau)^{{d+1}}}\, \int\D^{d} \bs{x}\,\Big(
\tau^2(\partial_{\tau}\vphi)^2-\tau^2 (\partial_i\vphi)^2 \nn\\&\qquad \qquad\qquad\qquad\qquad \qquad\qquad\quad-m_{dS}^2L_{dS}^2\vphi^2-\Big(\frac{L_{dS}}{\ell_P^{(dS)}}\Big)^2 2V_{int}(\vphi)\Big).
\end{align}
Notice the measure factor $\sqrt{-g} = (-\tau/L_{dS})^{-(d+1)}$ also contains a sign ensuring it remains positive in even boundary dimension $d$.
The asymptotic behaviour of the scalar field at late times is 
\begin{align}
\vphi(\tau,\x) = (-\tau)^{d-\Delta}\vphi_{(0)}(\x)+\ldots + (-\tau)^\Delta \vphi_{(\Delta)}(\x)+\ldots
,\qquad \tau\rightarrow 0^-.\label{dSasympt}
\end{align}
The equations of motion ensure there is an additional minus sign in the relation between mass and $\Delta$ relative to that in AdS \eqref{AdSmass},
\[
m_{dS}^2 =- \Delta(\Delta-d)L_{dS}^{-2}.
\]
The de Sitter wavefunction is given by the path integral
\[
\Psi_{dS}[\vphi_{(0)}] = \<\vphi_{(0)}(\x)|0\> = \int\mathcal{D}\vphi \,e^{iS_{dS}}
\]
subject to the boundary conditions
\[
\lim_{\tau\rightarrow -\infty} \vphi(\tau,\x)\ \ {\rm regular}, \qquad \lim_{\tau\rightarrow 0^-} (-\tau)^{\Delta-d}\vphi(\tau,
\x) = \vphi_{(0)}(\x)
\]
with the standard $i\ep$ prescription
\[\label{tauiep}
\tau \rightarrow \tau(1-i\ep), \qquad 0<\ep\ll 1.
\]
This corresponds to an infinitesimal rotation of the time integration contour for the classical action, where the direction is chosen so that $i$ times the Lorentzian action  continues to minus the corresponding Euclidean action, $i S_{dS}=-S_{dS}^E$.
In perturbation theory, the role of this $i\ep$ prescription is to
select the solution where $\vphi\sim e^{iq\tau}$ as $\tau\rightarrow -\infty$ so that we obtain the exponential suppression $e^{iq\tau(1-i\ep)}\rightarrow 0$.  Notice that in the in-in formalism, this mode where $\vphi\sim e^{iq\tau}$ as $\tau\rightarrow-\infty$ corresponds to that multiplying the {\it creation} operator in the Bunch-Davies vacuum. 
Instead of rotating the time contour, we can equivalently replace 
\[\label{qiep}
q \rightarrow q- i \ep,
\] 
in all momentum magnitudes $q=+\sqrt{\bs{q}^2}$.  This produces the same exponential suppression of propagators at early times since $e^{i\tau (q-i\ep)}$
 as $\tau\rightarrow -\infty$  is equivalent to 
 $e^{iq\tau(1-i\ep)}$.

The full dS wavefunction $\Psi_{dS}$ can be expanded perturbatively in powers of $\vphi_{(0)}$ to define the wavefunction coefficients $\psi_n$.  The most convenient definition for our purposes is 
\[
\Psi_{dS}[\vphi_{(0)}]  = \exp\Big(\sum_{n=2}^\infty \frac{(-1)^n}{n!}\int [\D \bs{q}_n] \psi_n(\bs{q}_1,\ldots, \bs{q}_n) \vphi_{(0)}(-\bs{q}_1)\ldots\vphi_{(0)}(-\bs{q}_n)\Big) \label{psiexp}
\]
where the measure is given in \eqref{measure} and we include an explicit factor of $(-1)^n$ multiplying the wavefunction coefficients for later convenience.   
In the saddle point approximation, the latter may be computed through a diagrammatic expansion analogous to Witten diagrams.  
First, we construct an analogue of the 1-point function in the presence of sources
\[\label{dS1ptfn}
\psi_s(\bs{q}) =-\frac{\delta \ln \Psi_{dS}}{\delta\vphi_{(0)}(-\bs{q})}= +i(2\Delta-d)
\Big(\frac{L_{dS}}{\ell_P^{(dS)}}\Big)^{d-1}\vphi_{(\Delta)}(\bs{q})
\]
such that the wavefunction coefficients are obtained by further functional differentiation,
\[
\psi_n(\bs{q}_1,\ldots,\bs{q}_n) = \prod_{i=2}^n \Big(-\frac{\delta}{\delta \vphi_{(0)}(-\bs{q}_i)}\Big)\psi_s(\bs{q}_1)\Big|_{\vphi_{(0)}\rightarrow 0}.
\]
Comparing \eqref{dS1ptfn} to the corresponding AdS formula \eqref{1ptfn}, the relative factor of $-i$ derives from the fact that  $\Psi_{dS} =\<e^{iS_{dS}}\>$ while $Z_{AdS}=\<e^{-S_{AdS}}\>$.  The full derivation is given in appendix \ref{1pt_app}.

To construct the bulk solution perturbatively, we again expand the bulk field in powers of the coupling as in \eqref{fieldexpinlambda}.
The solution of the bulk equation of motion 
\[\label{dSeom}
(-\Box+m_{dS}^2)\vphi=
L_{dS}^{-2}\Big(\tau^2\partial_\tau^2-(d-1)\tau\partial_\tau+\tau^2q^2-\Delta(\Delta-d)\Big)\vphi = -(\ell_P^{(dS)})^{-2}\partial_\vphi V_{int}.
\]
is then given by 
\begin{align}\label{dSintegraleqn}
\vphi(\bs{q}_1,\tau) &= \vphi_{\{0\}}(\bs{q}_1,z) + \lambda_k \vphi_{\{1\}}(\bs{q}_1,z) + \cdots \, ,\nn \\[2ex]
\vphi_{\{0\}}(\bs{q}_1,z)& =\mathcal{K}^{dS}_\Delta(q_1,\tau)\vphi_0(-\bs{q}_1), \\[2ex]
\vphi_{\{1\}}(\bs{q}_1,z)&=\frac{ -i }{(k-1)!}\Big(\frac{L_{dS}}{\ell_P^{(dS)}}\Big)^{d+1}\int^0_{-\infty}\frac{\D \tau'}{(-\tau')^{d+1}}\mathcal{G}^{dS}_\Delta(q_1;\tau,\tau')\nn\\&
\qquad \qquad \qquad \times
\Big(
\prod_{i=2}^k\int\frac{\D^d\bs{q}_i}{(2\pi)^d} 
\vphi_{\{0\}}(-\bs{q}_i,\tau')\Big)  (2\pi)^d\delta(\sum_{j=1}^k\bs{q}_j).\nn
\end{align}
Here, the bulk-to-boundary propagator $\mathcal{K}^{dS}_\Delta(q,\tau)$ satisfies
\[
(-\Box+m_{dS}^2)\mathcal{K}^{dS}_\Delta(q,\tau) = 0, 
\]
with boundary conditions
\[\label{dSbdycondsK}
\mathcal{K}^{dS}_\Delta(q,\tau)\rightarrow \begin{cases} (-\tau)^{d-\Delta} \,\, &\mathrm{as}\,\, \tau\rightarrow 0\\ 0\,\,&\mathrm{as}\,\,\tau\rightarrow -\infty(1-i\ep)
\end{cases}
\]
such that 
\[\label{dSK}
\mathcal{K}^{dS}_\Delta(q,\tau) = -i\pi\frac{ (-\tau)^{d/2}q^\beta}{2^{\beta}\Gamma(\beta)}H^{(2)}_\beta(-q \tau), \qquad \beta=\Delta-d/2.
\]
Note that at early times the Hankel function  $H^{(2)}_\beta(-q \tau)\sim (-q\tau)^{-1/2}e^{iq\tau}$.  As in AdS, the bulk-to-boundary propagators $\mathcal{K}^{dS}_\Delta(q,\tau)$ are taken to be independent of $\ell_P^{(dS)}$.

The bulk-to-bulk propagator $\mathcal{G}^{dS}_\Delta(q;\tau,\tau')$ satisfies
\[\label{dSGeqn}
(-\Box+m_{dS}^2)\mathcal{G}^{dS}_\Delta(q;\tau,\tau') =-i (\ell_P^{(dS)})^{d-1}\frac{1}{\sqrt{-g}}\delta(\tau-\tau')
\]
where the normalisation is such that  $\mathcal{G}^{dS}_\Delta(q;\tau,\tau')$ carries a factor of $(\ell_P^{(dS)})^{d-1}$ matching the normalisation of the bulk action. 
The boundary conditions are 
\[\label{GdSbdy}
\mathcal{G}^{dS}_\Delta(q;\tau,\tau')\rightarrow\begin{cases} \dfrac{+i}{2\beta}\,(-\tau)^\Delta\Big(\dfrac{L_{dS}}{\ell_P^{(dS)}}\Big)^{1-d}\mathcal{K}^{dS}_\Delta(q,\tau')\, \,&\mathrm{as}\,\, \tau\rightarrow 0\\[1ex] \quad 0 \,\,&\mathrm{as}\,\,\tau\rightarrow-\infty(1-i\ep)\end{cases}
\]
leading to the unique solution 
\begin{align}\label{dSG}
\mathcal{G}^{dS}_\Delta(q;\tau,\tau') &= +\frac{\pi}{4}\,
\Big(\frac{L_{dS}}{\ell_P^{(dS)}}\Big)^{1-d} (-\tau)^{d/2}(-\tau')^{d/2}\nn\\&\quad \times \Big[H_\beta^{(2)}(-q\tau)\Big(H_\beta^{(1)}(-q\tau')+H_\beta^{(2)}(-q\tau')\Big)\Theta(\tau'-\tau)+ (\tau\leftrightarrow\tau')\Big].
\end{align}
Here, the relative normalisation of the terms proportional to $H_\beta^{(1)}(-q\tau)$ and $H_\beta^{(2)}(-q\tau)$ for $\tau>\tau'$ is fixed by the required asymptotic behaviour $\mathcal{G}^{dS}_\Delta(q;\tau,\tau') \sim (-\tau)^{\Delta}$ as $\tau\rightarrow 0$.
The dependence on $L_{dS}$ 
enters via the inhomogeneous source term in \eqref{dSGeqn}, namely
\[\label{dSGreens}
\Big(\tau^2\partial_\tau^2-(d-1)\tau\partial_\tau+\tau^2q^2-\Delta(\Delta-d)\Big)\mathcal{G}_\Delta^{dS}(q;\tau,\tau') =-i \Big(\frac{L_{dS}}{\ell_P^{(dS)}}\Big)^{1-d}(-\tau)^{d+1}\delta(\tau-\tau').
\]
The overall sign in \eqref{dSG} is fixed by the junction condition following from \eqref{dSGreens}, namely
\[\lim_{\ep\rightarrow 0^+}
\Big[\partial_\tau\mathcal{G}^{dS}_\Delta(q;\tau,\tau')\Big]^{\tau=\tau'+\ep}_{\tau=\tau'-\ep} =-i \Big(\frac{L_{dS}}{\ell_P^{(dS)}}\Big)^{1-d}(-\tau')^{d-1}.
\]

To solve the  integral equation \eqref{dSintegraleqn} perturbatively, diagrams are then constructed from the bulk-to-bulk and bulk-to-boundary propagators, with vertices corresponding to conformal time integrals 
\[\label{dSvertex}
\mathcal{V}_k^{dS} =-i\frac{\lambda_k}{(k-1)!}\Big(\frac{L_{dS}}{\ell_P^{(dS)}}\Big)^{d+1}\int^0_{-\infty}\frac{\D \tau'}{(-\tau')^{d+1}}.
\]
As usual, momentum conservation is enforced at every vertex.

Examining \eqref{GdSbdy}, 
the factor of $i(2\beta)^{-1}(L_{dS}/\ell_P^{(dS)})^{1-d}$  in $\mathcal{G}^{dS}_\Delta(q;\tau,\tau')$ as $\tau\rightarrow 0$ cancels up to a sign with a corresponding factor appearing in \eqref{dS1ptfn}. 
The sum of $n$-point diagrams constructed using  
bulk-to-boundary propagators (with sources stripped off) for external legs therefore corresponds to $(-1)^n \psi_n$.
To construct the (log of) the full wavefunction $\Psi_{dS}$, we instead retain and integrate over the sources, multiplying the sum of $n$-point diagrams by $1/n!$.

\subsection{Analytic continuation}

The individual diagrams arising in the perturbative expansions of the AdS partition function and the dS wavefunction can be identified with one another through a suitable analytic continuation of the propagators.  
This ensures that to all orders in perturbation theory
\[\label{ZPsi}
Z_{AdS}\Big|_{\mathrm{analyt.\, cont.}} = \Psi_{dS}
\]
allowing AdS/CFT correlators to be related to wavefunction coefficients, and ultimately dS correlators as we will review.
The precise  form of the analytic continuation depends however on whether we choose to work in Planck units (fixing $\ell_P^{(AdS)}=\ell_P^{(dS)}=1$) or (A)dS units (fixing $L_{AdS}=L_{dS}=1$), although the physical content is the same in either case.

\subsubsection{Planck units}  

In Planck units where $\ell_P^{(AdS)}=\ell_P^{(dS)}=1$, the AdS and dS solutions are related by 
\[\label{Planckcont}
z=-i\tau, \qquad L_{AdS} = i L_{dS}, \qquad \vphi_{(0)}^{AdS} = (-i)^{d-\Delta}\vphi_{(0)}^{dS},
\]
with all other quantities the same for both.  
Here, the sign of the first continuation is fixed by matching the large-$z$ behaviour $e^{-qz}$ of the AdS propagators to the early-time $e^{iq\tau}$ behaviour of the dS propagators for the wavefunction. 
The continuation of the sources $\vphi_{(0)}^{(A)dS}$ then follows by matching the respective asymptotic expansions \eqref{AdSasympt} and \eqref{dSasympt}.
The continuation of the (A)dS radius is fixed by requiring $\sqrt{g_{AdS}} = \sqrt{-g_{dS}}$ and hence $(L_{AdS}/z)^{d+1} = (L_{dS}/(-\tau))^{d+1}$, ensuring that $-S_{AdS}=iS_{dS}$ in all dimensions $d$.\footnote{Following \cite{Maldacena:2002vr}, many authors use instead the measure 
$\sqrt{-g_{dS}}=(L_{dS}/\tau)^{d+1}$ leading to the  continuation $L_{AdS} = -iL_{dS}$.
This is fine for odd boundary dimensions (including the physically relevant case of $d=3$) but cannot be applied for even $d$ since the dS action becomes imaginary for $\tau<0$.  In contrast, the continuation \eqref{Planckcont} is valid in any dimension.}

Applying \eqref{Planckcont} to the AdS propagators \eqref{AdSK} and \eqref{AdSG}, and comparing to the dS propagators \eqref{dSK} and \eqref{dSG}, we find
\begin{align}\label{Kcontplanck}
\mathcal{K}_{\Delta}^{AdS}(q,z)&= (-i)^{\Delta-d} \mathcal{K}_{\Delta}^{dS}(q,\tau),\\[1ex]
\mathcal{G}^{AdS}_\Delta(q;z,z') &=
+\mathcal{G}^{dS}_\Delta(q;\tau,\tau').  \label{Gcontplanck}
\end{align}
Here we used the following analytic continuations, valid for $-\frac{\pi}{2}\le \mathrm{arg}\,x\le \pi$,
\begin{align}
K_\beta(x) &= -\frac{i\pi}{2}e^{-i\pi\beta/2}H_\beta^{(2)}(x e^{-i\pi/2}), \\
I_\beta(x) &= \frac{1}{2}e^{i\pi\beta/2}\Big(H_\beta^{(1)}(xe^{-i\pi/2})
+H_\beta^{(2)}(xe^{-i\pi/2})\Big),
\end{align}
with the replacement $\theta(z-z')\rightarrow \theta(|z|-|z'|)=\theta(|\tau|-|\tau'|)\rightarrow \theta(\tau'-\tau)$ and vice versa. 
From the result \eqref{Kcontplanck}, we see that the phase acquired by the bulk-boundary propagator cancels with that acquired by the source giving
\[\label{Kphicont}
\mathcal{K}_{\Delta}^{AdS}(q,z)\vphi_{(0)}^{AdS}(-\bs{q}) = \mathcal{K}_{\Delta}^{dS}(q,z)\vphi_{(0)}^{dS}(-\bs{q}).
\]
Next, we must consider the interactions.
Under \eqref{Planckcont}, the AdS classical equation of motion \eqref{AdSeom} continues directly to its dS counterpart \eqref{dSeom}.  Similarly,  the AdS vertex factor \eqref{AdSvertex}
continues to its dS counterpart  \eqref{dSvertex} as\footnote{Here, after performing the analytic continuation $z'=-i\tau'$, we reverse the limits of the time integration and rotate the contour sending $\int_0^{i\infty}\D\tau' =-\int^0_{i\infty}\D\tau' = -\int^0_{-\infty(1-i\ep)}\D\tau'$.  The rotation is permitted since the integral along the arc at infinity where $\vphi\sim e^{iq\tau'}$ vanishes in the upper half-plane. }
\[\label{vertexcont1}
\mathcal{V}_k^{AdS} = \mathcal{V}_k^{dS}.
\]
Overall, on continuing a diagram from AdS to dS via \eqref{Planckcont}, we thus find that: 
(i) bulk-boundary propagators acquire a factor of $(-i)^{\Delta-d}$ when stripped of boundary sources, but  have no factors when dressed according to   \eqref{Kphicont}; 
(ii) from \eqref{Gcontplanck} and \eqref{vertexcont1}, both bulk-bulk propagators and vertices continue exactly.

Each AdS diagram, with sources removed, therefore continues to $(-i)^{n(\Delta-d)}$ times the corresponding dS diagram.  Alternatively, the dS diagram is the continued AdS diagram times the inverse of this factor, $i^{n(\Delta-d)}$.
Since the sum of $n$-point AdS diagrams is $(-1)^n$ times the corresponding CFT correlator, and the sum of dS diagrams is $(-1)^n$ times the wavefunction coefficient, we find
\[\label{corrtopsi2}
 \psi_n(\bs{q}_1,\ldots,\bs{q}_n)= (-i)^{n(d-\Delta)}
\lla \O(\bs{q}_1)\ldots \O(\bs{q}_n)\rra \Big|_{L_{AdS} \rightarrow iL_{dS}}.
\]
This formula holds provided there are no divergences: we will return to discuss such cases later in section \ref{sec:regandren}.
If we construct the AdS partition function \eqref{ZAdSdef} and analytically continue via \eqref{Planckcont}, the factors from the  continuation of the sources cancel with those in \eqref{corrtopsi2} and we obtain
\[\label{ZtoPsi1}
\Psi_{dS}=
Z_{AdS}\Big|_{L_{AdS}\rightarrow iL_{dS},\,\,\vphi_{(0)}^{AdS}\rightarrow (-i)^{d-\Delta}\vphi_{(0)}^{dS}}
\]
to all loop orders.

\subsubsection{From Planck to AdS units via a Weyl transformation}
\label{Weylsec}

As we noted earlier, when working in Planck units with $\ell_P^{(AdS)}=\ell_P^{(dS)}=1$, the metric on which the CFT lives is $\D s^2 = L_{AdS}^2\, \D\x^2$.  When working in AdS units with $L_{(A)dS}=1$, however, we have just $\D s^2 = \D\x^2$.   From the perspective of the CFT, passing from Planck to AdS units is thus equivalent to performing a Weyl transformation  \cite{Garriga:2014fda}
\[
\gamma_{ij} \rightarrow L_{AdS}^{-2}\gamma_{ij}
\]
under which \[\label{Weyltrans}
q\rightarrow L_{AdS}q, \qquad
\vphi_{(0)}^{AdS}\rightarrow L_{AdS}^{d-\Delta} \vphi_{(0)}^{AdS}
\]
and the correlators transform as 
\[\label{Weyltrans2}
\lla \O(\bs{q}_1)\ldots \O(\bs{q}_n)\rra
\rightarrow L_{AdS}^{n(\Delta-d)+d}\lla \O(\bs{q}_1)\ldots \O(\bs{q}_n)\rra  
\]
according to their overall dimension $\Delta_t-(n-1)d=n(\Delta-d)+d$.

From the perspective of the Weyl-transformed theory in AdS units, the continuation \eqref{Planckcont} is then equivalent to continuing 
\[\label{AdSunitscont}
 q_{AdS} =i q_{dS}, \qquad  \vphi_{(0)}^{AdS}= \vphi_{(0)}^{dS},\qquad \ell_P^{(AdS)} = -i\ell_P^{(dS)}.
\]
The first two formulae here follow from \eqref{Weyltrans}, 
where in addition all correlators are multiplied by a factor $i^{n(\Delta-d)+d}$ from \eqref{Weyltrans2}.
The continuation of $\ell_P^{(AdS)}$ in the final formula derives from the presence of an overall power of $L_{AdS}/\ell_P^{(AdS)}$ in the bulk action, which translates to a power of the rank of the gauge group in the dual CFT. 
As in our conventions $L_{AdS}$ and $\ell_P^{(AdS)}$ only  appear in the ratio $L_{AdS}/\ell_P^{(AdS)}$,
the effect of the continuation $L_{AdS}=iL_{dS}$ in Planck units is thus reproduced by the continuation $\ell_P^{(AdS)} = -i\ell_P^{(dS)}$ in AdS units. 

The relation \eqref{corrtopsi2} between the wavefunction and CFT correlators  when working in Planck units can now be translated into an equivalent statement in AdS units.  In AdS units, we find
\begin{align}\label{corrtopsi3}
\psi_n(\bs{q}_1,\ldots,\bs{q}_n)
&= (-i)^{n(d-\Delta)}  
(-i)^{n(\Delta-d)+d}\lla \O(\bs{q}_1)\ldots \O(\bs{q}_n)\rra \Big|_{\ell_{P}^{(AdS)}\rightarrow -i\ell_P^{(dS)},\,\,q_{AdS} \rightarrow iq_{dS}}\nn\\
& = (-i)^{d} \lla \O(\bs{q}_1)\ldots \O(\bs{q}_n)\rra \Big|_{\ell_{P}^{(AdS)}\rightarrow -i\ell_P^{(dS)},\,\,q_{AdS} \rightarrow iq_{dS}}
\end{align}
where all correlators are those in the Weyl transformed theory on metric $\D s^2 = \D\x^2$. 
In the first line, the factor $ (-i)^{n(\Delta-d)+d}$ multiplying the transformed correlator is equivalent to the untransformed correlator appearing in \eqref{corrtopsi2}.  Again, this result holds in the absence of divergences; we will return to discuss these in section \ref{sec:regandren}.

\subsubsection{Pure (A)dS units perspective}

Instead of applying a Weyl transformation to the holographic formula in derived in Planck units, as we did in section \ref{Weylsec}, we can also recover the relation \eqref{corrtopsi3} between CFT correlators and dS wavefunction coefficients by working purely in (A)dS units.

From this perspective, setting $L_{(A)dS}=1$,  
the continuation \eqref{AdSunitscont} can be understood as follows. 
First,  we must have $q_{AdS}^2 = -q_{dS}^2$ and $(\ell_P^{(AdS)})^2=-(\ell_P^{(dS)})^2$ in order to map
the AdS equation of motion \eqref{AdSeom} to its dS counterpart \eqref{dSeom}.
The specific signs appearing in \eqref{AdSunitscont} are then fixed by enforcing $-S_{AdS}=iS_{dS}$.  In addition, this matches the $e^{-q_{AdS}z}$ behaviour  of the AdS propagators at large $z$ with the $e^{iq_{dS}\tau}$ behaviour of the propagator for the dS wavefunction at early times.\footnote{
In  \cite{McFadden:2009fg}, we instead continued the AdS solution to the dS mode function, namely, the coefficient of the annihilation operator in the in-in dS mode expansion.  In contrast, the wavefunction propagator corresponds to the coefficient of the {\it creation} operator in the dS mode expansion.  The signs in the  continuation \eqref{AdSunitscont}  therefore differ from those in \cite{McFadden:2009fg}, but the holographic formulae we derive are such that all results for observables  
remain the same.
In the present context, the continuation of \cite{McFadden:2009fg}
 amounts to mapping $Z_{AdS} = \Psi_{dS}^*$ which leads to the same results for observables since dS correlators are  constructed from $|\Psi_{dS}|^2$. \label{ft:an_cont}}
The sign here is also consistent with the direction of rotation implied by the dS $i\ep$ prescription  $q_{dS}\rightarrow q_{dS}-i\ep$ in \eqref{qiep}, since \eqref{AdSunitscont} implies $q_{dS} = -iq_{AdS}$ which also has a negative imaginary part.
Finally, since  the AdS and dS asymptotic expansions \eqref{AdSasympt} and \eqref{dSasympt} match directly under \eqref{AdSunitscont}, no  continuation of the sources is required and  $\vphi_{(0)}^{AdS} = \vphi_{(0)}^{dS}$.  Thus, all the continuations implied by the Weyl transformation argument in section \ref{Weylsec} can equivalently be recovered from consideration of the theory in (A)dS units alone.

Applying \eqref{AdSunitscont} to the AdS propagators \eqref{AdSK} and \eqref{AdSG}, and comparing to the dS propagators \eqref{dSK} and \eqref{dSG}, we find
\begin{align}
\mathcal{K}_{\Delta}^{AdS}(q_{AdS},z)&=\mathcal{K}_{\Delta}^{dS}(q_{dS},\tau),\\[1ex]
\mathcal{G}^{AdS}_\Delta(q_{AdS};z,z') &= (-i)^{d}
\mathcal{G}^{dS}_\Delta(q_{dS};\tau,\tau').
\end{align}
The vertex factors \eqref{AdSvertex} and \eqref{dSvertex} are proportional to $(\ell_P^{((A)dS)})^{-(d+1)}$, so overall we find
\[
\mathcal{V}_k^{AdS} = (-i)^{-d}\mathcal{V}_k^{dS}.
\]
In addition, we also need to take into account the continuation of the momentum integrals.  For the diagrams themselves, the only surviving momentum integrals come from loops, contributing an overall factor of $i^{dL}$.  
Since
\[\label{IVrel}
I-V= L-1
\]
for a connected diagram with $V$ vertices, $I$ internal propagators and $L$ loops,
 the continuation of an individual AdS diagram generates its dS counterpart multiplied by an overall factor of
\[
(-i)^{d(I-V-L)} =(-i)^{-d}.\nn
\]
Equivalently, to obtain the dS diagram we apply the continuation to the AdS diagram and multiply by the inverse of this factor.
As previously, the sum of $n$-point AdS diagrams corresponds to $(-1)^n$ times the  CFT correlator while the sum of dS diagrams gives $(-1)^n$ times the wavefunction coefficient. 
Putting  this together, we obtain the relation
\begin{align}\label{corrtopsi5}
\psi_n(\bs{q}_1,\ldots,\bs{q}_n)&=
 (-i)^{d}\lla \O(\bs{q}_1)\ldots \O(\bs{q}_n)\rra \Big|_{\ell_{P}^{(AdS)}\rightarrow -i\ell_P^{(dS)},\,q_{AdS} \rightarrow iq_{dS}}
\end{align}
This agrees with our earlier formula \eqref{corrtopsi3}.

\subsubsection{2-point function}

The 2-point function provides a quick check of the  continuations above.
Using the bulk-boundary propagators along with \eqref{1ptfn} and \eqref{dS1ptfn}, we find 
\begin{align}\label{2ptOO}
\lla\O(\bs{q})\O(-\bs{q})\rra = \frac{2\beta \Gamma(-\beta)}{4^\beta \Gamma(\beta)} \Big(\frac{L_{AdS}}{\ell_P^{(AdS)}}\Big)^{d-1} q_{AdS}^{2\beta} 
\end{align}
and
\begin{align}
\psi_2(q_{dS})= -ie^{i\pi\beta}  \frac{2\beta \Gamma(-\beta)}{4^\beta \Gamma(\beta)} \Big(\frac{L_{dS}}{\ell_P^{(dS)}}\Big)^{d-1} q_{dS}^{2\beta} ,
\end{align}
allowing us to verify the continuations \eqref{corrtopsi2} and \eqref{corrtopsi3} above. 
For the particular case of $d=\Delta=3$, note that 
\[
\lla\O(\bs{q})\O(-\bs{q})\rra = \Big(\frac{L_{AdS}}{\ell_P^{(AdS)}}\Big)^2 q_{AdS}^3,\qquad \psi_2(q) = - \Big(\frac{L_{dS}}{\ell_P^{(dS)}}\Big)^2 q_{dS}^3.
\]
The negative sign of $\psi_2(q_{dS})$ is consistent with the non-negativity of the dS 2-point function.\footnote{See \eqref{dSpsi1}, and also (5.8) in \cite{Maldacena:2002vr}; the sign is misprinted in (A.3) of \cite{Garriga:2014fda}.} 
Using 
\begin{align}
\ln Z_{AdS} = 
\frac{1}{2}\int\frac{\mathrm{d}^d \bs{q}}{(2\pi)^d}\,\lla\O(\bs{q})\O(-\bs{q})\rra\vphi_{(0)}^{AdS}(\bs{q})\vphi_{(0)}^{AdS}(-\bs{q})
\end{align}
and
\begin{align}
\ln \Psi_{dS}=\frac{1}{2}   \int\frac{\mathrm{d}^d \bs{q}}{(2\pi)^d}\psi_2(q_{dS}) \vphi_{(0)}^{dS}(\bs{q})\vphi_{(0)}^{dS}(-\bs{q}).
\end{align}
we can further check the relation \eqref{ZtoPsi1}.

\subsection{Holographic formulae for dS correlators}
\label{dScorrsec}

The  observables of interest in inflationary cosmology are the late-time correlation functions.  
These may be computed from the wavefunction coefficients via a path integral:
\begin{align} \label{eq: wavefn}
\<\vphi_{(0)}(\bs{x}_1)\ldots \vphi_{(0)}(\bs{x}_n)\>&=\lim_{\tau\rightarrow 0^-} (-\tau)^{n(\Delta-d)}\<\vphi(\bs{x}_1)\ldots \vphi(\bs{x}_n)\>\nn\\[1ex]&=\int\mathcal{D}\vphi_{(0)}\, \vphi_{(0)}(\bs{x}_1)\ldots \vphi_{(0)}(\bs{x}_n)
|\Psi[\vphi_{(0)}]|^2.
\end{align} 
Via standard path integral calculations we then obtain\footnote{Notice the signs in these formulae are dependent on those in the definition of the wavefunction coefficients \eqref{psiexp}, for which different conventions exist in the literature.} 
\begin{align}\label{dSpsi1}
\lla\vphi_{(0)}(\bs{q})\vphi_{(0)}(-\bs{q})\rra &=- \frac{1}{2}\frac{1}{\re\psi_2(q)},\\[2ex]\label{dSpsi2}
\lla\vphi_{(0)}(\bs{q}_1)\vphi_{(0)}(\bs{q}_2)\vphi_{(0)}(\bs{q}_3)\rra &=\frac{1}{4}\frac{\re\psi_3(\bs{q}_1,\bs{q}_2,\bs{q}_3)}{\prod_{i=1}^3\re\psi_2(q_i)}, \\[2ex]
\label{dSpsi3}
\lla\vphi_{(0)}(\bs{q}_1)\vphi_{(0)}(\bs{q}_2)\vphi_{(0)}(\bs{q}_3)\vphi_{(0)}(\bs{q}_4)\rra &
=\frac{1}{8}\Big[\frac{\re\psi_4(\bs{q}_1,\bs{q}_2,\bs{q}_3,\bs{q}_4)}{\prod_{i=1}^4\re\psi_2(q_i)}\nn
\\[2ex]& \qquad \hspace{-3cm}
-\Big(\frac{\re\psi_3(\bs{q}_1,\bs{q}_2,\bs{q}_{12})\re\psi_3(-\bs{q}_{12},\bs{q}_3,\bs{q}_4)}{\re\psi_2(q_{12})\prod_{i=1}^4\re\psi_2(q_i)}
+(2\leftrightarrow 3) + (2\leftrightarrow 4)\Big)\Big]
\end{align}
where $\bs{q}_{ij}=\bs{q}_i+\bs{q}_j$.

Working in (A)dS units, the analytic continuation between CFT correlators and wavefunction coefficients is given by \eqref{corrtopsi3}.  Combining this with the relations \eqref{dSpsi3} between wavefunction coefficients and dS correlators, we obtain the holographic formulae
\begin{align}\label{2ptholformula}
\lla\vphi_{(0)}(\bs{q})\vphi_{(0)}(-\bs{q})\rra &=- \frac{1}{2}\frac{1}{\re [ (-i)^d \lla\O(\bs{q})\O(-\bs{q})\rra]},\\[2ex]
\lla\vphi_{(0)}(\bs{q}_1)\vphi_{(0)}(\bs{q}_2)\vphi_{(0)}(\bs{q}_3)\rra &=\frac{1}{4}\frac{\re [(-i)^d\lla \O(\bs{q}_1)\O(\bs{q}_2)\O(\bs{q}_3)\rra]}{\prod_{i=1}^3\re [(-i)^d \lla\O(\bs{q}_i)\O(-\bs{q}_i)\rra]}, \\[2ex]
\lla\vphi_{(0)}(\bs{q}_1)\vphi_{(0)}(\bs{q}_2)\vphi_{(0)}(\bs{q}_3)\vphi_{(0)}(\bs{q}_4)\rra  &
=\frac{1}{8}\Big[\frac{\re [(-i)^d \lla\O(\bs{q}_1)\O(\bs{q}_2)\O(\bs{q}_3)\O(\bs{q}_4)\rra]}{\prod_{i=1}^4\re [(-i)^d \lla\O(\bs{q}_i)\O(-\bs{q}_i)\rra]}\nn
\\[2ex]&  \hspace{-6.0cm}
-\Big(\frac{\re [(-i)^d\lla\O(\bs{q}_1)\O(\bs{q}_2)\O(\bs{q}_{12})\rra] \re [(-i)^d\lla\O(-\bs{q}_{12})\O(\bs{q}_3)\O(\bs{q}_4)\rra] }{\re [(-i)^d \lla\O(\bs{q}_{12})\O(-\bs{q}_{12})\rra] \prod_{i=1}^4\re [(-i)^d \lla\O(\bs{q}_i)\O(-\bs{q}_i)\rra]}
+(2\leftrightarrow 3) + (2\leftrightarrow 4)\Big)\Big]
\label{4ptholformula}
\end{align}
where the real parts of all correlators on the right-hand sides are taken after continuing 
\[\label{acgen}
\ell_{P}^{(AdS)}=-i\ell_P^{(dS)},\qquad q_{AdS}= iq_{dS}.
\]
While our discussion has focused on a scalar field of a single type, the generalisation to multiple interacting scalar fields is straightforward: the continuations of propagators and vertices is the same and one arrives at the formulae above where the correlators on both sides are generalised so that each momentum is associated with a specific operator.

Finally, let us discuss the domain of validity of the various results above. 
While the relation \eqref{corrtopsi3} between CFT correlators and wavefunction coefficients holds for any diagram, and hence to all orders in perturbation theory, the relations \eqref{dSpsi1}-\eqref{dSpsi3} between wavefunction coefficients and dS correlators holds only to leading order in $\ell_P^{(dS)}/L_{dS}$.
   (Recall the dS action \eqref{dSaction} carries an overall factor of $(L_{dS}/\ell_P^{(dS)})^{d-1}$.) 
The holographic formula \eqref{2ptholformula}-\eqref{4ptholformula} are then valid to leading order in the large-$N$ expansion of the dual CFT.

While here we have been discussing a fixed gravitational background for simplicity, formulae analogous to \eqref{2ptholformula}-\eqref{4ptholformula} have been derived for  fully dynamical backgrounds at up to 3-points \cite{McFadden:2009fg,McFadden:2010vh,McFadden:2011kk}.  In this more general setting, 
when the dual CFT is strongly interacting, the bulk gravity 
 theory is weakly coupled ({\it i.e.,} Einstein gravity holds and higher-curvature corrections are suppressed).  Leading order in $\ell_P^{((A)dS)}/L_{(A)dS}$ is then equivalent to  tree level in the bulk loop expansion.  However, the holographic duality is also expected to hold in the opposite regime where the CFT is weakly interacting (though still at large $N$) and the bulk (super)gravity approximation breaks down.  Holographic formulae analogous to  \eqref{2ptholformula}-\eqref{4ptholformula} should still be applicable in this limit, though on the bulk side  leading order in $\ell_P^{((A)dS)}/L_{(A)dS}$ now corresponds to an expansion in string loops.

\section{Schwinger-Keldysh approach} \label{sec:SK_approach}

The Schwinger-Keldysh or in-in formalism provides an alternative approach for computing cosmological correlators.  
To arrive at this formalism, we first rewrite the wavefunction of the universe and its complex conjugate as
\[
\Psi[\vphi_{(0)}] = \int\mathcal{D}\vphi_+ e^{iS_+[\vphi_+]},\qquad
\Psi^*[\vphi_{(0)}] = \int\mathcal{D}\vphi_- e^{-iS_-[\vphi_-]},
\]
where $\vphi_+(\tau,\x)$ and $\vphi_-(\tau,\x)$ are regarded as independent  fields whose classical actions $S_{\pm}[\vphi_\pm]$ are evaluated subject to contour rotations $\tau\rightarrow \tau(1\mp i\ep)$ respectively. 
Both path integrals run over bulk field configurations subject to a common late-time boundary condition
\[ \label{def_phi0}
\lim_{\tau\rightarrow 0^-}\Big[ (-\tau)^{\Delta-d}\vphi_{\pm}(\tau,\x) \Big]=\vphi_{(0)}(\x).
\]
Performing a second path integral over the boundary field $\vphi_{(0)}(\x)$ then yields the de Sitter correlators as previously,
\begin{align}
\< \vphi_{(0)}(\x_1)\ldots\vphi_{(0)}(\x_n)\> &= \lim_{\tau\rightarrow 0^-}\Big[(-\tau)^{n(\Delta-d)}\< \vphi(\tau,\x_1)\ldots\vphi(\tau,\x_n)\> \Big]\nn\\&
=\int\mathcal{D}\vphi_{(0)} \,\Big(\prod_{i=1}^n\vphi_{(0)}(\x_i)\Big)\big|\Psi[\vphi_{(0)}]\big|^2.
\end{align}
In the Schwinger-Keldysh formalism these separate path integrals over bulk and boundary fields are merged into a single closed-time path integral
\begin{align}\label{SKformalism}
\< \vphi(\tau,\x_1)\ldots\vphi(\tau,\x_n)\> 
=\int\mathcal{D}\vphi_+\mathcal{D}\vphi_-\,\Big(\prod_{i=1}^n\vphi_{+}(\tau,\x_i)\Big)\exp\Big(iS_+[\vphi_+]-iS_{-}[\vphi_-]\Big),
\end{align}
where we path integrate over all field configurations subject only to the constraint that both fields coincide at late times
\[
\lim_{\tau\rightarrow 0^-}\vphi_+(\tau,\x)=\lim_{\tau\rightarrow 0^-}\vphi_-(\tau,\x).
\]
In  this closed-time path formalism $\vphi_+$ then corresponds to a field localised on the forward part of the contour while $\vphi_-$ is localised on the reverse part.
As previously, the asymptotic time-dependence of the correlator can be removed by multiplying by a factor of $(-\tau)^{n(\Delta-d)}$ and (in the absence of divergences) taking the limit $\tau\rightarrow 0^-$.

From the Schwinger-Keldysh perspective, we now obtain four different propagators according to the identity $\vphi_{\pm}$ of the two end-points, and vertices come in two different types (see \cite{Chen:2017ryl} for a review).
The pay-off for this increase in complexity is that the resulting diagrammatic expansion now computes correlators in de Sitter  directly, in contrast to the wavefunction formalism where an additional path integral over boundary fields is required to go from the wavefunction to the correlators.

In this section we show how the holographic formulae derived above using the wavefunction formalism can also be obtained by directly continuing diagrams in the Schwinger-Keldysh formalism.  
In particular, this requires continuing $\vphi_\pm$ in different directions \cite{Sleight:2020obc, Sleight:2021plv}.  However, the derivation of the holographic formulae is arguably less straightforward in this approach since, as noted by many authors \cite{Sleight:2020obc, Sleight:2021plv, DiPietro:2021sjt}, the AdS bulk-bulk propagator does not continue directly to the Schwinger-Keldysh bulk-bulk propagators.  In contrast, the AdS bulk-bulk propagator does continue directly to the bulk-bulk propagator used for computing the wavefunction coefficients as we saw above.  For this reason, we will restrict the Schwinger-Keldysh analysis in this section to tree-level correlators, unlike the wavefunction derivation which applied to all orders in perturbation theory.

\subsection{Amplitudes} 

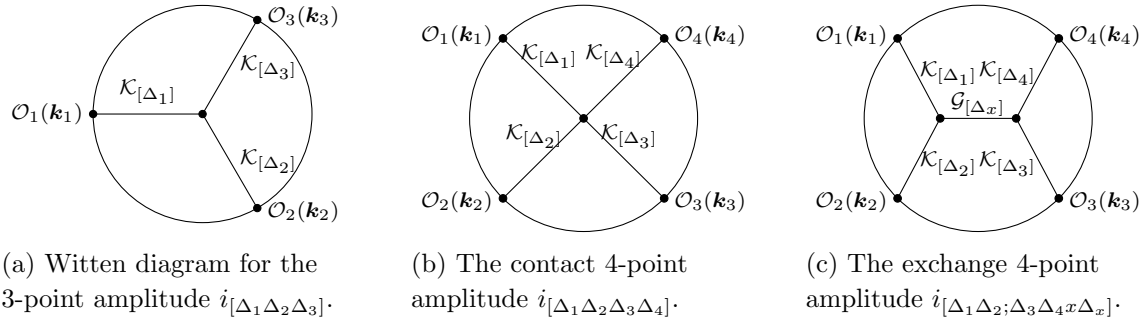
\begin{figure}[t]
\centering
\begin{subfigure}[t]{0.3\textwidth}
\centering
\begin{tikzpicture}[scale=0.48]
\draw (0,0) circle [radius=3];
\draw [fill=black] (-3,0) circle [radius=0.1];
\draw [fill=black] (1.5,2.598) circle [radius=0.1];
\draw [fill=black] (1.5,-2.598) circle [radius=0.1];
\draw [fill=black] ( 0, 0) circle [radius=0.1];
\draw (-3,0) -- (0,0) -- (1.5,2.598);
\draw (0,0) -- (1.5,-2.598);
\node [left] at (-3,0) {$\scriptstyle \O_1(\bs{k}_1)$}; 
\node [right] at (1.5,2.7) {$\scriptstyle \O_3(\bs{k}_3)$};
\node [right] at (1.5,-2.7) {$\scriptstyle \O_2(\bs{k}_2)$};
\node [above] at (-1.5,0) {$\scriptstyle \mathcal{K}_{[\Delta_1]}$};
\node [right] at (0.75,1.299) {$\scriptstyle \mathcal{K}_{[\Delta_3]}$};
\node [right] at (0.75,-1.299) {$\scriptstyle \mathcal{K}_{[\Delta_2]}$};
\end{tikzpicture}
\caption{Witten diagram for the\\ 3-point amplitude $\ino_{[\Delta_1 \Delta_2 \Delta_3]}$.
\label{fig:AdS3pt}}
\end{subfigure}
\hspace{6mm}
\begin{subfigure}[t]{0.3\textwidth}
\centering
\begin{tikzpicture}[scale=0.5]
\draw (0,0) circle [radius=3];
\draw [fill=black] (-2.121,-2.121) circle [radius=0.1];
\draw [fill=black] (-2.121, 2.121) circle [radius=0.1];
\draw [fill=black] ( 2.121,-2.121) circle [radius=0.1];
\draw [fill=black] ( 2.121, 2.121) circle [radius=0.1];
\draw [fill=black] ( 0, 0) circle [radius=0.1];
\draw (-2.121,-2.121) -- ( 2.121, 2.121);
\draw ( 2.121,-2.121) -- (-2.121, 2.121);	
\node [left] at (-2.121, 2.2) {$\scriptstyle \O_1(\bs{k}_1)$}; 
\node [left] at (-2.121,-2.2) {$\scriptstyle \O_2(\bs{k}_2)$}; 	
\node [right] at ( 2.121,-2.2) {$\scriptstyle \O_3(\bs{k}_3)$}; 
\node [right] at ( 2.121, 2.2) {$\scriptstyle \O_4(\bs{k}_4)$}; 	
\node [above] at (-0.9, 1.06) {$\scriptstyle \mathcal{K}_{[\Delta_1]}$};
\node [above] at (-1.3,-1.06) {$\scriptstyle \mathcal{K}_{[\Delta_2]}$};
\node [above] at ( 1.2,-1.06) {$\scriptstyle \mathcal{K}_{[\Delta_3]}$};
\node [above] at ( 0.8, 1.06) {$\scriptstyle \mathcal{K}_{[\Delta_4]}$};
\end{tikzpicture}
\caption{
The contact 4-point \\amplitude $\ino_{[\Delta_1 \Delta_2 \Delta_3 \Delta_4]}$.\label{fig:AdS4ptC}}
\end{subfigure}
\hspace{4mm}
\begin{subfigure}[t]{0.3\textwidth}
\centering
\begin{tikzpicture}[scale=0.5]
\draw (0,0) circle [radius=3];
\draw [fill=black] (-2.121,-2.121) circle [radius=0.1];
\draw [fill=black] (-2.121, 2.121) circle [radius=0.1];
\draw [fill=black] ( 2.121,-2.121) circle [radius=0.1];
\draw [fill=black] ( 2.121, 2.121) circle [radius=0.1];
\draw [fill=black] (-1, 0) circle [radius=0.1];
\draw [fill=black] ( 1, 0) circle [radius=0.1];
\draw (-2.121,-2.121) -- (-1,0) -- (-2.121, 2.121);
\draw ( 2.121, 2.121) -- ( 1,0) -- ( 2.121,-2.121);
\draw (-1,0) -- (1,0);
\node [left] at (-2.121, 2.2) {$\scriptstyle \O_1(\bs{k}_1)$}; 
\node [left] at (-2.121,-2.2) {$\scriptstyle \O_2(\bs{k}_2)$}; 	
\node [right] at ( 2.121,-2.2) {$\scriptstyle \O_3(\bs{k}_3)$}; 
\node [right] at ( 2.121, 2.2) {$\scriptstyle \O_4(\bs{k}_4)$}; 	
\node [right] at (-1.8, 1.2) {$\scriptstyle \mathcal{K}_{[\Delta_1]}$};
\node [right] at (-1.8, -1.2) {$\scriptstyle \mathcal{K}_{[\Delta_2]}$};
\node [left] at ( 1.8, -1.2) {$\scriptstyle \mathcal{K}_{[\Delta_3]}$};
\node [left] at ( 1.8, 1.2) {$\scriptstyle \mathcal{K}_{[\Delta_4]}$};
\node [above] at (0,-0.2) {$\scriptstyle \mathcal{G}_{[\Delta_x]}$};
\end{tikzpicture}
\caption{
The exchange 4-point \\amplitude $\ino_{[\Delta_1 \Delta_2; \Delta_3 \Delta_4 x \Delta_x]}$.\label{fig:AdS4ptX}}
\end{subfigure}
\caption{AdS Witten diagrams defining the corresponding AdS amplitudes. We use $\O_i$ to denote the dual operators, while $\mathcal{K}_{[\Delta]}$ and $\mathcal{G}_{[\Delta]}$ denote the bulk-to-boundary and bulk-to-bulk  propagators respectively. For precise expressions defining these amplitudes, see section 3 of \cite{Bzowski:2022rlz}.\label{fig:AdSamps}}
\end{figure}

Restricting our analysis to tree level enables some further simplification of the holographic formulae \eqref{2ptholformula}-\eqref{4ptholformula}. At tree level each correlator in the dual theory can be expressed as a sum of AdS Witten digrams multiplied by a number of constants, such as coupling constants, the AdS radius and the Planck length. By \emph{AdS amplitudes} we refer to the momentum-dependent part of each Witten diagram, with all couplings dropped (or set to one). In \cite{Bzowski:2022rlz} we defined and listed expressions for a number of interesting AdS amplitudes.  Figure \ref{fig:AdSamps} shows the notation used for each amplitude and its defining Witten diagram.

In our conventions,  every AdS diagram  carries an overall factor\footnote{Note \eqref{scaling} differs from  the usual expectation that  diagrams scale as $(L_{AdS}/\ell_P^{(AdS)})^{(d-1)(1-L)}$.  However, the latter  holds when the interaction potential is $L_{AdS}^{-2}V_{int}(\vphi)$ rather than 
$(\ell_P^{(AdS)})^{-2}V_{int}(\vphi)$ as here.} 
\[\label{scaling}
\Big(\frac{L_{AdS}}{\ell_P^{(AdS)}}\Big)^{(d+1)V+(1-d)I}
=\Big(\frac{L_{AdS}}{\ell_P^{(AdS)}}\Big)^{(d-1)(1-L)+2V},
\]
where $L$ is the number of loops, $V$ the number of vertices and $I$ the number of internal lines.  
The power $(d-1)(1-L)$  derives from the overall factor of $(L_{AdS}/\ell_P^{(AdS)})^{d-1}$ multiplying the bulk action \eqref{AdSaction1}, just as an ordinary Feynman diagram in QFT is of order $\hbar^{L-1}=(\hbar^{-1})^{1-L}$ since the action carries a factor of $\hbar^{-1}$ in the path integral.  The remaining power of $2V$ arises since the interaction potential  in \eqref{AdSaction1} carries a factor of $(L_{AdS}/\ell_P^{(AdS)})^{2}$ and we get one such factor per diagrammatic vertex. Thus, we can decompose each correlator in terms of AdS amplitudes. Schematically, we can write
\begin{align} \label{AdS_amp_def}
\lla \O(\bs{q}_1)\ldots \O(\bs{q}_n)\rra = \sum_V \Big(\frac{L_{AdS}}{\ell_P^{(AdS)}}\Big)^{(d-1) + 2V} \lambda_{AdS}^V \, \ino_V(\bs{q}_1, \ldots, \bs{q}_n).
\end{align}
Here $\ino_V$ represents a single Witten diagrams with $n$ external legs and $V$ vertices in the bulk. There is usually many such diagrams and they must be summed over. Similarly, each vertex comes with a coupling, and there may be many types of vertices/couplings. For simplicity we use only a single coupling here, denoted by $\lambda_{AdS}$.

Analogously, we can define the dS amplitudes, but carrying out the corresponding, tree-level calculations in de Sitter space. Thus, we can decompose the dS correlators as
\begin{align} \label{dS_amp_def}
\lla  \jsrc(\bs{q}_1) \ldots \jsrc(\bs{q}_n) \rra = \sum_V \Big(\frac{L_{dS}}{\ell_P^{(dS)}}\Big)^{(d-1) + 2V} \lambda_{dS}^V \, \dsno_V(\bs{q}_1, \ldots, \bs{q}_n).
\end{align}
We can now use the formulae \eqref{2ptholformula}-\eqref{4ptholformula} to relate the dS and AdS amplitudes on the tree level.

In (A)dS units $L_{(A)dS}=1$, 
we see now that the continuation $\ell_P^{(AdS)}\rightarrow -i \ell_P^{(dS)}$ generates a factor of $(-i)^{(1-d)(1-L)-2V}$ for a given diagram.  At tree level ($L=0$), this reduces to $-i (-i)^{-d} (-1)^V$ and the factor of $(-i)^{-d}$ cancels that in \eqref{corrtopsi3}. 
Thus in effect we can continue just the momentum alone:
\begin{align}
\re \psi_n^{tree}(\bs{q}_1,\ldots,\bs{q}_n)
 = (-1)^{V} \im \lla \O(\bs{q}_1)\ldots \O(\bs{q}_n)\rra \Big|_{q_{AdS} \rightarrow iq_{dS}}
\end{align}
The factor of $(-1)^V$ can then be introduced into the holographic formulae explicitly, since at tree level $V=1$ for contact diagrams and $V=2$ for 4-point exchanges. All in all, at tree level, we can continue only the amplitudes, keeping all remaining constants identified\footnote{Alternatively, this factor of $(-1)^V$ could be generated by continuing the couplings $\lambda^{AdS}_k \rightarrow -\lambda_k^{dS}$.},
\begin{align}
& L_{AdS} = L_{dS}, && \ell_P^{(AdS)} = \ell_P^{(dS)}, && \lambda_k^{AdS} = \lambda_k^{dS}.
\end{align}
The relation between the amplitudes becomes then
\begin{align}\label{holform1}
\dsno_{[\Delta \Delta]}(q) & = - \frac{1}{2 \im \ino_{[\Delta \Delta]}(\I q)}, \\
\dsno_{[\Delta_1 \Delta_2 \Delta_3]}(q_1, q_2, q_3) & = - \frac{1}{4} \frac{\im \ino_{[\Delta_1 \Delta_2 \Delta_3]}(\I q_1, \I q_2, \I q_3)}{\prod_{j=1}^3 \im \ino_{[\Delta_j \Delta_j]}(\I q_j)}, \\
\dsno_{[\Delta_1 \Delta_2 \Delta_3 \Delta_4]}(q_i) & = - \frac{1}{8} \frac{\im  \ino_{[\Delta_1 \Delta_2 \Delta_3 \Delta_4]}(\I q_i)}{\prod_{j=1}^4 \im \ino_{[\Delta_j \Delta_j]}(\I q_j)}, \\
\dsno_{[\Delta_1 \Delta_2; \Delta_3 \Delta_4 x \Delta_x]}(q_i, s) & = \frac{1}{8} \prod_{j=1}^4 \frac{1}{\im \ino_{[\Delta_j \Delta_j]}(\I q_j)} \left[ \im \ino_{[\Delta_1 \Delta_2; \Delta_3 \Delta_4 x \Delta_x]}(\I q_i, \I s) \right.\nn\\
& \qquad \left. - \, \frac{\im \ino_{[\Delta_1 \Delta_2 \Delta_x]}(\I q_1, \I q_2, \I s) \im \ino_{[\Delta_x \Delta_3 \Delta_4]}(\I s, \I q_3, \I q_4)}{\im \ino_{[\Delta_x \Delta_x]}(\I s)} \right],
\label{holform4}
\end{align}
where $\ino_{[\Delta \Delta]}(q)$ is the regulated AdS 2-point function; $\ino_{[\Delta_1 \Delta_2 \Delta_3]}(q_1, q_2, q_3)$ the AdS 3-point function; $\ino_{[\Delta_1\Delta_2 \Delta_3 \Delta_4]}(q_1,q_2,q_3,q_4)$ the AdS 4-point contact and $ \ino_{[\Delta_1 \Delta_2; \Delta_3 \Delta_4 x \Delta_x]}(q_j, s)$ the AdS 4-point $s$-channel exchange diagram for a field of dimension $\Delta_x$.
As $L_{(A)dS}$ and $\ell_P^{(A)dS}$ are no longer being continued, we can effectively set $L_{(A)dS}=1$ and $\ell_P^{(A)dS}=1$, with the only analytic continuations being those of the momenta indicated directly in the formulae.

\subsection{Free field} 

Let us begin with the analysis of the free scalar field $\j$ governed by the action,
\begin{align}
S_0 & = - \frac{1}{2} \int \D^{d+1} x \sqrt{-g} \left[ g^{\mu\nu} \partial_\mu \j \partial_\nu \j + m_{dS}^2 \j^2 \right], \label{S0} 
\end{align}
on the fixed $(d+1)$-dimensional de Sitter background with the metric
\begin{align} \label{gdS}
\D s^2 = \frac{1}{(-\tau)^2} ( - \D \tau^2 + \D \bs{x}^2 ),
\end{align}
where $\tau$ goes from $-\infty$ to $0$. We parameterise the mass as
\begin{align}
m_{dS}^2 = \frac{d^2}{4} - \beta^2 = - \Delta(\Delta - d)
\end{align}
and consider only the case of $0 < \beta \leq \frac{d}{2}$.

This is the free part of the system described by the de Sitter action \eqref{dSaction} with $L_{dS} = \ell_{P}^{(dS)} = 1$. One can recover the $L_{dS}$ and $\ell_{P}^{(dS)}$ dependence by decorating each Witten diagram (amplitude) with the factor \eqref{scaling}, where $V$ denotes the number of vertices in the diagram (and we consider only tree diagrams with $L = 0$). In particular by stripping the 2-point function $\lla \jsrc \jsrc \rra$ from its $L_{dS}$ and $\ell_{P}^{(dS)}$ we obtain the dS amplitude $\dsno_{[\Delta \Delta]}$ as its purely momentum-dependent part,
\begin{align} \label{2pt_to_ds}
\lla \jsrc(\bs{q}) \jsrc(-\bs{q}) \rra = \Big( \frac{L_{dS}}{\ell^{(dS)}_P} \Big)^{d-1} \dsno_{[\Delta \Delta]}(q) + O(\lambda_j).
\end{align}
If we added some bulk interactions, denoted here collectively by the coupling constants $\lambda_j$, the 2-point function would acquire additional contributions from bulk loops. Nevertheless, by $\dsno_{[\Delta \Delta]}$ we always denote the zeroth order, tree-level amplitude; in this case the free theory amplitude.

\subsubsection{Mode decomposition}

We write the mode decomposition
\begin{align} \label{Phi_decomp}
\j(\tau, \bs{x}) = \int \frac{\D^d \bs{q}}{(2 \pi)^d} \left[ a_{\bs{q}} \j_{\bs{q}}(\tau, \bs{x}) + a^{\ast}_{\bs{q}} \j^{\ast}_{\bs{q}}(\tau, \bs{x}) \right],
\end{align}
where each modes satisfies the Klein-Gordon equation $(-\Box + m_{dS}^2) \j_{\bs{q}} = 0$. The negative frequency solution $\j_{\bs{q}}$ reads
\begin{align}
\j_{\bs{q}}(\tau, \bs{x}) & = e^{\I \bs{q} \cdot \bs{x}} \varphi_{\bs{q}}(\tau), \\
\varphi_{\bs{q}}(\tau) & = \frac{\sqrt{\pi}}{2} e^{\frac{\I \pi}{2} (\beta + \frac{1}{2})} (-\tau)^{\frac{d}{2}} H^{(1)}_{\beta}(- q \tau),
\end{align}
where $H^{(1)}_{\beta}$ denotes the Hankel function and we chose the phase such that in the far past
\begin{align} \label{past_behaviour}
\lim_{\tau \rightarrow -\infty} \varphi_{\bs{q}}(\tau) \sim (-\tau)^{\frac{d-1}{2}} \frac{e^{-\I q \tau}}{\sqrt{2 q}}.
\end{align}
The normalisation is such that with respect to the Klein-Gordon scalar product
\begin{align}
(\j, \psi) = - \I (-\tau)^{1-d} \int \D^d \bs{x} \left[ \j \, \partial_\tau \psi^{\ast} - \partial_\tau \j \: \psi^{\ast} \right]
\end{align}
these modes are normalised as
\begin{align}
(\j_{\bs{q}}, \j_{\bs{q}'}) = (2 \pi)^d \delta(\bs{q} - \bs{q}').
\end{align}
The integral is taken over any constant-$\tau$ surface and is independent of the choice of $\tau$. This tells us that the coefficients in \eqref{Phi_decomp} are canonically normalised creation-annihilation operators after quantisation. The vacuum state, $| 0 \>$, annihilated by all $a_{\bs{q}}$, is the vacuum state in far past where the modes behave as in \eqref{past_behaviour}.

\subsubsection{Propagators} \label{sec:Propagators}

The Wightman functions $G_{-+}$ and $G_{+-}$ are defined as
\begin{align}
& G_{-+}(x_1, x_2) = \< 0 | \j(x_1) \j(x_2) | 0 \>_0, && G_{+-}(x_1, x_2) = \< 0 | \j(x_2) \j(x_1) | 0 \>_0,
\end{align}
where the subscript $\< - \>_0$ reminds us that the expectation value is evaluated in the free theory. In momentum space the overall delta function drops out,
\begin{align}
& G_{\mp \pm}(\tau_1, \bs{q}_1; \tau_2, \bs{q}_2) = (2 \pi)^d \delta(\bs{q}_1 + \bs{q}_2) G_{\mp \pm}(q_1, \tau_1, \tau_2),
\end{align}
and we obtain
\begin{align}
G_{+-}(q, \tau_1, \tau_2) & = \varphi^{\ast}_{\bs{q}}(\tau_1) \varphi_{-\bs{q}}(\tau_2) = \frac{\pi}{4} (-\tau_1)^{d/2} (-\tau_2)^{d/2}  H^{(2)}_{\beta}(- q \tau_1) H^{(1)}_{\beta}(- q \tau_2) \\
G_{-+}(q, \tau_1, \tau_2) & = \varphi_{\bs{q}}(\tau_1) \varphi^{\ast}_{-\bs{q}}(\tau_2)= G_{+-}^{\ast}(q, \tau_1, \tau_2), 
\end{align}
The time-ordered propagator $G_{++}$ and the anti-time-ordered propagator $G_{--}$ are
\begin{align}
G_{++}(q, \tau_1, \tau_2) & = \lla 0 | T \left[ \j(\tau_1, \bs{q}) \j(\tau_2, -\bs{q}) \right] | 0 \rra_0 \nn\\
& = \Theta(\tau_1 - \tau_2) G_{-+}(q, \tau_1, \tau_2) + \Theta(\tau_2 - \tau_1) G_{+-}(q, \tau_1, \tau_2), \\
G_{--}(q, \tau_1, \tau_2) & =  \lla 0 | \bar{T} \left[ \j(\tau_1, \bs{q}) \j(\tau_2, -\bs{q}) \right] | 0 \rra_0 \nn\\
& = \Theta(\tau_1 - \tau_2) G_{+-}(q, \tau_1, \tau_2) + \Theta(\tau_2 - \tau_1) G_{-+}(q, \tau_1, \tau_2) \nn\\
& = G_{++}^{\ast}(q, \tau_1, \tau_2).
\end{align}
This agrees with both the conventions of \cite{Birrell:1982ix} and \cite{Chen:2017ryl}.  Note that these propagators are invariant under $\beta \to -\beta$, as follows from $H^{(1)}_{-\beta} = e^{\beta \pi i} H^{(1)}_{\beta}$ and $ H^{(2)}_{-\beta} = e^{-\beta \pi i} H^{(2)}_{\beta}$.

The asymptotic behaviour of the free field near the boundary at $\tau = 0$ is given by \eqref{dSasympt}. Thus, the bulk-to-boundary propagators are defined as leading terms in the expansion of $G_{++}$ near the boundary,
\begin{align} \label{Gp}
G_{+}(q, \tau) & = \lim_{\tau_0 \rightarrow 0^{-}}G_{++}(q, \tau, \tau_0) = \lim_{\tau_0 \rightarrow 0^{-}}G_{+-}(q, \tau, \tau_0) \nn\\
& = - \I (-\tau_0)^{\frac{d}{2} - \beta} 2^{\beta - 2} \Gamma(\beta) q^{-\beta} (-\tau)^{\frac{d}{2}} H^{(2)}_{\beta} (-q \tau), \\
G_{-}(q, \tau) & = \left[ G_{+}(q, \tau) \right]^{\ast} \nn\\
& = \I (-\tau_0)^{\frac{d}{2} - \beta} 2^{\beta - 2} \Gamma(\beta) q^{-\beta} (-\tau)^{\frac{d}{2}} H^{(1)}_{\beta} (-q \tau). \label{Gm}
\end{align}
It is important here that this expression is valid for all $\beta > 0$. In particular there are no logarithmic terms when $\beta$ is integral. This follows from the definition of the Hankel functions as $H^{(\pm)}_{\beta} = J_{\beta} \pm \I Y_{\beta}$ with $+$ for $H^{(1)}$ and $-$ for $H^{(2)}$ and the power expansions of the Bessel functions involved.

\subsubsection{2-point function}

We can calculate the free late time 2-point function in the canonical formalism. The 2-point amplitude, \textit{i.e.}, its momentum-dependent part, free of the constants $L_{dS}$ and $\ell_{P}^{(dS)}$ reads
\begin{align} \label{ds2pt}
\dsno_{[\Delta \Delta]}(q) & = \lla 0 | \jsrc(\bs{q}) \jsrc(-\bs{q}) | 0 \rra_0 \nn\\
& = \lim_{\tau_0 \rightarrow 0^{-}} \left[ (-\tau_0)^{2(d - \Delta)} \lla 0 | \j(\tau_0, \bs{q}) \j(\tau_0, -\bs{q}) | 0 \rra_0 \right] \nn\\
& = \lim_{\tau_0 \rightarrow 0^{-}} \left[ (-\tau_0)^{2(d - \Delta)} \varphi_{\bs{q}}(\tau_0) \varphi^{\ast}_{-\bs{q}}(\tau_0) \right] \nn\\
& = \k_{\beta} q^{-2 \beta},
\end{align}
where we defined the constant
\begin{align} \label{kbeta}
\k_{\beta} = \frac{4^{\beta-1} \Gamma^2(\beta)}{\pi}.
\end{align}
This constant will appear repeatedly throughout the calculations. In order to stay consistent with the notation for the AdS amplitudes in \cite{Bzowski:2022rlz} we use the conformal weights $\Delta$ to distinguish various de Sitter amplitudes. The relations between $\beta$, $\Delta$ and de Sitter mass-squared $m_{dS}^2$ are given by
\begin{align}
& \beta = \Delta - \frac{d}{2}, && m_{dS}^2 L_{dS}^2 = - \Delta (\Delta - d) = \frac{d^2}{4} - \beta^2.
\end{align}
We are mostly interested in cases $\Delta = 2$ and $3$ in $d=3$ boundary spacetime dimensions, which correspond to $\beta = 1/2$ and $3/2$. From the point of view of the $4$-dimensional de Sitter bulk, these correspond to the conformally coupled ($m_{dS}^2 L_{dS}^2 = 2$) and massless ($m_{dS}^2 L_{dS}^2 = 0$) dual scalar fields respectively.

Clearly, the 2-point function matches the late-time limits of the propagators,
\begin{align}
\dsno_{[\Delta \Delta]}(q) & = \lim_{\tau_0 \rightarrow 0^{-}} (-\tau_0)^{2(d - \Delta)} G_{++}(q, \tau_0, \tau_0) \nn\\
& = \lim_{\tau_0 \rightarrow 0^{-}} (-\tau_0)^{2(d - \Delta)} G_{-+}(q, \tau_0, \tau_0)\nn\\
& = \lim_{\tau_0 \rightarrow 0^{-}} (-\tau_0)^{2(d - \Delta)} G_{+}(q, \tau_0).
\end{align}
As discussed in section \ref{sec:Propagators}, these results are valid for any real, positive $\beta$. In particular the dS 2-point function is always of the form $q^{-2\beta}$ and there are \emph{no logarithmic terms} present when $\beta$ is an integer.

The 2-point amplitude can be rewritten as
\begin{align}\label{dStoi2pt}
\dsno_{[\Delta \Delta]}(q) = - \frac{1}{2 \sin (\pi \beta) \ino_{[\Delta \Delta]}(q)},
\end{align}
where
\begin{align}\label{AdS2pt}
\ino_{[\Delta \Delta]}(q) & = - \frac{\Gamma(1 - \beta)}{2^{2\beta-1} \Gamma(\beta)} q^{2\beta}=-\frac{q^{2\beta}}{2\sin(\pi\beta)\k_\beta}.
\end{align}
is the properly normalised AdS 2-point function. For integral $\beta$'s the expression \eqref{dStoi2pt} must be understood as a limit. The singularity in the AdS 2-point function $\ino_{[\Delta \Delta]}$ is then cancelled by the sine in \eqref{dStoi2pt} and a finite dS 2-point function of the form $q^{-2\beta}$ is obtained for all $\beta > 0$. One can hide the sine even further by writing
\begin{align} \label{2pt}
\dsno_{[\Delta \Delta]}(q) = - \frac{1}{2 \im \ino_{[\Delta \Delta]}(\I q)}.
\end{align}
Now, however, the meaning of this expression for integral $\beta$'s is obscured. For integral betas this is still understood as the limit. Consequently, the dS 2-point function is immune to any logarithmic terms or renormalisation effects in the AdS 2-point function.

\subsection{Interactions, Schwinger-Keldysh and the generating functions}

Let us now consider the system in the presence of interactions, $S = S_0 + S_{int}$, where $S_0$ is the free action \eqref{S0}. We are interested in the boundary correlation functions with all operator insertions at the boundary at $\tau_0 = 0$. We define $\jsrc$ as the leading term in the expansion \eqref{dSasympt}, 
\begin{align} \label{defPhi0}
\jsrc(\bs{x}) = \lim_{\tau_0 \rightarrow 0} \left[ (-\tau_0)^{\Delta-d} \j(\tau, \bs{x}) \right].
\end{align}
In the presence of interactions $\jsrc$ remains well-defined except in the case of a massless particle, $m^2_{dS} = 0$, which corresponds to $\beta = d/2$. In such a case additional logarithmic terms, $\log^j (-\tau)$, appear in \eqref{defPhi0}, with $j$ denoting the order of the perturbative expansion as in \eqref{fieldexpinlambda}. In this paper, however, as in its AdS counterpart, \cite{Bzowski:2022rlz}, we work within the framework of dimensional regularisation and renormalisation. We treat dimensions $d$ and $\Delta$ as parameters and treat amplitudes (Witten diagrams) and correlation functions as their functions. We assume a generic near-boundary expansion \eqref{dSasympt} and derive the amplitudes. As the obtained amplitudes are valid in some open, non-empty set of the dimensions $d$ and $\Delta$, one can use analytic continuation to define them for almost all other values of dimensions. On the other hand, features such as the presence of secular, logarithmic terms in the near-boundary expansions is signaled by divergences (poles) for special values of the dimensions, such as $\Delta = d$ or, equivalently, $\beta = d/2$. For this reason, for now, we will consider the generic case of the expansion \eqref{dSasympt} and then carry out the renormalisation procedure in section \ref{sec:regandren}.

With $\jsrc$ defined in \eqref{defPhi0}, we define the cosmological correlators as
\begin{align} \label{n-pt}
\lla \jsrc(\bs{q}_1) \ldots \jsrc(\bs{q}_n) \rra = \lim_{\tau_0 \rightarrow 0^{-}} (-\tau_0)^{\Delta_t - n d} \lla \j(\tau_0, \bs{q}_1) \ldots \j(\tau_0, \bs{q}_n) \rra,
\end{align}
where $\Delta_t$ is the sum of dimensions associated with $\j$. Here, for simplicity, we take all fields to be identical (so $\Delta_t = n \Delta$)
in the interacting theory, evaluated perturbatively in the couplings $\lambda_3$ and $\lambda_4$. According to \cite{Chen:2017ryl} this can be done utilising the Schwinger-Keldysh formalism. The correlator is expressed as
\begin{align} \label{SK-formula}
& \< \j(0, \bs{q}_1) \ldots \j(0, \bs{q}_n) \> 
= \int \mathcal{D} \j_{+} \mathcal{D} \j_{-} \: \j_{+}(0, \bs{q}_1) \ldots \j_{+}(0, \bs{q}_n) \exp \left( \I S_{+}[\j_{+}] - \I S_{-}[\j_{-}] \right),
\end{align}
where the path integrals are taken over fields $\j_{-}$ and $\j_{+}$ whose values match at $\tau_0 = 0$. The actions $S_{+}$ and $S_{-}$ are identical with the full action $S$, except that now specific integration contours are specified,
\begin{align} \label{Spm}
S^{int}_{\pm}[\j] = - \int_{-\infty(1 \mp \I \ep)}^0 \D \tau \int \D^{d} \bs{x} \sqrt{-g} \, L[\j].
\end{align}

The path integral on the right-hand side of \eqref{SK-formula} can be thought of as evaluated in the auxiliary Schwinger-Keldysh theory represented by the partition function,
\begin{align} \label{Z_SK}
Z[J_{+}, J_{-}] & = \int \mathcal{D} \j_{+} \mathcal{D} \j_{-} \exp \left[ \I S_{+}[\j_{+}] - \I S_{-}[\j_{-}] + \I \int \D^{d+1} x \sqrt{-g} \left( J_{+} \j_{+} - J_{-} \j_{-} \right) \right].
\end{align}
This allows us to apply to \eqref{SK-formula} standard perturbation theory, Wick contractions \textit{etc.}, with the exception that we have now four propagators,
\begin{align}
\< \j_{\sigma_1}(x_1) \j_{\sigma_2}(x_2) \>_{SK} & = \frac{- \I \sigma_1}{\sqrt{-g(x_1)}} \frac{\delta}{\delta J_{\sigma_1}(x_1)} \frac{- \I \sigma_2}{\sqrt{-g(x_2)}} \frac{\delta}{\delta J_{\sigma_2}(x_2)} Z[J_{+}, J_{-}] \nn\\
& = G_{\sigma_1 \sigma_2}(x_1, x_2)
\end{align}
according to the choice of signs $\sigma_1, \sigma_2 = \pm$.

\subsection{3-point function} \label{sec:3pt}

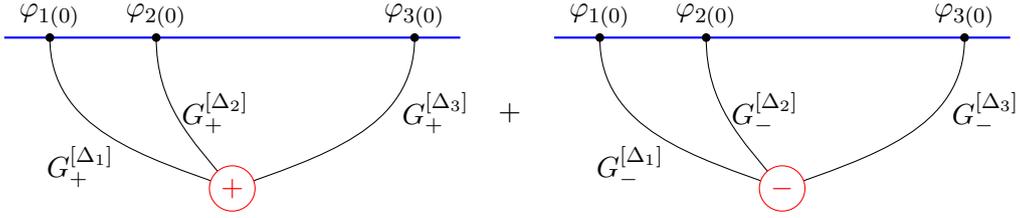
\begin{figure}[t]
\begin{tikzpicture}[scale=1.0]
\draw[thick, blue] (-3,1) -- (3,1);
\draw[fill=black] (-2.4,1) circle [radius=0.05];
\draw[fill=black] (-1.0,1) circle [radius=0.05];
\draw[fill=black] (2.4,1) circle [radius=0.05];
\draw (-2.4,1) to [out=-90,in=160] (0,-1);
\draw (-1.0,1) to [out=-90,in=130] (0,-1);
\draw (2.4,1) to [out=-90,in=20] (0,-1);
\draw[red,fill=white] (0,-1) circle [radius=0.3];
\node[red] at (0,-1) {$+$};
\node[above] at (-2.4,1) {$\j_{1(0)}$};
\node[above] at (-1.0,1) {$\j_{2(0)}$};
\node[above] at ( 2.4,1) {$\j_{3(0)}$};
\node[below] at (-2,-0.3) {$G_{+}^{[\Delta_1]}$};
\node[right] at (-0.8, 0) {$G_{+}^{[\Delta_2]}$};
\node[right] at ( 2.1, 0) {$G_{+}^{[\Delta_3]}$};
\end{tikzpicture}
\begin{tikzpicture}[scale=1.0]
\node[left] at (-3.3,0) {$+$};
\draw[thick, blue] (-3,1) -- (3,1);
\draw[fill=black] (-2.4,1) circle [radius=0.05];
\draw[fill=black] (-1.0,1) circle [radius=0.05];
\draw[fill=black] (2.4,1) circle [radius=0.05];
\draw (-2.4,1) to [out=-90,in=160] (0,-1);
\draw (-1.0,1) to [out=-90,in=130] (0,-1);
\draw (2.4,1) to [out=-90,in=20] (0,-1);
\draw[red,fill=white] (0,-1) circle [radius=0.3];
\node[red] at (0,-1) {$-$};
\node[above] at (-2.4,1) {$\j_{1(0)}$};
\node[above] at (-1.0,1) {$\j_{2(0)}$};
\node[above] at ( 2.4,1) {$\j_{3(0)}$};
\node[below] at (-2,-0.3) {$G_{-}^{[\Delta_1]}$};
\node[right] at (-0.8, 0) {$G_{-}^{[\Delta_2]}$};
\node[right] at ( 2.1, 0) {$G_{-}^{[\Delta_3]}$};
\end{tikzpicture}
\centering
\caption{Witten diagrams contributing to the de Sitter amplitude $\dsno_{[\Delta_1 \Delta_2 \Delta_3]}$. The blue line denotes the boundary surface of $\tau_0 = 0$.\label{fig:3pt}}
\end{figure}

Consider three scalar fields $\j_j$ for $j = 1,2,3$ governed by the action
\begin{align}
S_3 & = S_0 + S_{int}, \label{S3} \\
S_0 & = - \frac{1}{2} \int \D^{d} x \sqrt{-g} \sum_{j =1,2,3} \left[ \partial_\mu \j_j \partial^\mu \j_j + m^2_{j} \j_j^2 \right], \\
S_{int} & = - \lambda_{123} \int \D^{d} x \sqrt{-g} \, \j_1 \j_2 \j_3,
\end{align}
where $\lambda_{123}$ is a coupling constant. The boundary 3-point function can be presented in the form \eqref{dS_amp_def}, which in this case becomes
\begin{align} \label{ds3def}
\lla \j_{1 (0)}(\bs{q}_1) \j_{2 (0)}(\bs{q}_2) \j_{3 (0)}(\bs{q}_3) \rra = \Big( \frac{L_{dS}}{\ell_P^{(dS)}} \Big)^{d+1} \lambda_{123} \dsno_{[\Delta_1 \Delta_2 \Delta_3]}(q_1, q_2, q_3) + O(\lambda_{123}^2).
\end{align}
By $\dsno_{[\Delta_1 \Delta_2 \Delta_3]}$ we denote a momentum-dependent amplitude represented by the sum of the tree-level Witten diagrams presented in figure \ref{fig:3pt}. In general, the 3-point functions receives bulk loop corrections, all of higher order in the coupling $\lambda_{123}$. 

In order to evaluate it, notice that the formula \eqref{SK-formula} produces two terms, one from the expansion of $S_{+}$ and one from $S_{-}$. The two terms are complex conjugates of each other and we can think about them as obtained from the two vertices
\begin{align} \label{Vpm}
V_{\pm}[f_1, f_2, f_3] = \mp \I \lambda_{123} \int_{-\infty(1 \mp \I \ep)}^0 \frac{\D \tau}{(-\tau)^{d+1}} f_1(\tau) f_2(\tau) f_3(\tau).
\end{align}
The Schwinger-Keldysh formalism \eqref{SK-formula} tells us that we are allowed to plug only $+$ edges into $V_{+}$ and $-$ edges into $G_{-}$. With $\tau_0 = 0$ this means that
\begin{align} \label{3pt-int}
& \dsno_{[\Delta_1 \Delta_2 \Delta_3]}(q_1, q_2, q_3) =  2 \re \left[ - \I \int_{-\infty(1 - \I \ep)}^0 \frac{\D \tau}{(-\tau)^{d+1}} G^{[\Delta_1]}_{+}(q_1, \tau) G^{[\Delta_2]}_{+}(q_2, \tau) G_{+}^{[\Delta_3]}(q_3, \tau)  \right],
\end{align}
with $G^{[\Delta]}_{+}$ denoting the propagator \eqref{Gp} for the field whose mass is parameterised by $\Delta$.

There is an important assumption here, that the integral converges at the upper limit of $\tau = 0$. In other words that we can take the $\tau_0 \rightarrow 0^{-}$ limit before evaluating the integral. In such a case the most leading term in $\tau_0$ simply follows from the explicit factors in \eqref{Gp} and \eqref{Gm} and for the 3-point function is $(-\tau_0)^{\frac{3d}{2} - \beta_t} = (-\tau_0)^{3d - \Delta_t}$, where $\beta_t = \beta_1 + \beta_2 + \beta_3$ and $\Delta_t = \Delta_1 + \Delta_2 + \Delta_3$ . If the integral diverges at $\tau = 0$, however, there are more leading terms, which can be extracted from the near boundary behaviour of the integrand. For now we assume that all integrals converge.

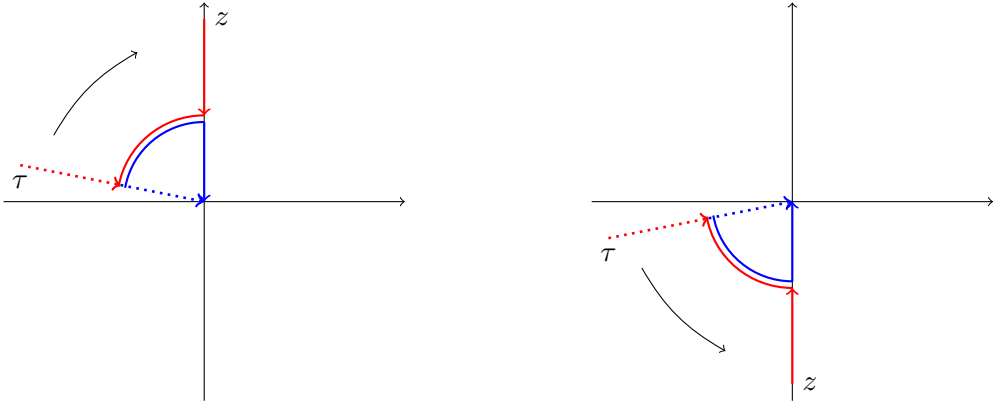
\begin{figure}[t]
\begin{tikzpicture}[scale=2.2]
\draw[->] (-1.2,0) -- (1.2,0);
\draw[->] (0,-1.2) -- (0,1.2);
\draw[->, dotted, red, line width = 0.35mm] (-1.1,0.22) -- (-0.5,0.1);
\draw[->, dotted, blue, line width = 0.35mm] (-0.5,0.1) -- (0,0);
\draw[->, red, thick] (0,1.1) -- (0,0.52);
\draw[red, thick] (0,0.52) arc (90:170:0.52);
\draw[blue, thick] (0,0.48) arc (90:170:0.48);
\draw[->, blue, thick] (0,0.48) -- (0,0);
\draw[->] (-0.9,0.4) to [out=60,in=-150] (-0.4,0.9);
\node[below] at (-1.1,0.22) {$\tau$};
\node[right] at (0,1.1) {$z$};
\end{tikzpicture}
\qquad\qquad\qquad
\begin{tikzpicture}[xscale=2.2, yscale=-2.2]
\draw[->] (-1.2,0) -- (1.2,0);
\draw[<-] (0,-1.2) -- (0,1.2);
\draw[->, dotted, red, line width = 0.35mm] (-1.1,0.22) -- (-0.5,0.1);
\draw[->, dotted, blue, line width = 0.35mm] (-0.5,0.1) -- (0,0);
\draw[->, red, thick] (0,1.1) -- (0,0.52);
\draw[red, thick] (0,0.52) arc (90:170:0.52);
\draw[blue, thick] (0,0.48) arc (90:170:0.48);
\draw[->, blue, thick] (0,0.48) -- (0,0);
\draw[->] (-0.9,0.4) to [out=60,in=-150] (-0.4,0.9);
\node[below] at (-1.1,0.22) {$\tau$};
\node[right] at (0,1.1) {$z$};
\end{tikzpicture}
\centering
\caption{Deformations of the two contours running from $\tau = -\infty$ to $\tau = 0$ to two the contours running over the imaginary axis. On the left the original dotted contour runs with slightly positive imaginary part, while the contour on the right has a slightly negative imaginary part. The arcs in the middle correspond to the split of the integration limits according to whether $\tau_1 < \tau_2$ or the opposite hold. Note that if the cut-off regularisation is imposed close to zero, one should also consider a small additional arc there.\label{fig:contours}}
\end{figure}

Just as in Minkowski field theory the integrals present in Feynman diagrams are carried out over slightly imaginary contours. In our case these contours are specified in \eqref{Spm}. Whether the contour goes over a slightly positive or negative imaginary times determines the direction of the contour rotation as presented in figure \ref{fig:contours}. For $S_{+}$, the contour runs over slightly imaginary, positive times $\tau$, and thus it can be deformed to the contour running from $+ \I \infty$ to $0$. Similarly, for $S_{-}$ the contour gets deformed to the contour running from $-\I \infty$ to $0$. Thus, we substitute
\begin{align}
& \text{for } + && \tau = \I z, && \int_{-\infty(1 - \I \ep)}^0 \frac{\D \tau}{(-\tau)^{d+1}} = e^{\frac{\I \pi d}{2}} \int_0^{\infty} \frac{\D z}{z^{d+1}}, \\
& \text{for } - && \tau = - \I z, && \int_{-\infty(1 + \I \ep)}^0 \frac{\D \tau}{(-\tau)^{d+1}} = e^{- \frac{\I \pi d}{2}} \int_0^{\infty} \frac{\D z}{z^{d+1}}.
\end{align}
This correspond to how Hankel functions are continued; both $H^{(1)}_{\beta}$ and $H^{(2)}_{\beta}$ continue to the Bessel-$K$ functions
\begin{align}
H^{(1)}_{\beta}(\I x) & = - \frac{2 \I}{\pi} e^{-\frac{\I \pi \beta}{2}} K_{\beta}(x), &
H^{(2)}_{\beta}(-\I x) & = \frac{2 \I}{\pi} e^{\frac{\I \pi \beta}{2}} K_{\beta}(x). \label{HtoK1}
\end{align}
valid for real $x > 0$ and real $\beta$. In this way we find
\begin{align}\label{Gpcont}
G_{\pm}(q, \pm \I z) = e^{\pm \frac{\I \pi}{2} ( \beta - \frac{d}{2})} (-\tau_0)^{\frac{d}{2} - \beta} \k_{\beta} q^{-2 \beta} \mathcal{K}_{\beta}(q, z)
\end{align}
where $\k_{\beta}$ is defined in \eqref{kbeta} and
\begin{align}\label{AdSkdef}
\mathcal{K}_{\beta}(q, z) & = \frac{q^{\beta} z^{\frac{d}{2}} K_{\beta}(q z)}{2^{\beta - 1} \Gamma(\beta)}.
\end{align}
Here we can think of $\mathcal{K}_{\beta}$ as the AdS bulk-to-boundary propagator. Just as for the 2-point function, these expressions are valid for all $\beta > 0$. Note that all factors except the explicit exponent are real.

Going back to the integral \eqref{3pt-int} we apply the analytic continuation to find
\begin{align} \label{3pt_in_dS}
& \dsno_{[\Delta_1 \Delta_2 \Delta_3]}(q_1, q_2, q_3)  \nn\\
& \qquad = 2 \lim_{\tau_0 \rightarrow 0^{-}} (-\tau_0)^{ \Delta_t-3d} \im \left[ e^{\frac{\I \pi d}{2}} \int_0^{\infty} \frac{\D z}{z^{d+1}} G_{+}(q_1, \I z) G_{+}(q_2, \I z) G_{+}(q_3, \I z) \right] \nn\\
& \qquad = 2 \prod_{j=1}^3 \left[ q_j^{-2 \beta_j} \k_{\beta_j} \right] \im \exp \left[ \frac{\I \pi}{2} \left(\beta_t - \frac{d}{2} \right) \right] \times\nn\\
& \qquad\qquad\qquad \times \int_0^{\infty} \frac{\D z}{z^{d+1}} \mathcal{K}_{\beta_1}(q_1, z) \mathcal{K}_{\beta_2}(q_2, z) \mathcal{K}_{\beta_3}(q_3, z) \nn\\
& \qquad = 2 \sin \left[ \frac{\pi}{2} ( \beta_t - \frac{d}{2}) \right] \prod_{j=1}^3 \left[ q_j^{-2 \beta_j} \k_{\beta_j} \right] \ino_{[\Delta_1 \Delta_2 \Delta_3]}(q_1, q_2, q_3),
\end{align}
where $\ino_{[\Delta_1 \Delta_2 \Delta_3]}$ denotes the AdS amplitude. The combination under the sine reads
\begin{align}
\beta_t - \frac{d}{2} = \Delta_t - 2 d,
\end{align}
the total dimension of the AdS amplitude. Thus for finite or regulated amplitudes one can substitute
\begin{align} \label{sin_to_im_i3}
\sin \left[ \frac{\pi}{2} ( \beta_t - \frac{d}{2}) \right] \ireg_{[\Delta_1 \Delta_2 \Delta_3]}(q_1, q_2, q_3) = \im \ireg_{[\Delta_1 \Delta_2 \Delta_3]}(\I q_1, \I q_2, \I q_3).
\end{align}
The 3-point function \eqref{3pt_in_dS} is valid for any finite $\beta$'s including integral ones. We can express $\k_{\beta} q^{-2 \beta}$ in terms of the AdS 2-point function by inverting \eqref{2pt}
\begin{align} \label{qBeta_to_2pt}
\k_{\beta} q^{-2 \beta} & = - \frac{1}{2 \im \ireg_{[\Delta \Delta]}(\I q)}.
\end{align}
The problem with this representation is that for integral $\beta$ the formula suggests that some logarithmic terms appear due to the divergence of the 2-point function. As discussed in the previous section, this is not the case as $\im \ireg_{[\Delta \Delta]}(\I q)$ is understood as the limit and is perfectly finite. Thus, we can consistently write the expression relating regulated AdS and dS amplitudes,
\begin{align} \label{3pt}
& \dsreg_{[\Delta_1 \Delta_2 \Delta_3]}(q_1, q_2, q_3) = - \frac{1}{4} \frac{\im \ireg_{[\Delta_1 \Delta_2 \Delta_3]}(\I q_1, \I q_2, \I q_3)}{\prod_{j=1}^3 \im \ireg_{[\Delta_j \Delta_j]}(\I q_j)}.
\end{align}

\subsection{4-point function}

Consider now five scalar bulk fields $\j_j$ for $j = 1,2,3,4,x$ governed by the action,
\begin{align}
S_4 & = S_0 + S_{int}, \label{S4} \\
S_0 & = - \frac{1}{2} \int \D^{d} x \sqrt{-g} \sum_{j =1,2,3,4,x} \left[ \partial_\mu \j_j \partial^\mu \j_j + m^2_{j} \j_j^2 \right], \\
S_{int} & = - \int \D^{d} x \sqrt{-g} \left[ \lambda_{12x} \j_1 \j_2 \j_x + \lambda_{34x} \j_x \j_3 \j_4 - \lambda_{1234} \j_1 \j_2 \j_3 \j_4 \right] \label{S4int}
\end{align}
on the fixed $(d+1)$-dimensional de Sitter background \eqref{gdS}. The idea behind is that such a theory contains as few symmetries as possible. In particular the leading terms in the 3-point functions $\lla \j_1 \j_2 \j_x \rra$ and $\lla \j_x \j_3 \j_4 \rra$ and the 4-point function $\lla \j_1 \j_2 \j_3 \j_4 \rra$ are given by single Witten diagrams (amplitudes). According to \eqref{dS_amp_def} we have two amplitudes to consider: the contact 4-point amplitude with $V = 1$, which we denote by $\dsno_{[\Delta_1 \Delta_2 \Delta_3 \Delta_4]}$, and the exchange 4-point amplitude with $V = 2$, which we denote by $\dsno_{[\Delta_1 \Delta_2; \Delta_3 \Delta_4 x \Delta_x]}$. We have
\begin{align} \label{ds4def}
& \lla \j_{1 (0)}(\bs{q}_1) \j_{2 (0)}(\bs{q}_2) \j_{3 (0)}(\bs{q}_3) \j_{4 (0)}(\bs{q}_4) \rra = \left( \frac{L_{dS}}{\ell_P^{(dS)}} \right)^{d+1} \lambda_{1234} \dsno_{[\Delta_1 \Delta_2 \Delta_3 \Delta_4]}(q_1, q_2, q_3, q_4) \nn\\
& \qquad\qquad + \left( \frac{L_{dS}}{\ell_P^{(dS)}} \right)^{d+3} \lambda_{12x} \lambda_{34x} \dsno_{[\Delta_1 \Delta_2; \Delta_3 \Delta_4 x \Delta_x]}(q_1, q_2, q_3, q_4, s) + \ldots
\end{align}
where the dropped terms are the bulk loop corrections, all higher order in the couplings. The sign choices in the action \eqref{S4int} follow the equation of (4.7) in \cite{Bzowski:2022rlz}, so that the dS and AdS interaction actions differ by the overall sign.

\subsubsection{Contact diagram}

\begin{figure}[t]
\begin{tikzpicture}[scale=1.0]
\draw[thick, blue] (-3,1) -- (3,1);
\draw[fill=black] (-2.4,1) circle [radius=0.05];
\draw[fill=black] (-1.0,1) circle [radius=0.05];
\draw[fill=black] (1.4,1) circle [radius=0.05];
\draw[fill=black] (2.4,1) circle [radius=0.05];
\draw (-2.4,1) to [out=-90,in=160] (0,-1);
\draw (-1.0,1) to [out=-90,in=130] (0,-1);
\draw (1.4,1) to [out=-90,in=40] (0,-1);
\draw (2.4,1) to [out=-90,in=20] (0,-1);
\draw[red,fill=white] (0,-1) circle [radius=0.3];
\node[red] at (0,-1) {$+$};
\node[above] at (-2.4,1) {$\j_{1(0)}$};
\node[above] at (-1.0,1) {$\j_{2(0)}$};
\node[above] at ( 1.4,1) {$\j_{3(0)}$};
\node[above] at ( 2.4,1) {$\j_{4(0)}$};
\node[below] at (-2,-0.3) {$G_{+}^{[\Delta_1]}$};
\node[right] at (-2.1, 0.5) {$G_{+}^{[\Delta_2]}$};
\node[left] at ( 1.4, 0.5) {$G_{+}^{[\Delta_3]}$};
\node[right] at ( 2.1, 0) {$G_{+}^{[\Delta_4]}$};
\end{tikzpicture}
\begin{tikzpicture}[scale=1.0]
\draw[thick, blue] (-3,1) -- (3,1);
\draw[fill=black] (-2.4,1) circle [radius=0.05];
\draw[fill=black] (-1.0,1) circle [radius=0.05];
\draw[fill=black] (1.4,1) circle [radius=0.05];
\draw[fill=black] (2.4,1) circle [radius=0.05];
\draw (-2.4,1) to [out=-90,in=160] (0,-1);
\draw (-1.0,1) to [out=-90,in=130] (0,-1);
\draw (1.4,1) to [out=-90,in=40] (0,-1);
\draw (2.4,1) to [out=-90,in=20] (0,-1);
\draw[red,fill=white] (0,-1) circle [radius=0.3];
\node[red] at (0,-1) {$-$};
\node[left] at (-3.5,0) {$+$};
\node[above] at (-2.4,1) {$\j_{1(0)}$};
\node[above] at (-1.0,1) {$\j_{2(0)}$};
\node[above] at ( 1.4,1) {$\j_{3(0)}$};
\node[above] at ( 2.4,1) {$\j_{4(0)}$};
\node[below] at (-2,-0.3) {$G_{-}^{[\Delta_1]}$};
\node[right] at (-2.1, 0.5) {$G_{-}^{[\Delta_2]}$};
\node[left] at ( 1.4, 0.5) {$G_{-}^{[\Delta_3]}$};
\node[right] at ( 2.1, 0) {$G_{-}^{[\Delta_4]}$};
\end{tikzpicture}
\centering
\caption{Witten diagrams contributing to the de Sitter amplitude $\dsno_{[\Delta_1 \Delta_2 \Delta_3 \Delta_4]}$. The blue line denotes the late-time boundary $\tau_0 = 0$.\label{fig:4ptc}}
\end{figure}
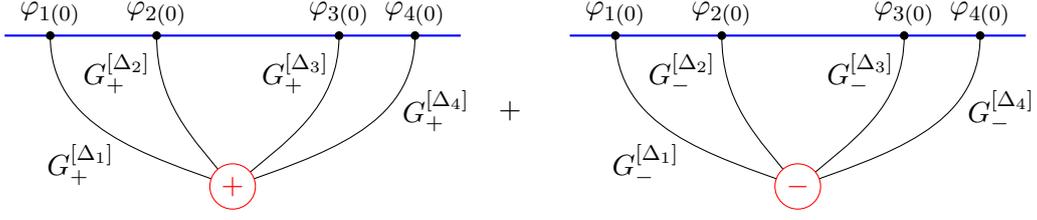

The situation here is analogous to the case of the 3-point function. The contact 4-point diagram is depicted in figure \ref{fig:4ptc}.

With the vertex
\begin{align}
V_{\pm}[f_1, f_2, f_3, f_4] & = \pm \I \lambda_{1234} \int_{-\infty(1 \mp \I \ep)}^0 \frac{\D \tau}{(-\tau)^{d+1}} f_1(\tau) f_2(\tau) f_3(\tau) f_4(\tau)
\end{align}
and $\beta_T = \beta_1 + \beta_2 + \beta_3 + \beta_4$ we get
\begin{align} \label{4pt_contact}
& \dsfin_{[\Delta_1 \Delta_2 \Delta_3 \Delta_4]}(q_1, q_2, q_3, q_4) = \nn\\
& \qquad =  2 \lim_{\tau_0 \rightarrow 0^{-}} (-\tau_0)^{ \Delta_T-4d} \re \left[ \I \int_{-\infty(1 - \I \ep)}^0 \frac{\D \tau}{(-\tau)^{d+1}} G_{+}(q_1, \tau) G_{+}(q_2, \tau) G_{+}(q_3, \tau)  G_{+}(q_4, \tau) \right] \nn\\
& \qquad = -2 \sin \left[ \frac{\pi}{2} ( \beta_T - d) \right] \prod_{j=1}^4 \left[ q_j^{-2 \beta_j} \k_{\beta_j} \right] \ifin_{[\Delta_1 \Delta_2 \Delta_3 \Delta_4]}(q_1, q_2, q_3, q_4) \nn\\
& \qquad = - \frac{1}{8} \frac{\im  \ifin_{[\Delta_1 \Delta_2 \Delta_3 \Delta_4]}(\I q_j)}{\prod_{j=1}^4 \im \ireg_{[\Delta_j \Delta_j]}(\I q_j)}. 
\end{align}

\subsubsection{4-point exchange}

\begin{figure}[t]
\begin{tikzpicture}[scale=1]
\draw[thick, blue] (-3,1) -- (3,1);
\draw[fill=black] (-2.4,1) circle [radius=0.05];
\draw[fill=black] (-1.0,1) circle [radius=0.05];
\draw[fill=black] (1.4,1) circle [radius=0.05];
\draw[fill=black] (2.4,1) circle [radius=0.05];
\draw (-2.4,1) to [out=-90,in=150] (-1.5,-1);
\draw (-1.0,1) to [out=-90,in=60] (-1.5,-1);
\draw (1.4,1) to [out=-90,in=100] (1.5,-1);
\draw (2.4,1) to [out=-90,in=30] (1.5,-1);
\draw (-1.5,-1) -- (1.5,-1);
\draw[red,fill=white] (-1.5,-1) circle [radius=0.3];
\node[red] at (-1.5,-1) {$+$};
\draw[red,fill=white] ( 1.5,-1) circle [radius=0.3];
\node[red] at ( 1.5,-1) {$+$};
\node[above] at (-2.4,1) {$\j_{1(0)}$};
\node[above] at (-1.0,1) {$\j_{2(0)}$};
\node[above] at ( 1.4,1) {$\j_{3(0)}$};
\node[above] at ( 2.4,1) {$\j_{4(0)}$};
\node[below] at (-2.5,-0.3) {$G_{+}^{[\Delta_1]}$};
\node[right] at (-2.1, 0.5) {$G_{+}^{[\Delta_2]}$};
\node[left] at ( 1.5, 0.5) {$G_{+}^{[\Delta_3]}$};
\node[right] at ( 2.3, 0) {$G_{+}^{[\Delta_4]}$};
\node[above] at (0,-1) {$G_{++}^{[\Delta_x]}$};
\end{tikzpicture}
\begin{tikzpicture}[scale=1]
\node[left] at (-3.3,0) {$+$};
\draw[thick, blue] (-3,1) -- (3,1);
\draw[fill=black] (-2.4,1) circle [radius=0.05];
\draw[fill=black] (-1.0,1) circle [radius=0.05];
\draw[fill=black] (1.4,1) circle [radius=0.05];
\draw[fill=black] (2.4,1) circle [radius=0.05];
\draw (-2.4,1) to [out=-90,in=150] (-1.5,-1);
\draw (-1.0,1) to [out=-90,in=60] (-1.5,-1);
\draw (1.4,1) to [out=-90,in=100] (1.5,-1);
\draw (2.4,1) to [out=-90,in=30] (1.5,-1);
\draw (-1.5,-1) -- (1.5,-1);
\draw[red,fill=white] (-1.5,-1) circle [radius=0.3];
\node[red] at (-1.5,-1) {$-$};
\draw[red,fill=white] ( 1.5,-1) circle [radius=0.3];
\node[red] at ( 1.5,-1) {$-$};
\node[above] at (-2.4,1) {$\j_{1(0)}$};
\node[above] at (-1.0,1) {$\j_{2(0)}$};
\node[above] at ( 1.4,1) {$\j_{3(0)}$};
\node[above] at ( 2.4,1) {$\j_{4(0)}$};
\node[below] at (-2.5,-0.3) {$G_{-}^{[\Delta_1]}$};
\node[right] at (-2.1, 0.5) {$G_{-}^{[\Delta_2]}$};
\node[left] at ( 1.5, 0.5) {$G_{-}^{[\Delta_3]}$};
\node[right] at ( 2.3, 0) {$G_{-}^{[\Delta_4]}$};
\node[above] at (0,-1) {$G_{--}^{[\Delta_x]}$};
\end{tikzpicture}
\begin{tikzpicture}[scale=1]
\draw[white](-1,2.5) -- (1,2.5);
\node[left] at (-3.3,0) {$+$};
\draw[thick, blue] (-3,1) -- (3,1);
\draw[fill=black] (-2.4,1) circle [radius=0.05];
\draw[fill=black] (-1.0,1) circle [radius=0.05];
\draw[fill=black] (1.4,1) circle [radius=0.05];
\draw[fill=black] (2.4,1) circle [radius=0.05];
\draw (-2.4,1) to [out=-90,in=150] (-1.5,-1);
\draw (-1.0,1) to [out=-90,in=60] (-1.5,-1);
\draw (1.4,1) to [out=-90,in=100] (1.5,-1);
\draw (2.4,1) to [out=-90,in=30] (1.5,-1);
\draw (-1.5,-1) -- (1.5,-1);
\draw[red,fill=white] (-1.5,-1) circle [radius=0.3];
\node[red] at (-1.5,-1) {$+$};
\draw[red,fill=white] ( 1.5,-1) circle [radius=0.3];
\node[red] at ( 1.5,-1) {$-$};
\node[above] at (-2.4,1) {$\j_{1(0)}$};
\node[above] at (-1.0,1) {$\j_{2(0)}$};
\node[above] at ( 1.4,1) {$\j_{3(0)}$};
\node[above] at ( 2.4,1) {$\j_{4(0)}$};
\node[below] at (-2.5,-0.3) {$G_{+}^{[\Delta_1]}$};
\node[right] at (-2.1, 0.5) {$G_{+}^{[\Delta_2]}$};
\node[left] at ( 1.5, 0.5) {$G_{-}^{[\Delta_3]}$};
\node[right] at ( 2.3, 0) {$G_{-}^{[\Delta_4]}$};
\node[above] at (0,-1) {$G_{+-}^{[\Delta_x]}$};
\end{tikzpicture}
\begin{tikzpicture}[scale=1.0]
\draw[white](-1,2.5) -- (1,2.5);
\node[left] at (-3.3,0) {$+$};
\draw[thick, blue] (-3,1) -- (3,1);
\draw[fill=black] (-2.4,1) circle [radius=0.05];
\draw[fill=black] (-1.0,1) circle [radius=0.05];
\draw[fill=black] (1.4,1) circle [radius=0.05];
\draw[fill=black] (2.4,1) circle [radius=0.05];
\draw (-2.4,1) to [out=-90,in=150] (-1.5,-1);
\draw (-1.0,1) to [out=-90,in=60] (-1.5,-1);
\draw (1.4,1) to [out=-90,in=100] (1.5,-1);
\draw (2.4,1) to [out=-90,in=30] (1.5,-1);
\draw (-1.5,-1) -- (1.5,-1);
\draw[red,fill=white] (-1.5,-1) circle [radius=0.3];
\node[red] at (-1.5,-1) {$-$};
\draw[red,fill=white] ( 1.5,-1) circle [radius=0.3];
\node[red] at ( 1.5,-1) {$+$};
\node[above] at (-2.4,1) {$\j_{1(0)}$};
\node[above] at (-1.0,1) {$\j_{2(0)}$};
\node[above] at ( 1.4,1) {$\j_{3(0)}$};
\node[above] at ( 2.4,1) {$\j_{4(0)}$};
\node[below] at (-2.5,-0.3) {$G_{-}^{[\Delta_1]}$};
\node[right] at (-2.1, 0.5) {$G_{-}^{[\Delta_2]}$};
\node[left] at ( 1.5, 0.5) {$G_{+}^{[\Delta_3]}$};
\node[right] at ( 2.3, 0) {$G_{+}^{[\Delta_4]}$};
\node[above] at (0,-1) {$G_{-+}^{[\Delta_x]}$};
\end{tikzpicture}
\centering
\caption{Witten diagrams contributing to the 4-point exchange de Sitter amplitude $\dsno_{[\Delta_1 \Delta_2; \Delta_3 \Delta_4 x \Delta_x]}$. The blue line denotes the late-time boundary  $\tau_0 = 0$.\label{fig:4ptx}}
\end{figure}
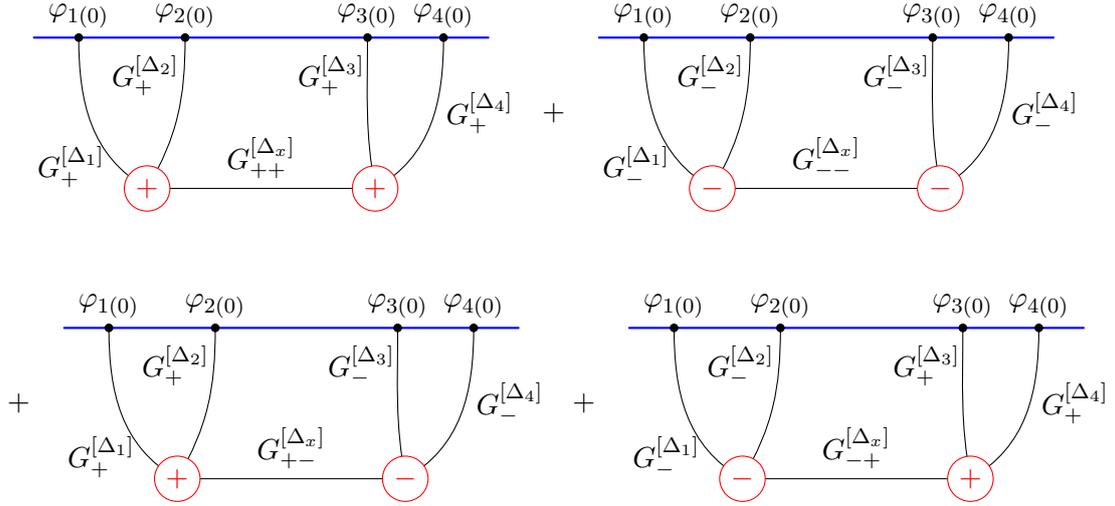

The 4-point exchange amplitude is depicted in figure \ref{fig:4ptx}. This amplitude possesses four terms arising from the four choices for the pairs $V_{\pm}$ of the left and right vertex,
\begin{align}
\dsno_{[\Delta_1 \Delta_2; \Delta_3 \Delta_4 x \Delta_x]} = \lim_{\tau_0 \rightarrow 0^{-}} (-\tau_0)^{ \Delta_T-4d} \left( I_{++} + I_{+-} + I_{-+} + I_{--} \right),
\end{align}
where
\begin{align}
I_{++} & = V_{+}[G_{+}(q_1), G_{+}(q_2), V_{+}[G_{++}(s), G_{+}(q_3), G_{+}(q_4)]] \nn\\
& = - \I \int_{-\infty(1-\I \ep)}^0 \frac{\D \tau_1}{(-\tau_1)^{d+1}} G_{+}(q_1, \tau_1) G_{+}(q_2, \tau_1) \times \nn\\
& \qquad\qquad\qquad\qquad \times (-\I) \int_{-\infty(1-\I \ep)}^0 \frac{\D \tau_2}{(-\tau_2)^{d+1}} G_{++}(s, \tau_1, \tau_2) G_{+}(q_3, \tau_2) G_{+}(q_4, \tau_2), \label{Ipp} \\[2ex]
I_{+-} & = V_{+}[G_{+}(q_1), G_{+}(q_2), V_{-}[G_{+-}(s), G_{-}(q_3), G_{-}(q_4)]] \nn\\
& = - \I \int_{-\infty(1-\I \ep)}^0 \frac{\D \tau_1}{(-\tau_1)^{d+1}} G_{+}(q_1, \tau_1) G_{+}(q_2, \tau_1) \times \nn\\
& \qquad\qquad\qquad\qquad \times \I \int_{-\infty(1+\I \ep)}^0 \frac{\D \tau_2}{(-\tau_2)^{d+1}} G_{+-}(s, \tau_1, \tau_2) G_{-}(q_3, \tau_2) G_{-}(q_4, \tau_2), \label{Ipm}
\end{align}
while $I_{--} = I_{++}^{\ast}$ and $I_{-+} = I_{+-}^{\ast}$. 

In order to calculate the integrals in \eqref{Ipp} and \eqref{Ipm} we have to carry out the appropriate analytic continuations. The dS bulk-to-bulk propagators are
\begin{align}
G_{+-}(q, \tau_1, \tau_2) & = \frac{\pi}{4} (-\tau_1)^{d/2} (-\tau_2)^{d/2}  H^{(2)}_{\beta}(- q \tau_1) H^{(1)}_{\beta}(- q \tau_2), \label{Gpm} \\
G_{++}(q, \tau_1, \tau_2) & = G_{--}^{\ast}(q, \tau_1, \tau_2) \nn\\
& = G_{-+}(q, \tau_1, \tau_2) \Theta(\tau_1 - \tau_2) + G_{+-}(q, \tau_1, \tau_2) \Theta(\tau_2 - \tau_1).
\end{align}
The analytic continuation of $G_{+-}$ is straightforward, since $\tau_1$ and $\tau_2$ are continued according to their labels, $\tau_1 = \I z_1$ and $\tau_2 = - \I z_2$. Using \eqref{HtoK1}, we have
\begin{align}\label{Gpmcont}
G_{+-}(q, \I z_1, - \I z_2) & = G_{-+}(q, -\I z_1, \I z_2) = \k_{\beta} q^{-2 \beta} \mathcal{K}_{\beta}(q, z_1) \mathcal{K}_{\beta}(q, z_2).
\end{align}
Note that these propagators are real. Furthermore, the factor of $q^{-2 \beta}$ has this form for all $\beta > 0$ and no logarithms are present.

For the analytic continuation of $G_{++}$ and $G_{--}$ we need more relations. This happens because for $G_{++}$ we continue both times in the same direction, $\tau_1 = \I z_1$ and $\tau_2 = \I z_2$. This means that we have $H_{\beta}^{(2)}(- \I q z_1)$, which continues according to \eqref{HtoK1}, but also $H_{\beta}^{(1)}(- \I q z_2)$, which now has the incorrect sign. Instead, for $x > 0$ we use the identities
\begin{align}
H_{\beta}^{(1)}(- \I x) &= - \frac{2 \I}{\pi} e^{-\frac{\I \pi \beta}{2}} K_{\beta}(e^{-\I \pi } x), & H_{\beta}^{(2)}(\I x) &=  \frac{2 \I}{\pi} e^{\frac{\I \pi \beta}{2}} K_{\beta}(e^{\I \pi } x),
\end{align}
which then can be composed with
\begin{align}
K_{\beta}(e^{- \I \pi} x) & = e^{\I \pi \beta} K_{\beta}(x) + \I \pi I_{\beta}(x), & K_{\beta}(e^{\I \pi} x) & = e^{-\I \pi \beta} K_{\beta}(x) - \I \pi I_{\beta}(x)
\end{align}
to obtain
\begin{align} \label{HtoK2}
H_{\beta}^{(1)}(- \I x) &= 2 e^{-\frac{\I \pi \beta}{2}} I_{\beta}(x) - \frac{2 \I}{\pi} e^{\frac{\I \pi \beta}{2}} K_{\beta}(x), &
H_{\beta}^{(2)}(\I x) &= 2 e^{\frac{\I \pi \beta}{2}} I_{\beta}(x) + \frac{2 \I}{\pi} e^{-\frac{\I \pi \beta}{2}} K_{\beta}(x).
\end{align}
This is how we will now get Bessel-$I$'s into the game.

We also have to deal with the Heaviside thetas. We deform the contour as presented in figure \ref{fig:contours}. Since the integrand is continuous, the contributions from both arcs cancel and we have to substitute
\begin{align}
\Theta(\tau_1 - \tau_2) \longmapsto \Theta(z_2 - z_1).
\end{align}

To write the analytic continuations of the propagators we define
\begin{align}\label{AdSIdef}
\mathcal{I}_{\beta}(q, z) & = 2^{\beta - 1} \Gamma(\beta) q^{-\beta} z^{\frac{d}{2}} I_{\beta}(q z). \end{align}
We find
\begin{align}
G_{+-}(q, \I z_1, \I z_2) & = e^{\I \pi (\beta - \frac{d}{2})} q^{-2 \beta} \k_{\beta} \mathcal{K}_{\beta}(q, z_1) \mathcal{K}_{\beta}(q, z_2) \nn\\
& \qquad\qquad + \I e^{- \frac{\I \pi d}{2}} \mathcal{K}_{\beta}(q, z_1) \mathcal{I}_{\beta}(q, z_2)
\end{align}
and
\begin{align}
G_{+-}(q, \I z_1, \I z_2) = \left[ G_{-+}(q, -\I z_1, -\I z_2) \right]^{\ast} = G_{-+}(q, \I z_2, \I z_1) = \left[ G_{+-}(q, - \I z_2, - \I z_1) \right]^{\ast}.
\end{align}
This leads to
\begin{align}\label{Gppcont}
G_{\pm \pm}(q, \pm \I z_1, \pm \I z_2) & = e^{\pm \I \pi ( \beta - \frac{d}{2})} q^{-2 \beta} \k_{\beta} \mathcal{K}_{\beta}(q, z_1)\mathcal{K}_{\beta}(q, z_2) \nn\\
& \qquad \pm \I e^{\mp \frac{\I \pi d}{2}} \mathcal{G}_{\beta}(q, z_1, z_2),
\end{align}
where $\mathcal{G}_{\beta}$ denotes the AdS bulk-to-bulk propagator,
\begin{align}\label{AdSGdef}
\mathcal{G}_{\beta}(q, z_1, z_2) = \mathcal{K}_{\beta}(q, z_1) \mathcal{I}_{\beta}(q, z_2) \Theta(z_1 - z_2) + \mathcal{K}_{\beta}(q, z_2) \mathcal{I}_{\beta}(q, z_1) \Theta(z_2 - z_1).
\end{align}

With all these analytic continuations we can finally evaluate \eqref{Ipp} and \eqref{Ipm}. We obtain the following amplitudes
\begin{align}\label{Ippampl}
I_{++} & = - e^{\I \pi d} \int_0^{\infty} \frac{\D z_1}{z_1^{d+1}} G_{+}(q_1, \I z_1) G_{+}(q_2, \I z_1) \times \nn\\
& \qquad\qquad\qquad\qquad \times \int_0^{\infty} \frac{\D z_2}{z_2^{d+1}} G_{++}(s, \I z_1, \I z_2) G_{+}(q_3, \I z_2) G_{+}(q_4, \I z_2) \nn\\
& = - (-\tau_0)^{2d-\beta_T}  e^{\frac{\I \pi}{2} (\beta_T - d)} \prod_{j=1}^4 \k_{\beta_j} q_j^{-2 \beta_j} \times \left[ \I \times \ifin_{[\Delta_1 \Delta_2; \Delta_3 \Delta_4 x \Delta_x]} + \right.\nn\\
& \qquad\qquad\qquad\qquad \left. + e^{\I \pi \beta_x} \k_{\beta_x} s^{-2 \beta_x} \ifin_{[\Delta_1 \Delta_2 \Delta_x]}(q_1, q_2, s) \ifin_{[\Delta_x \Delta_3 \Delta_4]}(s, q_3, q_4) \right], \\
\label{Ipmampl}
I_{+-} & = (-\tau_0)^{2d-\beta_T} e^{\frac{\I \pi}{2} ( \beta_1 + \beta_2 - \beta_3 - \beta_3)} \prod_{j=1}^4 \k_{\beta_j} q_j^{-2 \beta_j} \times\nn\\
& \qquad\qquad \times \k_{\beta_x} s^{-2 \beta_x} \ifin_{[\Delta_1 \Delta_2 \Delta_x]}(q_1, q_2, s) \ifin_{[\Delta_x \Delta_3 \Delta_4]}(s, q_3, q_4),
\end{align}
where $\ino_{[\Delta_1 \Delta_2; \Delta_3 \Delta_4 x \Delta_x]}$ denotes the AdS 4-point exchange amplitude as defined in \cite{Bzowski:2022rlz}. All in all the 4-point function amplitude reads
\begin{align}\label{4ptindSsin}
& \dsno_{[\Delta_1 \Delta_2; \Delta_3 \Delta_4 x \Delta_x]} = 2 \lim_{\tau_0 \rightarrow 0^{-}} (-\tau_0)^{ \Delta_T-4d} \re \left[ I_{++} + I_{+-} \right] \nn\\
& \qquad = 2 \prod_{j=1}^4 \k_{\beta_j} q_j^{-2 \beta_j} \left[ \ifin_{[\Delta_1 \Delta_2; \Delta_3 \Delta_4 x \Delta_x]} \times \sin \left( \frac{\pi}{2} (\beta_T - d) \right) \right. \nn\\
& \qquad\qquad\qquad + 2 \k_{\beta_x} s^{-2 \beta_x} \ifin_{[\Delta_1 \Delta_2 \Delta_x]}(q_1, q_2, s) \ifin_{[\Delta_x \Delta_3 \Delta_4]}(s, q_3, q_4)  \times\nn\\
& \qquad\qquad\qquad\qquad \left. \times \sin \left( \frac{\pi}{2} (\beta_1 + \beta_2 + \beta_x - \frac{d}{2} ) \right) \sin \left( \frac{\pi}{2} (\beta_x + \beta_3 + \beta_4 - \frac{d}{2} ) \right)  \right].
\end{align}
The factors under the sines work out correctly in such a way that, for finite or regulated amplitudes one can substitute
\begin{align}
\sin \left[ \frac{\pi}{2} ( \beta_T - d) \right] \ireg_{[\Delta_1 \Delta_2; \Delta_3 \Delta_4 x \Delta_x]}(q_1, q_2, q_3, q_4, s) & = \im \ireg_{[\Delta_1 \Delta_2; \Delta_3 \Delta_4 x \Delta_x]}(\I q_1, \I q_2, \I q_3, \I q_4, \I s), \\
\sin \left( \frac{\pi}{2} (\beta_1 + \beta_2 + \beta_x - \frac{d}{2} ) \right) \ireg_{[\Delta_1 \Delta_2 \Delta_x]}(q_1, q_2, s) & = \im \ireg_{[\Delta_1 \Delta_2 \Delta_x]}(\I q_1, \I q_2, \I s).
\end{align}
Finally, we can rewrite the factors of $q_j^{-2 \beta_j}$ and $s^{-2 \beta_j}$ using \eqref{qBeta_to_2pt}. In this way one finds
\begin{align} \label{4pt_exchange}
\dsreg_{[\Delta_1 \Delta_2; \Delta_3 \Delta_4 x \Delta_x]} & = \frac{1}{8} \prod_{j=1}^4 \frac{1}{\im \ireg_{[\Delta_j \Delta_j]}(\I q_j)} \left[ \im \ireg_{[\Delta_1 \Delta_2; \Delta_3 \Delta_4 x \Delta_x]}(\I q_j, \I s) \right.\nn\\
& \qquad\qquad \left. - \, \frac{\im \ireg_{[\Delta_1 \Delta_2 \Delta_x]}(\I q_1, \I q_2, \I s) \im \ireg_{[\Delta_x \Delta_3 \Delta_4]}(\I s, \I q_3, \I q_4)}{\im \ireg_{[\Delta_x \Delta_x]}(\I s)} \right].
\end{align}

\section{Regularisation and renormalisation}
\label{sec:regandren}

\subsection{Outline of renormalisation}

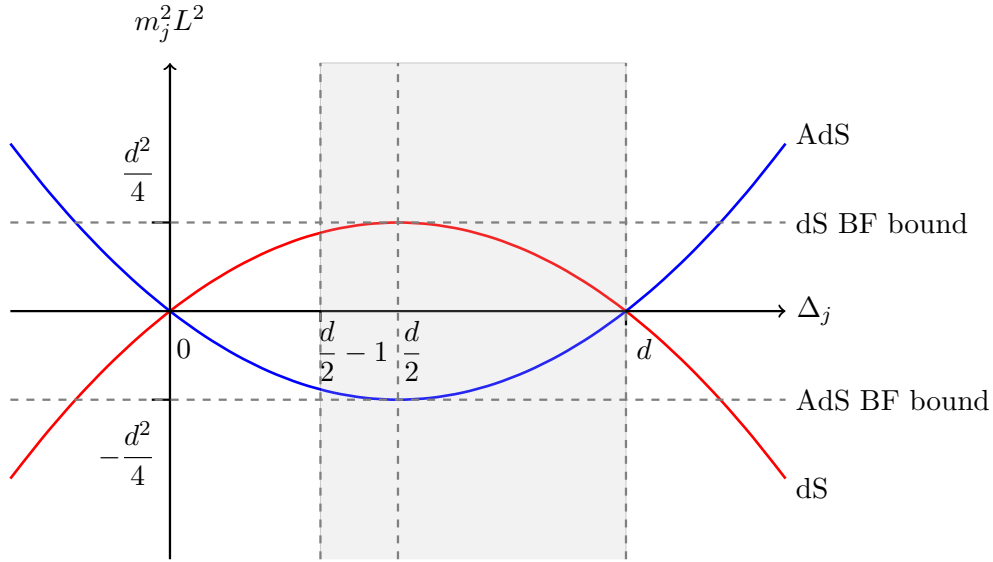
\begin{figure}[t]

\hspace{1.6cm}\begin{tikzpicture}[domain=-0.35:1.3, xscale = 6.0,yscale=4.7]
\draw[red, line width = 0.35mm]   plot[smooth,domain=-0.35:1.35] (\x, {\x-\x*\x});
\draw[blue, line width = 0.35mm]   plot[smooth,domain=-0.35:1.35] (\x, {-\x+\x*\x});
\draw[->, line width=0.3mm] (-0.35,0)--(1.35,0);
\draw[->, line width=0.3mm] (0,-0.7)--(0,0.7);
\draw[line width=0.3mm] (1,-0.04)--(1,0);
\draw[fill=gray!50!white,opacity=0.2]  (0.33,-0.7)--(0.33,0.7)--(1,0.7)--(1,-0.7);
\node[below] at (1.04,-0.05) {$d$};
\node[below] at (0.03,-0.05) {$0$};
\node[right] at (1.352,0) {$\Delta_j$};
\node[above] at (0,0.73) {$m_j^2L^2$};
\node[right] at (1.35,0.5) {AdS};
\node[right] at (1.35,-0.5) {dS};

\draw[line width=0.3mm] (0,0.25)--(-0.04,0.25);
\draw[line width=0.3mm] (0,-0.25)--(-0.04,-0.25);
\node[left] at (-0.01,0.4) {$\dfrac{d^2}{4}$};
\node[left] at (-0.01,-0.4) {$-\dfrac{d^2}{4}$};
\node[below] at (0.4,0.00) {$\dfrac{d}{2}-1$};
\node[below] at (0.53,0.0) {$\dfrac{d}{2}$};
\draw[line width=0.3mm] (0.33,-0.03)--(0.33,0);
\draw[dashed, gray, line width=0.3mm] (0.33,-0.7)--(0.33,0.7);
\draw[dashed, gray, line width=0.3mm] (0.5,-0.7)--(0.5,0.7);
\draw[dashed, gray, line width=0.3mm] (1,-0.7)--(1,0.7);
\draw[dashed, gray, line width=0.3mm] (-0.35,0.25)--(1.35,0.25);
\draw[dashed, gray, line width=0.3mm] (-0.35,-0.25)--(1.35,-0.25);
\node[right] at (1.35,-0.25) {AdS BF bound};
\node[right] at (1.35,0.25) {dS BF bound};
\end{tikzpicture}

\caption{The squared mass as a function of the operator dimension $\Delta_j$ in AdS (blue) and dS (red).  The shaded region indicates the unitarity bound $d/2-1\le \Delta_j\le d$.}
\end{figure}

In the previous section we expressed 2-, 3-, and 4-point regulated de Sitter amplitudes in terms of regulated AdS amplitudes. In general some of the presented expressions are divergent. In the cosmology literature the spacetime dimension $d$ and the masses $m_j^2$ (equivalently the dimensions  $\Delta_j$) are usually considered as fixed parameters  and the de Sitter correlation functions are calculated with a cut-off $\tau_0$ on the time coordinate $\tau \leq \tau_0 < 0$, see for example  \cite{Falk:1992sf, Zaldarriaga:2003my, Seery:2008qj, Creminelli:2011mw}.  Recall that we are considering here light scalars,  $0\leq m_j^2 \leq d^2/4L_{dS}^2$. The upper limit in the this inequality is the de Sitter analogue 
\cite{Skenderis:2006fb} of the Breitenlohner-Freedman (BF) bound \cite{Breitenlohner:1982bm, Breitenlohner:1982jf}. With masses in this range, the dual operators are relevant or marginal with dimension above the unitarity bound, $d/2-1 \leq \Delta_j \leq d$ \footnote{The range of dimensions $d/2-1 \leq \Delta_j < d/2$ (realised in the mass range $d^2/4-1\leq m^2 L_{dS}^2<d^2/4$) is special, as discussed in the AdS context in \cite{Klebanov:1999tb}. In this paper we restrict ourselves to $\Delta_j > d/2$.}.
Irrelevant operators $\Delta_j >d$ are mapped to negative masses in dS, while massive fields with positive mass outside the above range correspond to tachyonic fields with mass below the BF bound in AdS. Such fields correspond to operators with complex dimensions and the systematic discussion of such cases from the perspective of holography is still an open question, see \cite{Isono:2020qew} for work in this direction. 

When $d/2 < \Delta_j < d$  the asymptotic expansion of the de Sitter bulk field $\j$ has the form \eqref{dSasympt}, with $\jsrc$ being the most leading coefficient. When $\Delta_j=d/2$, the solution of the free equation of motion $\varphi_{\{0\}}$ (see \eqref{fieldexpinlambda}-(\ref{AdSinteqn}))
already contains logarithms, see \cite{Bianchi:2001de, Bianchi:2001kw}, 
and similarly when $\Delta_j = d$ the perturbative solution to the Klein-Gordon equation \eqref{dSeom} contains more leading, logarithmic terms, and the perturbative solution, $\j_{\{k\}}$, of order $k$ may contain $\log^k (-\tau)$ terms, see \cite{Bzowski:2015pba}. In order to define the limit $\tau_0 \to 0$ and, consequently, the boundary correlation functions, one must renormalise the theory. This can be done similarly to holographic renormalisation in AdS/CFT, with suitable counterterms added at a cut-off surface at $\tau_0 < 0$. Then, the limit $\tau_0 \rightarrow 0$ can be taken. 

Let us recall that holographic renormalisation in AdS involves both the addition of local counterterms and renormalisation of the sources. While the addition of counterterms was considered since the early days of AdS/CFT  \cite{Henningson:1998gx, Henningson:1998ey}, the need to renormalise the sources was realised much later \cite{Bzowski:2015pba}\footnote{One reason for this is that such renormalisation is needed only for 3-point functions and higher.} and is less well-known. Let us briefly summarise how this works. Let $\varphi^i_{(0)}(\x)$ be the fields parametrising the boundary conditions of the bulk fields $\varphi^i(z, \x)$, as in \eqref{AdSasympt},  where $i$ is an index enumerating bulks field (in general of different mass and interactions). Then to derive renormalised correlators we need a counterterm action $S_{\rm ct}[\varphi_{(0)}^i; \varepsilon]$, which is local in $\varphi^i_{(0)}$, where $\varepsilon$ is (any) regulator, and in addition we need to renormalise the sources, schematically as
\begin{equation} \label{ren_sou}
\varphi^i_{(0)} = \varphi^i_{(0)}[\phi_{(0)}^j,\varepsilon]= \phi_{(0)}^i + \frac{1}{\varepsilon} \Box^{k_1} \phi_{(0)}^{j_1}   \Box^{k_2} \phi_{(0)}^{j_2} + \cdots
\end{equation}
where $k_1, k_2$ are integers and the derivatives are along the boundary directions (in general all possible contractions of the derivatives must considered). In this schematic expression  the pole is a placeholder for singularity -- the singularity may also logarithmic. 
Such renormalisation is possible only when the spectrum of the theory contains operators of suitable dimension such that the second term on the right-hand side of \eqref{ren_sou} has the same scaling dimension as the first one. This is precisely one of the conditions for the appearance of short-distance singularities in 3-point functions \cite{Bzowski:2013sza, Bzowski:2015pba}. The dots indicate additional contributions that may be needed to renormalise short-distance singularities in higher-point functions. Then renormalised correlators are obtained by functionally differentiating with respect to $\phi^i_{(0)}$ the renormalised on-shell action,
\begin{equation}
S_{\rm ren}[\phi^i_{(0)}] = \lim_{\varepsilon \to 0} \left(S_{\rm on-shell}[ \varphi^i_{(0)}[\phi_{(0)}^j,\varepsilon]; \varepsilon] + S_{\rm ct}[ \varphi^i_{(0)}[\phi_{(0)}^j,\varepsilon]; \varepsilon] \right)\, .
\end{equation}
Now recall that in the AdS/CFT correspondence the field parametrising the boundary conditions, $\varphi^i_{(0)}$, is considered as the source that couples to the corresponding gauge invariant operator $\O$, namely in the CFT we have the coupling
\begin{equation} \label{O_coupling}
S_{\rm CFT}[\varphi^i_{(0)}, \O_i] = \int d^d x \varphi^i_{(0)} \O_i\, .
\end{equation}
Then the renormalisation of the sources amounts to additional counterterms on the CFT side that renormalise these couplings of the schematic form,
\begin{equation} \label{CFT_ct}
S^{\rm ct}_{\rm CFT} [\phi^i_{(0)},\O_i; \varepsilon] = \frac{1}{\varepsilon} \int d^d x  \Box^{k_1} \phi_{(0)}^{j_1}   \Box^{k_2} \phi_{(0)}^{j_2}
\O_i 
\end{equation} 
As we will see below, the usual AdS counterterms $S_{\rm ct}[\varphi_{(0)}^i; \varepsilon]$ are projected out when we go from AdS to dS but the analogues of \eqref{CFT_ct} survive.
 
 In this paper we will not use a time cut-off. Rather, just as in its AdS counterparts \cite{Bzowski:2022rlz}, we will work within dimensional regularisation and renormalisation.  This means that we set $\tau_0 = 0$ \emph{from the beginning} and keep $d$ and $\Delta_j$ as free parameters. As the amplitudes are analytic in $d$ and $\Delta_j$, one may generically define the amplitudes by starting from values of $d, \Delta_j$ where the amplitudes are finite and then analytically continue to the desired $d, \Delta_j$.  Only for specific values of the dimensions amplitudes cannot be defined in this way and require explicit subtractions. This happens on a hyperplane (or, in more special cases, intersections of hyperplanes) in the space spanned by $d$ and $\Delta_j$. 
In such a case boundary counterterms should be added to the effective action to yield the renormalised amplitude finite.

We regulate the theory by shifting the dimensions $d$ and $\Delta_j$ away from the singular point. As in case of AdS calculations, the regulator will be denoted by $\ep$ and we use the \emph{beta-scheme} where the regulated dimensions $\dreg$ and $\Dreg_j$ are
\begin{align} \label{half-beta_scheme}
& d \, \longmapsto \, \dreg = d + 2 \ep, && \Delta_j \, \longmapsto \, \Dreg_j = \Delta_j + \ep.
\end{align}
The idea behind the beta-scheme is that the combinations appearing as the orders of the Bessel functions in expressions such as \eqref{AdSK} and \eqref{dSK} remain unregulated,
\begin{align}
\beta_j = \Delta_j - \frac{d}{2} \, \longmapsto \, \reg{\beta}_j = \beta_j.
\end{align}
The beta-scheme regulates all correlation functions except for 2-point functions. However, for the case of $d = 3$ analyzed in this paper, the 2-point functions under consideration are all finite.

In order to renormalise\footnote{Note that this renormalisation is about IR divergences at tree-level. We are not addressing UV issues and IR issues at loop level in this paper.} the de Sitter amplitudes computed using \eqref{Z_SK} we want to add a counterterm action at $\tau=0$. The counterterm action $S_{\rm ct}= S_{ct}[\jsrc, J_+, J_-; \mu, \act_j;\epsilon]$ is a functional of the boundary value $\jsrc$ of the bulk field and the sources $J_+, J_-$. It also depends on the renormalisation constant $\mu$ and a set of renormalisation constants $\act_j$ parameterising scheme dependence. It is entirely contained within the boundary at $\tau = 0$ and it only contributes a boundary term to correlation functions. Two complex conjugate copies of the boundary counterterm must be then included in the Schwinger-Keldysh generating functional \eqref{Z_SK} by replacing
\begin{align} \label{StoSct}
& S_{+}[\varphi_{+}] \longmapsto S_{+}[\varphi_{+}] + S_{\rm ct}[\jsrc, J_{+}], && S_{-}[\varphi_{-}] \longmapsto S_{-}[\varphi_{-}] + S_{\rm ct}[\jsrc, J_{-}]
\end{align}
When the contribution from the counterterms is added to the regulated correlation functions, the limit $\ep \rightarrow 0$ limit should exist.

The counterterm action $S_{\rm ct}$ must be real, $S_{\rm ct} = S^{\ast}_{\rm ct}$, just as the original de Sitter action. 
Any counterterm that only depends on $\varphi_{(0)}$ will cancel between the forward and backward path of the Schwinger-Keldysh path integral. Such terms correspond to the traditional AdS counterterms. Another way to see that the contribution of the traditional counterterms vanishes is to note that they are analytic in momenta and analytically continuing from AdS to dS 
and taking the imaginary part projects these terms out. Counterterms that only depend on $\varphi_{(0)}$ are responsible for conformal anomalies in AdS \cite{Henningson:1998gx}. The fact that these terms cancel means that there are no conformal anomalies in de Sitter. This point was also made recently in  \cite{Chakraborty:2023yed, Chakraborty:2023los}.

Now, note that the coupling $J_+$ and $J_-$ to $\varphi_{(0)}$ at $\tau=0$ 
in \eqref{Z_SK} is the same as the coupling of $\varphi_{(0)}$  to $O$ in \eqref{O_coupling}.\footnote{Recall that $\lim_{\tau \to 0^-} \varphi_+ (\tau, \x) =\lim_{\tau \to 0^-} \varphi_- (\tau, \x)= \varphi_{(0)}(\x)$.} Then the AdS results suggest that we need a counterterm of the form \eqref{CFT_ct}. Thus we expect that we should be able to renormalise the de Sitter amplitudes using a counterterm of the form,
\begin{equation} \label{dS_ct}
S_{\rm ct}^{dS}[\varphi_{(0)}, (J_+-J_-); \mu, \act_j;\epsilon] = \int d^d x (J_+-J_-) f\left(\varphi_{(0)}; \mu, \act_j; \epsilon \right)
\end{equation}
for a suitable local function $f$ of $\varphi_{(0)}$ (that also depends on the regulator $\epsilon$, a renormalisation scale $\mu$ and constants $\act_j$ parametrising the scheme dependence). We will indeed see that such counterterm is sufficient to renormalise all cases. In a sense holographic renormalisation in dS is much simpler than in AdS.

Usually to carry out the renormalisation procedure the action of the regulated theory, $S$, must be specified. Then, it is the correlation functions in this theory that get renormalised. Here, however, we will follow \cite{Bzowski:2022rlz} and renormalise tree-level amplitudes instead. As discussed in \cite{Bzowski:2022rlz}, each such amplitude is a conformal correlator in some holographic CFT described by a suitable bulk action.
As far as 3- and 4-point functions are concerned, we will consider the ``asymmetric'' theories given by the actions \eqref{S3} and \eqref{S4} and apply the renormalisation procedure. When this is achieved, we effectively renormalise the amplitudes $\dsno_{[\Delta_1 \Delta_2 \Delta_3]}$ as well as $\dsno_{[\Delta_1 \Delta_2 \Delta_3 \Delta_4]}$ and $\dsno_{[\Delta_1 \Delta_2; \Delta_3 \Delta_4 x \Delta_x]}$ in \eqref{ds3def} and \eqref{ds4def}. As these are obtained in the least symmetric theory, any other 3- and 4-point functions in more symmetric theories can be obtained as linear combinations of these amplitudes. Thus, any renormalised correlation function can then be expressed in terms of the renormalised amplitudes.

\subsubsection{Results}

The procedure outlined above leads to renormalised de Sitter amplitudes. These objects, denoted by $\dsren$, depend on the renormaalisation scale $\mu_{dS}$ as well as some scheme-dependent renormalisation constants $\act_i^{dS}$, in addition to their momentum dependence, $\dsren = \dsren(q_i; \mu_{dS}, \act^{dS}_i)$. The main question we want to address in this section is whether the renormalised de Sitter amplitudes, $\dsren$, can be obtained from renormalised anti-de Sitter amplitudes, $\iren$, by applying the same analytic continuations as in \eqref{2pt}, \eqref{3pt}, \eqref{4pt_contact} and \eqref{4pt_exchange}. In other words, we want to write
\begin{align}
\dsren_{[\Delta \Delta]}(q) & = - \frac{1}{2} \frac{1}{\im \iren_{[\Delta \Delta]}(\I q)}, \label{amp2} \\[1ex]
\dsren_{[\Delta_1 \Delta_2 \Delta_3]}(q_i) & = - \frac{1}{4} \frac{\im \iren_{[\Delta_1 \Delta_2 \Delta_3]}(\I q_i)}{\prod_{j=1}^3 \im \iren_{[\Delta_j \Delta_j]}(\I q_j)}, \\[1ex]
\dsren_{[\Delta_1 \Delta_2 \Delta_3 \Delta_4]}(q_i) & = - \frac{1}{8} \frac{\im  \iren_{[\Delta_1 \Delta_2 \Delta_3 \Delta_4]}(\I q_i)}{\prod_{j=1}^4 \im \iren_{[\Delta_j \Delta_j]}(\I q_j)}, \\[1ex]
\dsren_{[\Delta_1 \Delta_2; \Delta_3 \Delta_4 x \Delta_x]}(q_i, s) & = \frac{1}{8} \prod_{j=1}^4 \frac{1}{\im \iren_{[\Delta_j \Delta_j]}(\I q_j)} \left[ \im \iren_{[\Delta_1 \Delta_2; \Delta_3 \Delta_4 x \Delta_x]}(\I q_i, \I s) \right.\nn\\
& \qquad\qquad \left. - \, \frac{\im \iren_{[\Delta_1 \Delta_2 \Delta_x]}(\I q_1, \I q_2, \I s) \im \iren_{[\Delta_x \Delta_3 \Delta_4]}(\I s, \I q_3, \I q_4)}{\im \iren_{[\Delta_x \Delta_x]}(\I s)} \right]. \label{amp4x}
\end{align}
The two sides depend on different sets of renormalisation data. On the left-hand side, the amplitudes depend on the renormalisation scale $\mu_{dS}$ and a set of scheme-dependent constants $\act_i^{dS}$. On the right-hand side, the renormalised AdS amplitudes, $\iren$, depend on the AdS renormalisation scale $\mu_{AdS}$ and another set of scheme-dependent constants $\bct_i^{AdS}$. Can we equate the variables on both sides so that \eqref{amp2}-\eqref{amp4x} hold?
 
We can always equate the renormalisation scales
\begin{align} \label{mu_eq}
\mu = \mu_{dS} = \mu_{AdS}
\end{align}
and from now on we use $\mu$ only. While the scheme-dependent constants $\act_i^{dS}$ and $\bct_i^{AdS}$ generally do not match, there always exists a choice of the scheme-dependent constants $\bct_i^{AdS}$ on the AdS side such that the formulae \eqref{amp2}-\eqref{amp4x} hold.  
More precisely, there exists polynomial functions which map AdS scheme-dependent constants to dS constants, $\act^{dS}_i = \act^{dS}_i (\bct^{AdS}_j)$, such that the holographic formulae \eqref{amp2}-\eqref{amp4x} hold. These functions do not depend on the momenta, nor the renormalisation scale, nor any other constants (such as the coupling constants) present in the system.  For the precise formulation, see subsection \ref{sec:sub_renorm}.

\subsection{Divergences}

All de Sitter 2-point amplitudes $\dsreg_{[\Delta \Delta]}$ are finite. As discussed earlier, this differs from the AdS case, where the amplitudes $\ireg_{[\Delta \Delta]}$ are divergent whenever $\beta = \Delta - \frac{d}{2}$ is a non-negative integer. Most of the de Sitter amplitudes are less singular than their AdS counterparts, but not all. The orders of the poles at $\ep = 0$ are listed in the tables below.

The dS 3-point amplitudes under consideration are either finite or exhibit a linear divergence as $\ep \rightarrow 0$. Their degrees of divergence are listed in the second column of table \ref{fig:deg3}, while the third column contains the degrees of divergence of the corresponding AdS amplitudes.

\begin{table}[t]
\begin{tabular}{|c|c|c|}
\hline
3-point amplitude & dS amplitude $\dsreg_{[\Delta_1 \Delta_2 \Delta_3]}$ & AdS amplitude $\ireg_{[\Delta_1 \Delta_2 \Delta_3]}$ \\  \hline
$[222]$ & 0 & 1 \\ \hline
$[322]$ & 1 & 1 \\ \hline
$[332]$ & 0 & 1 \\ \hline
$[333]$ & 1 & 1 \\ \hline
\end{tabular}
\centering
\caption{Degrees of divergence of 3-point dS and AdS amplitudes.\label{fig:deg3}}
\end{table}

The exchange 4-point amplitudes may exhibit up to a double pole in $\ep$ as presented in table \ref{fig:deg4}. Notice that while usually de Sitter amplitudes are less singular than their AdS counterparts, this is not always the case. For example, $\dsreg_{[22;22x3]}$ exhibits a double pole, while $\ireg_{[22;22x3]}$ is finite.
For a more general discussion of where singularities arise in (A)dS contact and exchange diagrams, see appendix \ref{app:sings}.

\begin{table}[t]
\begin{tabular}{|c|c|c|c|c|c|c|}
\hline
& \multicolumn{3}{c|}{de Sitter amplitudes} & \multicolumn{3}{c|}{Anti-de Sitter amplitudes} \\ \hline
External operators & Contact & $\Delta_x = 2$ & $\Delta_x = 3$ & Contact & $\Delta_x = 2$ & $\Delta_x = 3$ \\ \hline
$[22;22]$ & 0 & 0 & 2 & 0 & 0 & 0 \\ \hline
$[32;22]$ & 0 & 1 & 1 & 1 & 2 & 1 \\ \hline
$[33;22]$ & 1 & 1 & 2 & 1 & 1 & 2 \\ \hline
$[32;32]$ & 1 & 2 & 1 & 1 & 2 & 1 \\ \hline
$[33;32]$ & 0 & 1 & 1 & 1 & 2 & 2 \\ \hline
$[33;33]$ & 1 & 1 & 2 & 1 & 1 & 2 \\ \hline
\end{tabular}
\centering
\caption{Degrees of divergence of the contact and exchange 4-point (A)dS amplitudes.\label{fig:deg4}}
\end{table}

\subsection{Renormalisation} \label{sec:sub_renorm}

To prove that the formulae \eqref{amp2}-\eqref{amp4x} hold, we must renormalise both sides separately and compare the results. First, are the de Sitter amplitudes renormalisable at all? The de Sitter action \eqref{dSaction} is the analytic continuation of the AdS action \eqref{EAdSaction}, which suggests that we can analytically continue the AdS counterterm action to obtain a corresponding dS counterterm action. As we discussed, this suggests that the de Sitter counterterms are of the form \eqref{dS_ct}, and we indeed find that such action is sufficient to renormalise the dS amplitudes.

The renormalised dS amplitudes are then defined as
\begin{align} \label{imren2}
\dsren(q_i; \mu, \act^{dS}_i) & = \lim_{\ep \rightarrow 0} \left[ \dsreg(q_i;\epsilon) + \dsct(q_i; \mu, \act_i;\epsilon) \right],
\end{align}
where (as usual) the counterterm contribution, $\dsct(q_i; \mu, \act_i)$, depends on the renormalisation scale $\mu$ and a set of scheme-dependent constants, $\act_i$. 
Equations \eqref{amp2}-\eqref{amp4x} contain imaginary parts of analytically continued renormalised AdS amplitudes  on the right-hand side. These are by definition
\begin{align} \label{imren1}
\im \iren(\I q_i; \mu, \bct_i) & = \im \lim_{\ep \rightarrow 0} \left[ \ireg(\I q_i;\epsilon) + \ict(\I q_i; \mu, \bct_i;\epsilon) \right],
\end{align}
where $\ireg$ is the AdS regulated amplitude and $\ict$ the AdS counterterm contribution (as mentioned earlier, we take without loss of generality  $\mu_{dS} = \mu_{AdS}=\mu$).
The AdS counterterms are devised in such a way that the counterterm contribution $\ict$ cancels the divergences of the regulated amplitude, $\ireg$. The question is then whether the scheme-dependence matches or not.  As already discussed, there are more counterterms in AdS than in dS, so in general there are more scheme-dependent constants in AdS than in dS. In the following we show that the scheme-dependence of the correlators also matches provided  the scheme-dependent constant are mapped to each other via polynomial map, $\bct_k = \bct_k (\act_i)$.

In order to establish the exact form of the map  we must consider how the renormalisation procedures in dS and AdS are related. Essentially, the problem boils down to the question whether the $\ep \rightarrow 0$ limit and taking the imaginary part commute,
\begin{align} \label{imren3}
\im \lim_{\ep \rightarrow 0} \left[ \ireg(\I q_i;\epsilon) + \ict(\I q_i; \mu, \bct_i;\epsilon) \right] \overset{?}{=} \lim_{\ep \rightarrow 0} \left[ \im \ireg(\I q_i;\epsilon) + \im \ict(\I q_i; \mu, \bct_i;\epsilon) \right].
\end{align}
In other words, is the dS counterterm contribution $\dsct(q_j; \mu, \act_i)$ equal to the imaginary part of the analytically continued AdS contribution? In general, the answer is no for the scheme-dependent part, but we find that this part can also be made to agree provided the scheme-dependent constants are connected via a non-trivial map. 

In summary, the relation between renormalised amplitudes is 
\begin{align}
&\dsren_{[\Delta \Delta]}(q)  = - \frac{1}{2} \frac{1}{\im \iren_{[\Delta \Delta]}(\I q; \mu, \bct)}, \label{Amp2} \\[1ex]
& \dsren_{[\Delta_1 \Delta_2 \Delta_3]}(q_i; \mu, \act_i) = - \frac{1}{4} \frac{\im \iren_{[\Delta_1 \Delta_2 \Delta_3]}(\I q_i)}{\prod_{j=1}^3 \im \iren_{[\Delta_j \Delta_j]}(\I q_j; \mu, \bct_k(\act_i))}, \label{Amp3} \\[1ex]
& \dsren_{[\Delta_1 \Delta_2 \Delta_3 \Delta_4]}(q_i; \mu, \act_i) = - \frac{1}{8} \frac{\im  \iren_{[\Delta_1 \Delta_2 \Delta_3 \Delta_4]}(\I q_i; \mu, \bct_k(\act_i))}{\prod_{j=1}^4 \im \iren_{[\Delta_j \Delta_j]}(\I q_j)}, \\[1ex]
& \dsren_{[\Delta_1 \Delta_2; \Delta_3 \Delta_4 x \Delta_x]}(q_i, s; \mu, \act_i) = \frac{1}{8} \prod_{j=1}^4 \frac{1}{\im \iren_{[\Delta_j \Delta_j]}(\I q_j)} \left[ \im \iren_{[\Delta_1 \Delta_2; \Delta_3 \Delta_4 x \Delta_x]}(\I q_i, \I s; \mu, \bct_k(\act_i)) \right.\nn\\
& \qquad\qquad \left. - \, \frac{\im \iren_{[\Delta_1 \Delta_2 \Delta_x]}(\I q_1, \I q_2, \I s; \mu, \bct_k(\act_i)) \im \iren_{[\Delta_x \Delta_3 \Delta_4]}(\I s, \I q_3, \I q_4; \mu, \bct_k(\act_i))}{\im \iren_{[\Delta_x \Delta_x]}(\I s)} \right], \label{Amp4x}
\end{align}
where $\bct_k(\act_i)$ is a map between the scheme-dependent constants that we discuss below.
In the following, we present a derivation of these formulae.

\subsubsection{2-point amplitudes}

At two points, de Sitter amplitudes are finite for any $\beta > 0$ and are given by \eqref{ds2pt}. Thus, there is no scheme-dependence for dS 2-point functions. 
The AdS amplitudes, on the other hand, become singular at integral, non-negative $\beta$'s. Since their renormalisation requires a counterterm containing two CFT sources and no operator, we are in the case where the dS amplitude must remain finite. It would be convenient, however, to be able to relate the (trivially renormalised) de Sitter amplitude to the renormalised AdS amplitude, so that we can use renormalised expressions on both sides of \eqref{amp2}. 

Let $\beta = n$ be a non-negative integer. The renormalised AdS amplitude reads
\begin{align}
\iren_{[n + \frac{d}{2}, n + \frac{d}{2}]}(q; \mu, \bct) & = \frac{(-1)^n 4^{1-n}}{(n-1)!^2} q^{2n} \left( - \log \frac{q}{\mu} + \bct \right),
\end{align}
where $\mu$ is the renormalisation scale and $\bct$ is a scheme-dependent constant adjustable by a change in $\mu$. By plugging this expression into \eqref{amp2} we find that
\begin{align}
- \frac{1}{2 \im \iren_{[n + \frac{d}{2}, n + \frac{d}{2}]}(\I q; \mu)} = \frac{4^{n-1} (n-1)!^2}{\pi} q^{-2n} = \dsren_{[n + \frac{d}{2}, n + \frac{d}{2}]}(q).
\end{align}
The renormalisation scale and the scheme-dependent constant drops out and thus for all $\beta > 0$, including integral ones, we can write
\begin{align}
\dsren_{[\Delta \Delta]}(q) = - \frac{1}{2 \im \iren_{[\Delta \Delta]}(\I q; \mu, \bct)}.
\end{align}
The left-hand side is scheme independent. 

\subsubsection{Higher-point functions}

In this subsection we discuss the higher-point functions. To keep technicalities to a minimum while still discussing all non-trivial issues, we will  carry out the renormalisation procedure for a single scalar field of dimension $\Delta = 3$. We carry out the procedure directly in de Sitter spacetime and compare it to the AdS result obtained via formulae \eqref{amp2}-\eqref{amp4x}. We consider a single massless scalar field, $\j$, and the regulated de Sitter action
\begin{align} \label{SdSall3}
S_{dS} = - \int \D^{4+2 \ep} x \sqrt{-g} \left[ \frac{1}{2} \partial_{\mu} \j \partial^{\mu} \j + \frac{1}{2} \ep (3 + \ep) \j^2 + \frac{1}{6} \lambda_{[333]} \j^3 - \frac{1}{24} \lambda_{[3333]}  \j^4 \right].
\end{align}
The 3- and 4-point functions are expressed in terms of amplitudes as
\begin{align}
& \lla \jsrc(\bs{q}_1) \jsrc(\bs{q}_2) \jsrc(\bs{q}_3) \rra = \lambda_{[333]} \dsno_{[333]}(q_j) + O(\lambda_{[333]}^2), \\
& \lla \jsrc(\bs{q}_1) \jsrc(\bs{q}_2) \jsrc(\bs{q}_3) \jsrc(\bs{q}_4) \rra = \lambda_{[3333]} \dsno_{[3333]}(q_j) \nn\\
& \qquad\qquad \lambda_{[333]}^2 \left[ \dsno_{[33;33x3]}(q_j, s) + (t-\text{channel}) + (u-\text{channel}) \right] + \ldots
\end{align}

Looking at tables \ref{fig:deg3} and \ref{fig:deg4}, 
we see that the amplitudes $\dsreg_{[333]}$ and $\dsreg_{[3333]}$ are linearly divergent, while $\dsreg_{[33;33x3]}$ exhibits double pole. This is the same situation as in AdS case and thus we consider the dS counterterm action inspired by the AdS case,
\begin{align} \label{SdSctall3}
S^{dS}_{\text{ct}}[\j_{(0)}, J] = \int \D^{\dreg} \bs{x} \left[ \frac{1}{2} \lambda_{[333]} \mathfrak{r}_{[333]} \j_{(0)}^2 J + \left( \frac{1}{2} \lambda_{[333]}^2 \mathfrak{r}_{[33;33x3]} + \frac{1}{6} \lambda_{[3333]} \mathfrak{r}_{[3333]} \right) \j_{(0)}^3 J \right],
\end{align}
where $\mathfrak{r}_{[333]}$, $\mathfrak{r}_{[3333]}$ and $\mathfrak{r}_{[33;33x3]}$ are counterterm constants that we will fix shortly.

\subsubsection*{Comparison to AdS}

The regulated AdS action corresponding to \eqref{SdSctall3} reads
\begin{align}
S_{AdS} = \int \D^{4+2 \ep} x \sqrt{g} \left[ \frac{1}{2} \partial_{\mu} \j \partial^{\mu} \j - \frac{1}{2} \ep (3 + \ep) \j^2 + \frac{1}{6} \lambda_{[333]} \j^3 - \frac{1}{24} \lambda_{[3333]} \j^4 \right].
\end{align}
With this convention, the 3- and 4-point functions of the operator $\O$ dual to the AdS field $\j$ read
\begin{align}
& \lla \O(\bs{q}_1) \O(\bs{q}_2) \O(\bs{q}_3) \rra = \lambda_{[333]} \ino_{[333]}(q_j) + O(\lambda_{[333]}^2), \\
& \lla \O(\bs{q}_1) \O(\bs{q}_2) \O(\bs{q}_3) \O(\bs{q}_4) \rra = \lambda_{[3333]} \ino_{[3333]}(q_j) \nn\\
& \qquad\qquad + \lambda^2_{[333]} \left[ \ino_{[33;33x3]}(q_j, s) + (t-\text{channel}) + (u-\text{channel}) \right] + \ldots
\end{align}
where $\ino_{[333]}$ and $\ino_{[33;33x3]}$ can represent both regulated and renormalised AdS amplitudes. 

The renormalisation of the theory requires the addition of the boundary counterterm actions. The AdS counterterm action is given by the $Z_{[0]}$ part of equation (4.143) (with $Z_{[0]}$ given in (4.145)) of \cite{Bzowski:2022rlz} and reads
\begin{align} \label{SAdSctall3}
S^{AdS}_{\text{ct}}[\phi, \O] = \int \D^{3 + 2 \ep} \bs{x} \left[ \frac{1}{2} \lambda_{[333]} \mathfrak{s}_{[333]} \phi^2 \O + \left( \frac{1}{2} \lambda_{[333]}^2 \mathfrak{s}_{[33;33x3]} + \frac{1}{6} \lambda_{[3333]} \mathfrak{s}_{[3333]} \right) \phi^3 \O \right],
\end{align}
where $\phi$ is the source for $\O$. The values of the counterterm constants are given by equations (4.30) and (4.77)  of \cite{Bzowski:2022rlz} and read
\begin{align} \label{s333}
\mathfrak{s}_{[333]} & = \frac{1}{3} \Gamma(\ep) \mu^{-\ep} \left[ 1 + \ep \,  \mathfrak{b}^{(1)}_{[333]} + \ep^2 \mathfrak{b}^{(2)}_{[333]} + O(\ep^3) \right], \\
\mathfrak{s}_{[3333]} & = - \frac{1}{3} \Gamma(2\ep) \mu^{-2\ep} \left[ 1 + \ep \, \mathfrak{b}^{(1)}_{[3333]} + O(\ep^2) \right], \label{s3333} \\
\mathfrak{s}_{[33;33x3]} & = \frac{1}{18} \Gamma^2(\ep) \mu^{-2\ep} \left[ 1 + \ep \, \left( 2 \mathfrak{b}^{(1)}_{[333]} - \tfrac{1}{3} \right) + \ep^2 \mathfrak{b}^{(2)}_{[33;33x3]} + O(\ep^3) \right].
\end{align}
The constants $\mathfrak{b}^{(1)}_{[333]}, \mathfrak{b}^{(2)}_{[333]}, \mathfrak{b}^{(1)}_{[3333]}$ and $\mathfrak{b}^{(2)}_{[33;33x3]}$ parametrise the scheme-dependence. The omitted higher order terms do not contribute to 3- and 4-point functions. It is crucial, however, that the first subleading terms in both terms is related, \textit{i.e.}, $\mathfrak{b}^{(1)}_{[33;33x3]} = 2 \mathfrak{b}^{(1)}_{[333]} - \tfrac{1}{3}$.

With the presented counterterm action the renormalised AdS amplitudes become
\begin{align} \label{iren333}
\iren_{[333]}(q_i; \mu, \mathfrak{b}^{(1)}_{[333]}) & = \lim_{\ep \rightarrow 0} \Big[ \ireg_{[333]}(q_i) - \mathfrak{s}_{[333]} \sum_{j=1}^3 \ireg_{[33]}(q_j) \Big], \\
\iren_{[3333]}(q_i; \mu, \mathfrak{b}^{(1)}_{[3333]}) & = \lim_{\ep \rightarrow 0} \Big[ \ireg_{[3333]}(q_i) + \mathfrak{s}_{[3333]} \sum_{j=1}^4 \ireg_{[33]}(q_j) \Big],
\end{align}
and
\begin{align} \label{iren3333x3}
& \iren_{[33;33x3]}(q_i, s; \mu, \mathfrak{b}^{(1)}_{[333]}, \mathfrak{b}^{(2)}_{[333]}, \mathfrak{b}^{(2)}_{[33;33x3]}) = \lim_{\ep \rightarrow 0} \Big[ \ireg_{[33;33x3]}(q_i, s)  \nn\\
& \qquad\qquad  - \, \mathfrak{s}_{[333]} \left( \ireg_{[333]}(q_1, q_2, s) + \ireg_{[333]}(s, q_3, q_4) \right) + \mathfrak{s}^2_{[333]} \ireg_{[33]}(s) + \mathfrak{s}_{[33;33x3]} \sum_{j=1}^4 \ireg_{[33]}(q_j) \Big].
\end{align}
We can substitute these to the (naive) holographic formula \eqref{amp2}-\eqref{amp4x} and compare with the actual  renormalised dS amplitude. Let us denote such amplitudes by $\overline{\dsno}^{\,\text{ren}}$:
\begin{align}
\overline{\dsno}^{\,\text{ren}}_{[333]}(q_i; \mu, \bct_i) & = - \frac{1}{4} \frac{\im \iren_{[333]}(\I q_i; \mu, \bct_i)}{\prod_{j=1}^3 \im \iren_{[33]}(\I q_j)}, \\
\overline{\dsno}^{\,\text{ren}}_{[33;33x3]}(q_i, s; \mu, \bct_i) & = \frac{1}{8} \prod_{j=1}^4 \frac{1}{\im \iren_{[33]}(\I q_j)} \left[ \im \iren_{[33;33x3]}(\I q_i, \I s; \mu, \bct_i) \right.\nn\\
& \qquad \left. - \, \frac{\im \iren_{[333]}(\I q_1, \I q_2, \I s; \mu, \bct_i) \im \iren_{[333]}(\I s, \I q_3, \I q_4; \mu, \bct_i)}{\im \iren_{[33]}(\I s)} \right].
\end{align}
Are these equal to the genuine renormalised de Sitter amplitudes?

\subsubsection*{3-point function}

We include the counterterm action \eqref{SdSctall3} in the generating functional \eqref{Z_SK}. By taking functional derivatives with respect to $J$ we obtain additional insertions of higher powers of $\j_{(0)}$, which lead to the renormalised dS correlations functions. As far as the 3-point amplitude goes, we have
\begin{align}
& \< \jsrc(\bs{x}_1) \jsrc(\bs{x}_2) \jsrc(\bs{x}_3) \>_{\text{ren}} = \nn\\
& \qquad = \lim_{\ep \rightarrow 0} \< \left[ \jsrc + \frac{1}{2} \lambda_{[333]} \mathfrak{r}_{[333]} \jsrc^2 \right] (\bs{x}_1) \left[ \jsrc + \frac{1}{2} \lambda_{[333]} \mathfrak{r}_{[333]} \jsrc^2 \right] (\bs{x}_2) \times \nn\\
& \qquad\qquad\qquad\qquad \times \left[ \jsrc + \frac{1}{2} \lambda_{[333]} \mathfrak{r}_{[333]} \jsrc^2 \right] (\bs{x}_3) \>_{\text{reg}} + O(\lambda_{[333]}^2) \nn\\
& \qquad = \lim_{\ep \rightarrow 0} \left[ \< \jsrc(\bs{x}_1) \jsrc(\bs{x}_2) \jsrc(\bs{x}_3) \>_{\text{reg}} \right.\nn\\
& \qquad\qquad\qquad + \lambda_{[333]} \mathfrak{r}_{[333]} \left(  \< \jsrc(\bs{x}_1) \jsrc(\bs{x}_2) \>_{\text{reg}} \< \jsrc(\bs{x}_1) \jsrc(\bs{x}_3) \>_{\text{reg}} \right. \nn\\
& \qquad\qquad\qquad\qquad \left.\left. + \, (\bs{x}_1 \leftrightarrow \bs{x}_2) + (\bs{x}_1 \leftrightarrow \bs{x}_3) \right) \right] + O(\lambda_{[333]}^2).
\end{align}
After  Fourier transforming we obtain,
\begin{align}
\dsren_{[333]}(q_i) & = \lim_{\ep \rightarrow 0} \left[ \dsreg_{[333]}(q_i) + \mathfrak{r}_{[333]} \left( \dsreg_{[33]}(q_1) \dsreg_{[33]}(q_2) + \rm cyclic\ in\ q_1 \to q_2 \to q_3
\right) \right] \nn\\
& = - \frac{1}{4} \lim_{\ep \rightarrow 0} \left[ \frac{\im \ireg_{[333]}(\I q_i) - \mathfrak{r}_{[333]} \sum_{j=1}^3 \im \ireg_{[33]}(\I q_j)}{\prod_{j=1}^3 \im \ireg_{[33]}(\I q_j)} \right].
\end{align}
The limit of the 2-point functions is finite and thus we can replace them with the renormalised expression $\im \iren_{[33]}(\I q_j)$. 
Using that the scaling dimensions $\reg{D}_{[333]}$ and $\reg{D}_{[33]}$ of the regulated AdS amplitudes $\ireg_{[333]}$ and $\ireg_{[33]}$, respectively, are $\reg{D}_{[333]}  = 3 - \ep,\reg{D}_{[33]}  = 3$, we obtain
\begin{align}
& \lim_{\ep \rightarrow 0} \left[ \im \ireg_{[333]}(\I q_i) - \mathfrak{r}_{[333]} \sum_{j=1}^3 \im \ireg_{[33]}(\I q_j) \right] = \nn\\
& \qquad = \lim_{\ep \rightarrow 0} \left[ \sin \left( \frac{\pi}{2} \reg{D}_{[333]} \right) \ireg_{[333]}(q_i) - \mathfrak{r}_{[333]} \sin \left( \frac{\pi}{2} \reg{D}_{[33]} \right) \sum_{j=1}^3 \ireg_{[33]}(\I q_j) \right] \nn\\
& \qquad = - \lim_{\ep \rightarrow 0} \left[ \left(1 - \frac{\pi^2 \ep^2}{8} + O(\ep^4) \right) \ireg_{[333]}(q_i) - \mathfrak{r}_{[333]} \sum_{j=1}^3 \ireg_{[33]}(q_j) \right].
\end{align}
Since the regulated amplitude is linearly divergent, the term of order $\ep^2$ is irrelevant when the limit $\ep \rightarrow 0$ is taken. Thus, if we take $\mathfrak{r}_{[333]} = \mathfrak{s}_{[333]}$ the expression reduces to \eqref{iren333}. On the other hand the renormalised AdS amplitude $\iren_{[333]}$ takes form
\begin{align}
\iren_{[333]}(q_i; \mu) = f_0(q_i) + f_1(q_i) \log \left( \frac{q_i}{\mu} \right),
\end{align}
where $f_0$ and $f_1$ are homogeneous functions of degree $3$. Thus,
\begin{align}
\im \iren_{[333]}(\I q_i; \mu) & = - \iren_{[333]}(q_i; \mu) \nn\\
& = \lim_{\ep \rightarrow 0} \left[ \im \ireg_{[333]}(\I q_i) - \mathfrak{s}_{[333]} \sum_{j=1}^3 \im \ireg_{[33]}(\I q_j) \right] 
\end{align}
and we arrive at the expected equality,
\begin{align}
\dsren_{[333]}\left(q_i; \mu, \act_{[333]}^{(1)}=\bct_{[333]}^{(1)}\right) = \overline{\dsno}^{\text{ren}}_{[333]}(q_i; \mu, \bct_{[333]}^{(1)}).
\end{align}
where we explicitly included how the scheme-dependent constant matches. 

\subsubsection*{Exchange 4-point function}

The situation is more complicated for the 4-point function. By including both terms in the counterterm action \eqref{SdSctall3}, we obtain
\begin{align}
& \< \jsrc(\bs{x}_1) \jsrc(\bs{x}_2) \jsrc(\bs{x}_3) \jsrc(\bs{x}_4) \>_{\text{ren}} \nn\\[1ex]
& = \lim_{\ep \rightarrow 0} \left\{ \< \jsrc(\bs{x}_1) \jsrc(\bs{x}_2) \jsrc(\bs{x}_3) \jsrc(\bs{x}_4) \>_{\text{reg}} \right.\nn\\
& + \lambda_{[333]} \mathfrak{r}_{[333]} \< \jsrc(\bs{x}_1) \jsrc(\bs{x}_2) \>_{\text{reg}} \< \jsrc(\bs{x}_1) \jsrc(\bs{x}_3) \jsrc(\bs{x}_4) \>_{\text{reg}} + 11 \text{ perms.} \nn\\
& + \lambda_{[333]}^2 \mathfrak{r}_{[333]}^2 \< \jsrc(\bs{x}_1) \jsrc(\bs{x}_2) \>_{\text{reg}} \< \jsrc(\bs{x}_2) \jsrc(\bs{x}_3) \>_{\text{reg}} \< \jsrc(\bs{x}_3) \jsrc(\bs{x}_4) \>_{\text{reg}} + 11 \text{ perms.} \nn\\
& \left. + \, 3 \lambda_{[333]}^2 \mathfrak{r}_{[33;33x3]} \< \jsrc(\bs{x}_1) \jsrc(\bs{x}_2) \>_{\text{reg}} \< \jsrc(\bs{x}_1) \jsrc(\bs{x}_3) \>_{\text{reg}} \< \jsrc(\bs{x}_1) \jsrc(\bs{x}_4) \>_{\text{reg}} + 3 \text{ perms.} \right\} \nn\\
& + O(\lambda_{[333]}^3).
\end{align}
We keep terms of order $\lambda_{[333]}^2$, split all the terms into those corresponding to $s$-, $t$- and $u$-channels, and Fourier transform to momentum space. This yields the renormalised dS amplitude
\begin{align}
& \dsren_{[33;33x3]}(q_i, s; \mu, \act_i) = \lim_{\ep \rightarrow 0} \left\{ \dsreg_{[33;33x3]}(q_i, s) \right.\nn\\
& \qquad + \mathfrak{r}_{[333]} \left[ \dsreg_{[333]}(q_3, q_4, s) \left( \dsreg_{[33]}(q_1) + \dsreg_{[33]}(q_2) \right) \right. \nn \\
& \qquad \qquad \quad \left.+ \dsreg_{[333]}(q_1, q_2, s) \left( \dsreg_{[33]}(q_3) + \dsreg_{[33]}(q_4) \right) \right] \nn\\
& \qquad + \mathfrak{r}_{[333]}^2 \dsreg_{[33]}(s) \left[ \dsreg_{[33]}(q_1) \dsreg_{[33]}(q_3) + \dsreg_{[33]}(q_1) \dsreg_{[33]}(q_4) \right. \nn \nn \\
& \qquad \qquad \qquad \qquad \quad \left. + \dsreg_{[33]}(q_2) \dsreg_{[33]}(q_3) + \dsreg_{[33]}(q_2) \dsreg_{[33]}(q_4) \right] \nn\\
& \qquad \left. + \, \mathfrak{r}_{[33;33x3]} \left[ \dsreg_{[33]}(q_1) \dsreg_{[33]}(q_2) \dsreg_{[33]}(q_3) + 3 \text{ perms.} \right] \right\} \nn\\
& = - \frac{1}{8} \prod_{j=1}^4 \frac{1}{\im \iren_{[33]}(\I q_j)} \lim_{\ep \rightarrow 0} \Big\{ \im \ireg_{[33;33x3]}(\I q_i, \I s) + \mathfrak{r}_{[33;33x3]} \sum_{j=1}^4 \im \ireg_{[33]}(\I q_j) \nn\\
& \qquad - \frac{1}{\im \ireg_{[33]}(\I s)} \left[ \im \ireg_{[333]}(\I q_1, \I q_2, \I s) - \mathfrak{r}_{[333]} \left( \im \ireg_{[33]}(\I q_1) + \im \ireg_{[33]}(\I q_2) \right) \right] \times\nn\\
& \qquad\qquad\qquad \left. \times \left[ \im \ireg_{[333]}(\I s, \I q_3, \I q_4) - \mathfrak{r}_{[333]} \left( \im \ireg_{[33]}(\I q_3) + \im \ireg_{[33]}(\I q_4) \right) \right] \right\}.
\end{align}
The terms in the square brackets lack one term to be re-composed into \eqref{iren333}. With the value of $\mathfrak{r}_{[333]}$ already fixed by the renormalisation of the 3-point function and equal to $\mathfrak{s}_{[333]}$ in \eqref{s333}, we can rewrite this expression as
\begin{align} \label{towards_ds3333x3}
& \dsren_{[33;33x3]}(q_i, s; \mu, \act_i) = - \frac{1}{8} \prod_{j=1}^4 \frac{1}{\im \iren_{[33]}(\I q_j)} \left\{ A(q_i, s; \mu, \act_{[333]}^{(1)}) \right.\nn\\
& \qquad  + \, \lim_{\ep \rightarrow 0} \Big[ \im B_{\ep}(\I q_i, \I s; \mu, \bct_k(\act_i)) + \left( \mathfrak{s}_{[333]}^2 - \mathfrak{s}_{[33;33x3]} + \mathfrak{r}_{[33;33x3]} \right) \sum_{j=1}^4 \im \ireg_{[33]}(\I q_j) \Big] \Big\},
\end{align}
where
\begin{align}
A(q_i, s; \mu, \act_{[333]}^{(1)}) & = - \frac{\im \iren_{[333]}(\I q_1, \I q_2, \I s; \mu, \act_{[333]}^{(1)}) \im \iren_{[333]}(\I s, \I q_3, \I q_4; \mu, \act_{[333]}^{(1)})}{\im \ireg_{[33]}(\I s)}
\end{align}
and $B_{\epsilon}$ denotes the combination in \eqref{iren3333x3} entering the renormalisation of the AdS 4-point amplitude,
\begin{align}
B_{\epsilon}(q_i, s; \mu, \bct_k(\act_i) ) & = \ireg_{[33;33x3]}(q_i, s) - \mathfrak{s}_{[333]} \left( \ireg_{[333]}(q_1, q_2, s) + \ireg_{[333]}(s, q_3, q_4) \right) \nn\\
& \qquad\qquad + \mathfrak{s}^2_{[333]} \ireg_{[33]}(s) + \mathfrak{s}_{[33;33x3]} \sum_{j=1}^4 \ireg_{[33]}(q_j)
\end{align}
so that $\iren_{[33;33x3]}(q_i, s; \mu, \bct_k) = \lim_{\ep \rightarrow 0} B_{\ep}(q_i, s; \mu, \bct_k)$. At this point, the dependence between the dS counterterm constants $\act_i$ and the AdS constants $\bct_i$ is not yet fixed, with the exception of $\act_{[333]}^{(1)} = \bct_{[333]}^{(1)}$. The imaginary parts lead to factors of $\sin(\pi \reg{D}/2)$, where the regulated dimensions of the amplitudes under consideration are
\begin{align}
\reg{D}_{[33;33x3]} & = 3 - 2 \ep, & \reg{D}_{[333]} & = 3 - \ep, & \reg{D}_{[33]} & = 3.
\end{align}
Thus, the subleading terms in the expansion of the sines do contribute. Indeed, we find
\begin{align}
 \lim_{\ep \rightarrow 0} \im B_{\epsilon}(\I q_i, \I s; \mu, \bct_k) &= - \Big[ \left( 1 - \frac{\pi^2 \ep^2}{2} \right) \ireg_{[33;33x3]}(q_i, s) \nn\\
& \qquad\quad - \, \left( 1 - \frac{\pi^2 \ep^2}{8} \right) \mathfrak{s}_{[333]} \left( \ireg_{[333]}(q_1, q_2, s) + \ireg_{[333]}(s, q_3, q_4) \right) \nn\\
& \qquad\quad  + \mathfrak{s}^2_{[333]} \ireg_{[33]}(s) + \mathfrak{s}_{[33;33x3]} \sum_{j=1}^4 \ireg_{[33]}(q_j) + O(\ep) \Big] \nn\\
&  = - \iren_{[33;33x3]}(q_i, s; \mu, \bct_k) + \frac{\pi^2}{72} \left( q_1^3 + q_2^3 + q_3^3 + q_4^3 + 2 s^2 \right).
\end{align}
On the other hand the renormalised AdS amplitude, $\iren_{[33;33x3]}$, contains square of the logarithm of the renormalisation scale, $\log^2 \mu$. This means that under the analytic continuation we find
\begin{align}
& \im \iren_{[33;33x3]}(\I q_i, \I s; \mu, \bct_k) = - \iren_{[33;33x3]}(q_i, s; \mu, \bct_k) + \frac{\pi^2}{72} \left( q_1^3 + q_2^3 + q_3^3 + q_4^3 + 2 s^2 \right).
\end{align}
Thus, when put together,
\begin{align}
& \lim_{\ep \rightarrow 0} \im B_{\epsilon}(\I q_i, \I s; \mu, \bct_k ) =  \lim_{\ep \rightarrow 0} \im \left[ \ireg_{[33;33x3]}(q_i, s) + \ict_{[33;33x3]}(q_i, s; \mu, \bct_k) \right] \nn\\
& \qquad = \im \iren_{[33;33x3]}(\I q_i, \I s; \mu, \bct_k).
\end{align}
Substituting back to \eqref{towards_ds3333x3}, we end up with
\begin{align}
\dsren_{[33;33x3]}(q_i, s; \mu, \act_i) & = - \frac{1}{8} \prod_{j=1}^4 \frac{1}{\im \iren_{[33]}(\I q_j)} \Big[ \im \iren_{[33;33x3]}(\I q_i, \I s; \mu, \bct_k(\act_i) ) \nn\\
& \qquad - \frac{\im \iren_{[333]}(\I q_1, \I q_2, \I s; \mu, \act_{[333]}^{(1)}) \im \iren_{[333]}(\I s, \I q_3, \I q_4; \mu, \act_{[333]}^{(1)})}{\im \ireg_{[33]}(\I s)} \nn\\
& \qquad  - \left( \mathfrak{s}_{[333]}^2 - \mathfrak{s}_{[33;33x3]} + \mathfrak{r}_{[33;33x3]} \right) \sum_{j=1}^4 q_j^3 \Big].
\end{align}
In order to match our desired formula \eqref{Amp4x}, we must cancel the last term, \textit{i.e.}, the dS renormalisation constant $\mathfrak{r}_{[33;33x3]}$ is not equal to $\mathfrak{s}_{[33;33x3]}$, but rather
\begin{align}
\mathfrak{r}_{[33;33x3]} = \mathfrak{s}_{[33;33x3]} - \mathfrak{s}_{[333]}^2 + O(\ep).
\end{align}
Thus the scheme-dependent constants between the dS and AdS expressions are non-trivially related.

In total we find is that the scheme-dependence matches between dS and AdS for contact diagrams, since
\begin{align}
\mathfrak{r}_{[333]}(\act_{[333]}^{(1)}) & = \mathfrak{s}_{[333]}(\bct_{[333]}^{(1)}), && \text{with } \bct_{[333]}^{(1)} = \act_{[333]}^{(1)}, \\
\mathfrak{r}_{[3333]}(\act_{[3333]}^{(1)}) & = \mathfrak{s}_{[3333]}(\bct_{[3333]}^{(1)}), && \text{with } \bct_{[3333]}^{(1)} = \act_{[3333]}^{(1)},
\end{align}
where the AdS constants $\mathfrak{s}_{[333]}$ and $\mathfrak{s}_{[3333]}$ are given by \eqref{s333} and \eqref{s3333}. For the exchange diagram we have $\mathfrak{r}_{[33;33x3]} = \mathfrak{s}_{[33;33x3]}$, provided that the relation between dS and AdS scheme-dependent constants reads
\begin{align} \label{Arel2}
\mathfrak{b}^{(2)}_{[333]} & = \mathfrak{a}^{(2)}_{[333]}, \qquad\qquad \mathfrak{a}^{(1)}_{[33;33x3]} = 2 \mathfrak{a}^{(1)}_{[333]} + \frac{1}{3} = 2 \mathfrak{b}^{(1)}_{[333]} + \frac{1}{3}, \\
\bct^{(2)}_{[33;33x3]} & = - \act^{(2)}_{[33;33x3]} + 2 (\act_{[333]}^{(1)})^2 + 4 \act_{[333]}^{(2)}.
\end{align}

\subsubsection{Concluding remarks}

The analysis carried out here concerns only the 3- and 4-point renormalised amplitudes $\dsren_{[333]}$, $\dsren_{[3333]}$ and $\dsren_{[33;33x3]}$. We have further analysed all the remaining 3- and 4-point functions under consideration to show that the formulae \eqref{Amp2}-\eqref{Amp4x} hold in all cases. We evaluated explicitly the constants $a^{[0]}_1$ and $a^{[0]}_2$ from equation \eqref{intro:f} in the theory described by the action \eqref{SdSall3}. In the beta scheme, we found
\begin{align}
a^{[0]}_1 & = \frac{1}{2} \lambda_{[333]} \mathfrak{r}_{[333]}, \label{eq: a1} \\
a^{[0]}_2 & = \frac{1}{2} \lambda_{[333]}^2 \mathfrak{r}_{[33;33x3]} + \frac{1}{6} \lambda_{[3333]} \mathfrak{r}_{[3333]}, \label{eq: a_2}
\end{align}
where
\begin{align} 
\mathfrak{r}_{[333]} & = \frac{1}{3} \Gamma(\ep) \mu^{-\ep} \left[ 1 + \ep \,  \mathfrak{a}^{(1)}_{[333]} + \ep^2 \mathfrak{a}^{(2)}_{[333]} + O(\ep^3) \right], \label{eq:333}\\
\mathfrak{r}_{[3333]} & = - \frac{1}{3} \Gamma(2\ep) \mu^{-2\ep} \left[ 1 + \ep \, \mathfrak{a}^{(1)}_{[3333]} + O(\ep^2) \right], \label{eq: 3333}\\
\mathfrak{r}_{[33;33x3]} & = \frac{1}{18} \Gamma^2(\ep) \mu^{-2\ep} \left[ 1 + \ep \, \left( 2 \mathfrak{a}^{(1)}_{[333]} + \tfrac{1}{3} \right) + \ep^2 \mathfrak{a}^{(2)}_{[33;33x3]} + O(\ep^3) \right]. \label{eq: 3333x3}
\end{align}

\subsection{Results}

Here we list all renormalised de Sitter amplitudes. In all cases, \eqref{Amp2}-\eqref{Amp4x} are satisfied.

\subsubsection{Preliminaries}

Consider a real homogeneous function $f = f(\bs{q})$ of some momenta $\bs{q}$. Let $D$ be the homogeneity degree of $f$ (\textit{e.g.}, $f = q^D$ has the degree equal to $D$.) Then
\begin{align}
\im f(\I q) = f(q) \sin \left( \frac{\pi}{2} D \right).
\end{align}
With $D$ being an integer, this expression vanishes for even $D$'s and equals $\pm f(q)$ for odd $D$'s. For integral $D$'s we have
\begin{align} \label{sin_cos_int}
& \sin \left( \frac{\pi}{2} D \right) = \left\{  \begin{array}{cl}
0 & \text{ if } D \text{ is even}, \\
+1 & \text{ if } D \equiv 1 \text{ mod } 4, \\
-1 & \text{ if } D \equiv 3 \text{ mod } 4,
\end{array} \right. && \cos \left( \frac{\pi}{2} D \right) = \left\{  \begin{array}{cl}
0 & \text{ if } D \text{ is odd}, \\
+1 & \text{ if } D \equiv 0 \text{ mod } 4, \\
-1 & \text{ if } D \equiv 2 \text{ mod } 4.
\end{array} \right.
\end{align}

The homogeneity degree $D$ of the $n$-point AdS amplitude of operators of dimensions $\Delta_1, \ldots, \Delta_n$ reads
\begin{align} \label{homo_deg}
& D = \Delta_t - (n-1) d, && \Delta_t = \sum_{j=1}^n \Delta_j.
\end{align}
Only if the amplitude is divergence free, and thus no renormalisation is required, is it represented by a truly homogeneous function of degree $D$. Otherwise, renormalisation introduces scale-violating logarithms of the form $\log(q/\mu)$, where $\mu$ is the renormalisation scale. In such cases, the scale-violating logarithms and their powers multiply homogeneous functions of degree $D$. 

By placing the branch cuts of logarithms on the negative real axis we have
\begin{align}
\log \left( \frac{\I q}{\mu} \right) & = \log \left( \frac{q}{\mu} \right) + \frac{\I \pi}{2}.
\end{align}
Thus, depending on the homogeneity degree $D$ such logarithms may or may not survive the analytic continuation. Indeed, for AdS amplitudes with single logarithms, the dS amplitude will be scale-free if the naive degree is even. The precise form of the dS amplitude depends on the degree of divergence of the regulated AdS amplitude. Since the AdS amplitudes in this paper  exhibit at most a double pole in the regulator, we consider three cases as follows.

\paragraph{No divergence.} The AdS amplitude is scale-independent, $f = f(\bs{q})$ and we have
\begin{align}
\im f(\I q) = f(q) \sin \left( \frac{\pi}{2} D \right),
\end{align}
where $D$ is given by \eqref{homo_deg} and the sine simplifies according to \eqref{sin_cos_int}.

\paragraph{Linear divergence.} If the regulated AdS amplitude is linearly divergent, then it can be written in the form
\begin{align}
f(\bs{q}_j; \mu) = f_0(\bs{q}_j) + f_1(\bs{q}_j) \log \left( \frac{q_t}{\mu} \right),
\end{align}
where $f_0$ and $f_1$ are homogeneous functions of degree $D$. It follows that
\begin{align} \label{anal_cont_deg1}
\im f(\I \bs{q}; \mu) = f(\bs{q}; \mu) \sin \left( \frac{\pi}{2} D \right) + \frac{\pi}{2} \cos \left( \frac{\pi}{2} D \right) f_1(\bs{q}).
\end{align}
The sines and cosines can be simplified according to \eqref{sin_cos_int}.

\paragraph{Double pole.} Similarly, if the regulated amplitude exhibits a double pole, the renormalised amplitude takes form
\begin{align}
f(\bs{q}_j; \mu) = f_0(\bs{q}_j) + f_1(\bs{q}_j) \log \left( \frac{q_t}{\mu} \right)  + f_2(\bs{q}_j) \log^2 \left( \frac{q_t}{\mu} \right),
\end{align}
where $f_0, f_1, f_2$ are homogeneous functions of degree $D$. In such a case
\begin{align}
\im f(\I \bs{q}; \mu) & = \sin \left( \frac{\pi}{2} D \right) \left[ f(\bs{q}; \mu) - \frac{\pi^2}{4} f_2(\bs{q}) \right] \nn\\
& \qquad + \frac{\pi}{2} \cos \left( \frac{\pi}{2} D \right) \left[ f_1(\bs{q}) + 2 f_2(\bs{q}) \log \left( \frac{q_t}{\mu} \right) \right].
\end{align}

\subsubsection{2-point amplitudes}

The 2-point dS amplitudes read
\begin{align}
\dsren_{[22]}  = \frac{1}{2 q}, \qquad\qquad
\dsren_{[33]}  = \frac{1}{2 q^3}.
\end{align}

\subsubsection{3-point amplitudes}

All regulated AdS 3-point amplitudes are linearly divergent and thus each renormalised AdS amplitude contains a scale-violating logarithm containing the renormalisation scale $\mu$. Whether the dS amplitude contains a scale-violating logarithm depends on the dimensions $\Delta_j$. According to \eqref{anal_cont_deg1} and \eqref{sin_cos_int} the logarithm drops from $\dsren_{[222]}$ and $\dsren_{[332]}$, while remains present in $\dsren_{[322]}$ and $\dsren_{[333]}$. To be precise,
\begin{align}\label{ds222}
\dsren_{[222]} & = - \frac{\pi}{8 q_1 q_2 q_3}, \\[2ex]
\label{ds322}
\dsren_{[322]} & = \frac{\iren_{[322]}}{4 q_1^3 q_2 q_3} \nn\\
& = \frac{1}{4 q_1^3 q_2 q_3} \left\{ - q_1 + (q_2 + q_3) \left[ \log \left( \frac{q_t}{\mu} \right) + \act_{[322]}^{(1)} - 1 \right] \right\}, 
\end{align}
\begin{align}
\label{ds332}
\dsren_{[332]} & = - \frac{\pi}{16} \frac{q_1^2 + q_2^2 - q_3^2}{q_1^3 q_2^3 q_3}, \\[2ex]
\dsren_{[333]} & = - \frac{\iren_{[333]}}{4 q_1^3 q_2^3 q_3^3} \nn\\
& = - \frac{1}{12 q_1^3 q_2^3 q_3^3} \left\{ q_1^2 q_2 + q_2^2 q_1 + q_1^2 q_3 + q_3^2 q_1 + q_2^2 q_3 + q_3^2 q_2 - q_1 q_2 q_3 \right.\nn\\
& \qquad \left. - \, ( q_1^3 + q_2^3 + q_3^3)  \left[ \log \left( \frac{q_t}{\mu} \right) + \act_{[333]}^{(1)} - \frac{4}{3} \right] \right\}. \label{ds333}
\end{align}

\subsubsection{4-point contact amplitudes}

Here the situation is analogous to the 3-point functions. Only in three amplitudes the AdS 4-point contact diagram contributes non-trivially to the non-local part after analytic continuation:
\begin{align}
\label{dsren2222}
\dsren_{[2222]} & = \frac{\ino_{[2222]}}{8 q_1 q_2 q_3 q_4} = \frac{1}{8 q_1 q_2 q_3 q_4 q_T}, \\[2ex]
\label{dsren3322}
\dsren_{[3322]} & = - \frac{\iren_{[3322]}}{8 q_1^3 q_2^3 q_3 q_4} \nn\\
& = - \frac{1}{8 q_1^3 q_2^3 q_3 q_4} \left\{ (q_3 + q_4) \left[ \log \left( \frac{q_T}{\mu} \right) + \frac{1}{2} \act_{[3322]}^{(1)} \right] + \frac{q_1 q_2}{q_T} - q_T \right\}, \\[2ex]
\label{dsren3333}
\dsren_{[3333]} & = \frac{\iren_{[3333]}}{8 q_1^3 q_2^3 q_3^3 q_4^3} \nn\\
& = - \frac{1}{8 q_1^3 q_2^3 q_3^3 q_4^3} \left\{ \frac{1}{3} (q_1^3 + q_2^3 + q_3^3 + q_4^3) \left[ \log \left( \frac{q_T}{\mu} \right) + \frac{1}{2} \act_{[3333]}^{(1)} \right] \right.\nn\\
& \qquad\qquad\qquad \left. - \frac{\s{4}{1234}}{q_T} + q_T \s{2}{1234} - \frac{4}{9} q_T^3 \right\},
\end{align}
where $\s{k}{1234}$ is the $k$-th symmetric polynomial on $q_1, q_2, q_3, q_4$, see Appendix \ref{sec:notation}. In the remaining two amplitudes the imaginary part of the AdS 4-point contact amplitude is local and we get,
\begin{align}
\label{dsren3222}
\dsren_{[3222]} & = \frac{\pi}{16 q_1^3 q_2 q_3 q_4}, \\[2ex]
\label{dsren3332}
\dsren_{[3332]} & = \frac{\pi}{32} \frac{q_1^2+q_2^2+q_3^2-q_4^2}{q_1^3 q_2^3 q_3^3 q_4}. 
\end{align}
All analytic continuations have been performed in these formulae. For example $\ino_{[2222]}$ in \eqref{dsren2222} is the standard AdS amplitude  $\ino_{[2222]} = \ino_{[2222]}(q_i)=1/q_T$.

\subsubsection{4-point exchange amplitudes}
 
From \eqref{homo_deg} and \eqref{sin_cos_int} we see that if the total dimension of the external operators, $\Delta_T = \Delta_1 + \Delta_2 + \Delta_3 + \Delta_4$, is odd, the finite part containing the dilogarithm vanishes from the de Sitter amplitude. In such cases only logarithms and scheme-dependent terms survive. For the remaining cases the sine equals $\pm 1$, see table \ref{tab:sin}. The dimension of the exchange operator is irrelevant.
\begin{table}[t] 
\centering
\begin{tabular}{|c|c|}
\hline
\textbf{Amplitude} & $\bm{\sin ( \frac{\pi D}{2} )}$ \\ \hline
$22,22$ and $33,33$ & $-1$ \\ \hline
$33,22$ and $32,32$ & $+1$ \\ \hline
$32,22$ and $32,33$ & $0$ \\ \hline
\end{tabular}
\caption{Values of the sine present in \eqref{sin_cos_int} for various amplitudes.\label{tab:sin}}
\end{table}
There is no more analytic continuation in the formulae below, \textit{i.e.}, the AdS amplitudes on the right-hand sides do not involve analytically continued momenta. 
The definitions of the symbols used in the expressions below are presented in appendix \ref{sec:notation}. 

\paragraph{Exchange dimension $\bs{\Delta_x = 2}$.} With the exchange scalar of dimension $\Delta_x = 2$ we have the following amplitudes,
\begin{align} \label{ds2222x2}
\dsren_{[22,22x2]} & = \frac{1}{8 q_1 q_2 q_3 q_4} \left[ - \iren_{[22,22x2]} + \frac{\pi^2}{4 s} \right] \nn\\
& = - \frac{1}{8 q_1 q_2 q_3 q_4} \left. \iren_{[22,22x2]} \right|_{\PLp \mapsto \PLp + \frac{\pi^2}{2}} \nn\\
& = \frac{1}{16 q_1 q_2 q_3 q_4 s} \left[ \PLp + \frac{\pi^2}{2} \right].\\[2ex]
\dsren_{[33,22x2]} & = \frac{1}{8 q_1^3 q_2^3 q_3 q_4} \left[ \iren_{[33,22x2]} + \frac{\pi^2}{8 s} (q_1^2 + q_2^2 - s^2) \right] \nn\\
& = \frac{1}{8 q_1^3 q_2^3 q_3 q_4} \left. \iren_{[33,22x2]} \right|_{\PLp \mapsto \PLp + \frac{\pi^2}{2}} \nn\\
& = \frac{1}{16 q_1^3 q_2^3 q_3 q_4} \left\{ \frac{q_1^2 + q_2^2 - s^2}{2 s} \left( \PLp + \frac{\pi^2}{2} \right) + (q_1 + q_2) \left[ \log \left( \frac{\p{34}}{q_T} \right) + 1 \right]\right. \nn\\
& \qquad\qquad\qquad \left. - \, (q_3 + q_4) \left[ \log \left( \frac{q_T}{\mu} \right) + \frac{1}{2} \act_{[33,22x2]}^{(1)} - \frac{7}{4} \right] \right\}.\\[2ex]
\dsren_{[33,33x2]} & = \frac{1}{8 q_1^3 q_2^3 q_3^3 q_4^3} \left[ - \iren_{[33,33x2]} + \frac{\pi^2}{16 s} ( q_1^2 + q_2^2 - s^2) ( q_3^2 + q_4^2 - s^2) \right] \nn\\
& = - \frac{1}{8 q_1^3 q_2^3 q_3^3 q_4^3} \left. \iren_{[33,33x2]} \right|_{\PLp \mapsto \PLp + \frac{\pi^2}{2}}.
\end{align}
Since these amplitudes are related to each other by the raising/lowering operators, they share the same structure. As we can see these three dS amplitudes can be obtained from their AdS counterparts by the substitution $\PLp \mapsto \PLp + \frac{\pi^2}{2}$. Not all amplitudes with even $\Delta_T$ share this feature, though. In particular, we have the following amplitude,
\begin{align}
\dsren_{[32,32x2]} & = \frac{\iren_{[32,32x2]}}{8 q_1^3 q_2 q_3^3 q_4} + \frac{1}{8 q_1^3 q_2 q_3^3 q_4 s} \left\{ \left[ (s + q_2) \left( \log \left( \frac{\p{12}}{\mu} \right) - 1 + \act_{[322]}^{(1)} \right) - q_1 \right] \right. \times\nn\\
& \qquad\qquad\qquad \times \left[ (s + q_4) \left( \log \left( \frac{\p{34}}{\mu} \right) - 1 + \act_{[322]}^{(1)} \right) - q_3 \right] \nn\\
& \qquad\qquad \left. + \, \frac{\pi^2}{8} s ( 2 s + q_2 + q_4) \right\}.
\end{align}
Finally, in two amplitudes with odd $\Delta_T$ the analytic continuation of the AdS exchange diagrams is local and the non-trivial part of the amplitudes originates from the product of 3-point functions in \eqref{Amp4x},
\begin{align}
\dsren_{[32,22x2]} & = - \frac{\pi}{16 q_1^3 q_2 q_3 q_4 s} \left\{ q_2 \left[ \log \left( \frac{\p{12}}{\mu} \right) - 1 + \act_{[322]}^{(1)} \right] + s \log \left( \frac{\p{12}}{\p{34}} \right) - q_1 \right\},
\end{align}
\begin{align}
\dsren_{[32,33x2]} & = - \frac{\pi}{32 q_1^3 q_2 q_3^3 q_4^3 s} \left\{ (q_3^2 + q_4^2 - s^2) \left[ q_2 \left( \log \left( \frac{\p{12}}{\mu} \right) + \act_{[322]}^{(1)} \right) + s \log \left( \frac{\p{12}}{\p{34}} \right) \right] \right.\nn\\
& \qquad \left. + s^2 (q_1 + q_2 - q_3 - q_4) + \frac{s}{2} \left[ -q_1^2 + q_2^2 + (q_3 + q_4)^2 \right] - (q_1 + q_2)(q_3^2 + q_4^2) \right\},
\end{align}

\paragraph{Exchange dimension $\bs{\Delta_x = 3}$.} For the exchange $\Delta_x = 3$ scalar expressions are more complicated. The only dS amplitude, which can be obtained from the corresponding AdS amplitude by the substitution $\PLp \mapsto \PLp + \frac{\pi^2}{2}$ is
\begin{align}
\dsren_{[32,32x3]} & = \frac{1}{8 q_1^3 q_2 q_3^3 q_4} \left[ \iren_{[32,32x3]} + \frac{\pi^2}{16 s^3} (s^2 + q_1^2 - q_2^2) (s^2 + q_3^2 - q_4^2) \right] \nn\\
& = \frac{1}{8 q_1^3 q_2 q_3^3 q_4} \left. \iren_{[32,32x3]}  \right|_{\PLp \mapsto \PLp + \frac{\pi^2}{2}}.
\end{align}
The remaining amplitudes of even total degree $\Delta_T$ do not exhibit this feature and become more complicated,
\begin{align} \label{ds2222x3}
\dsren_{[22,22x3]} & = - \frac{\iren_{[22,22x3]}}{8 q_1 q_2 q_3 q_4} + \frac{1}{8 q_1 q_2 q_3 q_4 s^3} \left\{ (q_1 + q_2) \left[ \log \left( \frac{\p{12}}{\mu} \right) - 1 + \act_{[322]}^{(1)} \right]  - s \right\} \times\nn\\
& \qquad\qquad \times \left\{ (q_3 + q_4) \left[ \log \left( \frac{\p{34}}{\mu} \right) - 1 + \act_{[322]}^{(1)} \right]  - s \right\} \nn\\
& = - \frac{1}{8 q_1 q_2 q_3 q_4} \left\{ \frac{(q_1 + q_2)(q_3 + q_4)}{2 s^3} \left[ \PLp + \right.\right. \nn\\
& \qquad\qquad\qquad\qquad\qquad \left. - 2 \left( \log \left( \frac{\p{12}}{\mu} \right) - 1 + \act_{[322]}^{(1)} \right) \left( \log \left( \frac{\p{34}}{\mu} \right) - 1 + \act_{[322]}^{(1)} \right) \right] \nn\\
& \qquad\qquad \left. + \, \frac{q_T}{s^2} \left( \log \left( \frac{q_T}{\mu} \right) - 1 + \act_{[322]}^{(1)} \right) \right\}.
\end{align}
\begin{align}
\dsren_{[33,22x3]} & = \frac{\iren_{[33,22x3]}}{8 q_1^3 q_2^3 q_3 q_4} + \frac{1}{24 q_1^3 q_2^3 q_3 q_4 s^3} \left\{ \left[ (s^3 + q_1^3 + q_2^3) \left( \log \left( \frac{\p{12}}{\mu} \right)  - \frac{4}{3} + \act_{[333]}^{(1)} \right) \right.\right.\nn\\
& \qquad\qquad\qquad\qquad \left. - \s{1}{12s} \s{2}{12s} + 4 \s{3}{12s} \right] \times\nn\\
& \qquad\qquad \left. \times \left[ (q_3 + q_4) \left( \log \left( \frac{\p{34}}{\mu} \right) - 1 + \act_{[322]}^{(1)} \right) - s \right] + \frac{\pi^2}{8} s^3 (q_3 + q_4) \right\}.
\end{align}
\begin{align}
\dsren_{[33,33x3]} & = - \frac{\iren_{[33,33x3]}}{8 q_1^3 q_2^3 q_3^3 q_4^3} + \frac{1}{72 q_1^3 q_2^3 q_3^3 q_4^3 s^3} \times \nn\\
& \qquad\qquad \left\{ \frac{1}{s^3} \left[ (s^3 + q_1^3 + q_2^3) \left( \log \left( \frac{\p{12}}{\mu} \right) - \frac{4}{3} + \act_{[333]}^{(1)} \right) - \s{1}{12s} \s{2}{12s} + 4 \s{3}{12s} \right] \times \right. \nn\\
& \qquad\qquad\qquad \times \left[ (s^3 + q_3^3 + q_4^3) \left( \log \left( \frac{\p{34}}{\mu} \right) - \frac{4}{3} + \act_{[333]}^{(1)} \right) - \s{1}{s34} \s{2}{s34} + 4 \s{3}{s34} \right] \nn\\
& \qquad\qquad\qquad \left. + \frac{\pi^2}{8} \left( q_1^3 + q_2^3 + q_3^3 + q_4^3 + 2 s^3 \right) \right\}.
\end{align}
The remaining amplitudes of odd total dimension $\Delta_T$ receive non-trivial contribution only from the product of the 3-point function in \eqref{Amp4x},
\begin{align}
\dsren_{[32,22x3]} & = - \frac{\pi}{32 q_1^3 q_2 q_3 q_4 s^3} \left\{ (s^2 + q_1^2 - q_2^2) \left[ (q_3 + q_4) \left( \log \left( \frac{\p{34}}{\mu} \right) - 1 + \act_{[322]}^{(1)} \right) - s \right] + s^3  \right\},
\end{align}
\begin{align}
\dsren_{[32,33x3]} & = - \frac{\pi}{96 q_1^3 q_2 q_3^3 q_4^3} \left\{ \frac{1}{s^3} ( s^2 + q_1^2 - q_2^2)(s^3 + q_3^3 + q_4^3) \left[ \log \left( \frac{\p{34}}{\mu} \right) - \frac{4}{3} + \act_{[333]}^{(1)} \right] \right. \nn\\
& \qquad\qquad - (s^2 + q_1^2 - q_2^2) \left[ \log \left( \frac{\p{12}}{\mu} \right) - \frac{4}{3} + \act_{[333]}^{(1)} \right] \nn\\
& \qquad\qquad + s (q_1 - q_2 - q_3 - q_4) - \frac{1}{2} q_T (q_1 + q_2 - q_3 - q_4) \nn\\
& \qquad\qquad\qquad - \, \frac{(q_1^2 - q_2^2 + q_3 q_4) (q_3 + q_4)}{s} - \frac{(q_1^2 - q_2^2) (q_3^2 - q_3 q_4 + q_4^2)}{s^2} \nn\\
& \qquad\qquad\qquad \left. - \, \frac{(q_1^2 - q_2^2) q_3 q_4 (q_3 + q_4)}{s^3} \right\}.
\end{align}

\subsection{Comparison to the literature}

Relatively few exact expressions for de Sitter amplitudes  are available in the existing literature.
As far as  exchange 4-point functions are concerned, only the finite amplitude  $\dsno_{[22,22x2]}$ is abundantly present: see, for example, equations
(5.75) of \cite{Arkani-Hamed:2015bza}; 
(4.54) of \cite{Arkani-Hamed:2018kmz}; 
(4.52) of \cite{Sleight:2019mgd}; 
(4.84) of \cite{Sleight:2019hfp}; 
(3.41) of \cite{Baumann:2021fxj}. 
All these expressions match \eqref{ds2222x2}.

To our knowledge, the only other de Sitter exchange 4-point amplitude with non-derivative vertices to have been calculated previously is  $\dsren_{[22;22x3]}$.
This amplitude was first calculated in equation (4.90) of \cite{Sleight:2019hfp},
however the result obtained disagrees with ours here. 
This is not unexpected given the regulated  amplitude $\dsreg_{[22;22x3]}$ is divergent and requires renormalisation. Indeed,  our  result \eqref{ds2222x3} contains the renormalisation scale $\mu$ and a scheme-dependent constant, $\act_{[322]}^{(1)}$, while equation (4.90) of \cite{Sleight:2019hfp} does not. 
Our result is therefore anomalous under scale transformations while that of \cite{Sleight:2019hfp} transforms homogeneously.  
To see the mismatch more clearly, let $\dsno_{[22;22x3]}|_{*}$ denote   (4.90)  of \cite{Sleight:2019hfp}. It is easy to check that, in our conventions, it reads
\begin{align}
\dsno_{[22;22x3]}\Big|_{*} & = - \frac{1}{8 q_1 q_2 q_3 q_4} \left. \iren_{[22;22x3]}  \right|_{\PLp \mapsto \PLp + \frac{\pi^2}{2}} \nn\\
& = - \frac{1}{8 q_1 q_2 q_3 q_4} \left[ \iren_{[22;22x3]} + \frac{\pi^2}{4} \frac{(q_1 + q_2)(q_3 + q_4)}{s^3} \right]
\end{align}
where we fixed the overall normalisation constant to match our conventions. The resulting expression is scale independent and misses the terms proportional to the product of the 3-point functions in \eqref{amp4x}. By comparing with \eqref{ds2222x3}, we see that the difference is 
\begin{align}
\dsren_{[22;22x3]} - \dsno_{[22;22x3]}\Big|_{*} & = \frac{1}{8 q_1 q_2 q_3 q_4 s^3} \left[ \iren_{[223]}(q_1, q_2, s; \mu) \iren_{[322]}(s, q_3, q_4; \mu) \right.\nn\\
& \qquad\qquad\qquad\qquad \left. + \, \frac{\pi^2}{4} (q_1 + q_2)(q_3 + q_4) \right].
\end{align}

More recently, this same amplitude $\dsren_{[22;22x3]}$ was recomputed through bootstrap considerations in section 3.3 of \cite{Wang:2022eop}, see equations (3.35), (3.41), (3.42) and (3.45).
After accounting for differences in notation and the overall normalisation, our result \eqref{ds2222x3} agrees 
with that of \cite{Wang:2022eop} upon setting the scheme-dependent constant
\[
\act_{[322]}^{(1)} = \gamma_E+\log(-\tau_0\mu)
\]
where $\tau_0$ (denoted $\eta_0$ in \cite{Wang:2022eop}) is the late-time cut-off in conformal time.

\subsubsection{Shift operators and derivative vertices}

In \cite{Arkani-Hamed:2018kmz}, a family of shift operators $\mathcal{W}_{ij}^{\pm \pm}$ were introduced. When acting on a given AdS Witten diagram, these shift operators increase or decrease the value of the external dimensions $\Delta_i$ and $\Delta_j$ by one. The resulting expressions, however, represent Witten diagrams where the interactions contain additional derivatives acting on the fields of the form $\vphi(\partial\vphi)^2$ \cite{Bzowski:2022rlz}.\footnote{For AdS shift operators that map contact diagrams to contact diagrams, and exchange diagrams to exchange diagrams {\it without} introducing derivative vertices, see \cite{Caloro:2023cep}.} In the presence of such interactions, the propagators (bulk-to-boundary and bulk-to-bulk) may contain derivatives acting on them. In this subsection we make a few comments about such amplitudes -- for a detailed analysis, see \cite{Bzowski:2023jwt}.

Let  us consider tree-level exchange diagrams based on an action with the only derivative interaction being $\vphi_i(\partial\vphi_{[3]})^2$,  where $\vphi_{[3]}$ is the bulk field dual to a dimension 3 operator and $\vphi_i$ is either $\vphi_{[3]}$ or $\vphi_{[2]}$, the  bulk field dual to a dimension 2 operator,  plus the standard non-derivative cubic interactions involving fields dual to operators of dimension 2 and 3.  One may distinguish the new exchange diagrams constructed using the derivative interactions from a standard exchange diagrams based on non-derivative interactions by indicating which propagator has a derivative acting on it. We will do this by putting a hat to indicate which propagator contains a derivative. 
For example, $\ino_{[\hat{3}\hat{3},33 x3]}$ is constructed using one derivative vertex $\vphi_{[3]}(\partial\vphi_{[3]})^2$ and one non-derivative vertex, $\vphi_{[3]}^3$ and the derivatives act on two bulk-to-boundary propagator; $\ino_{[\hat{3} 2,\hat{3} 2 x\hat{3}]}$ indicates the exchange diagram constructed using two derivative vertices $\vphi_{[2]}(\partial\vphi_{[3]})^2$, and each of the bulk-to-boundary propagators of $\vphi_{[3]}$ has a derivative acting on it, and two derivatives act on the bulk-to-bulk propagator.

Consider the AdS exchange diagram $\ino_{[\hat{3}\hat{3},\hat{3}\hat{3} x3]}$ for a massless scalar.
As discussed in section 6.3.2 of \cite{Bzowski:2022rlz}, this diagram is connected with $\ino_{[22,22x3]}$ via the action of  
weigh-shifting operator:
\[\label{WWideriv}
\ino_{[\hat{3}\hat{3},\hat{3}\hat{3} x3]} = \mathcal{W}_{12}^{++}\mathcal{W}_{34}^{++}\ino_{[22,22x3]}.
\]
The integral $\ino_{[22,22x3]}$ is finite and hence the right-hand side of \eqref{WWideriv} is also finite.  More non-trivially, we find that the dilogarithms present in $\ino_{[22,22x3]}$ all cancel out after repeated application of the standard dilogarithm identities leaving $\ino_{[\hat{3}\hat{3},\hat{3}\hat{3} x3]}$ a rational function. The result is
\begin{align}
\ireg_{[\hat{3}\hat{3},\hat{3}\hat{3} x3]} & = \frac{s^3}{4} + \frac{s^2}{4} \left[ \frac{2 \sigma_4}{q_T^3} + \frac{\sigma_3}{q_T^2} + \frac{\sigma_2}{q_T} - q_T \right] + \frac{\tau \sigma_4}{2 q_T^3} + \frac{\tau \sigma_3}{4 q_T^2} + \frac{-\tau^2 + 2 \tau \sigma_2 + 2 \sigma_4}{4 q_T} - \frac{\sigma_3}{2}.
\end{align}
where $\sigma_k = \s{k}{1234}$ is the $k$-th symmetric polynomial on $q_1, q_2, q_3, q_4$ and
\begin{align}
\tau = (q_1 + q_2)(q_3 + q_4).
\end{align}
The five polynomials, $\sigma_k$ for $k=1,2,3,4$ and $\tau$ form the basis of minimal dimension of the polynomials invariant under $q_1 \leftrightarrow q_2$, $q_3 \leftrightarrow q_4$ as well as simultaneous $(q_1,q_2) \leftrightarrow (q_3,q_4)$.

\begin{table}[t]
\centering
\begin{tabular}{|c|c|c|}
\hline
Amplitude & Degree of divergence & Degree of transcendence \\ \hline
$\ireg_{[\hat{3}\hat{3},22x2]}$ & $0$ & $\Li_2$ \\ \hline
$\ireg_{[\hat{3}\hat{3},32x2]}$ & $2$ & $\Li_2$ \\ \hline
$\ireg_{[\hat{3}\hat{3},33x2]}$ & $0$ & $\Li_2$ \\ \hline
$\ireg_{[\hat{3}\hat{3},22x3]}$ & $0$ & $\log$ \\ \hline
$\ireg_{[\hat{3}\hat{3},32x3]}$ & $1$ & $\log$ \\ \hline
$\ireg_{[\hat{3}\hat{3},33x3]}$ & $1$ & $\log$ \\ \hline
$\ireg_{[\hat{3}\hat{3},\hat{3}\hat{3} x2]}$ & $0$ & $\Li_2$ \\ \hline
$\ireg_{[\hat{3}\hat{3},\hat{3}\hat{3} x3]}$ & $0$ & rational \\ \hline
$\ireg_{[\hat{3}2,\hat{3}2 x \hat{3}]}$ & $0$ & $\Li_2$ \\ \hline
$\ireg_{[3\hat{3},3\hat{3} x \hat{3}]}$ & $0$ & rational \\ \hline
\end{tabular}
\caption{Degrees of divergence and transcendence of AdS amplitudes with derivative interaction vertices. The indices $2$ and $3$ indicate the fields dual to operators of dimension $2$ and $3$ respectively and without derivatives on their interaction vertices. By $\hat{3}$ we indicate the operator of dimension $3$ with the derivative interaction acting on the corresponding propagator. Degree of transcendence indicates whether the function is rational in momenta, contains logarithms, or dilogarithms. Notice the degrees of divergence and transcendence do not match precisely, as, for example, the amplitude $\ireg_{[\hat{3}\hat{3},\hat{3}\hat{3} x2]}$ is finite (and by extension scale-invariant) and yet it does contain dilogarithms. \label{fig:deriv}}
\end{table}

Similar expressions can be obtained for other AdS amplitudes involving interactions with derivatives \cite{Bzowski:2023jwt}. They are less singular than the corresponding amplitudes without derivative interactions, but not all of them are finite.  In table \ref{fig:deriv}, we summarise the degrees of divergence and transcendentality of the exchange 4-point functions involving derivative vertices.

\section{Shadow CFT description and its breakdown \label{sec: shadow}}

An interesting possibility is that de Sitter correlators are directly dual to CFT correlators of 
the shadow dimensions 
\[\label{shadowdef}
\bar{\Delta}_i = d- \Delta_i
\]
with respect to the canonical AdS/CFT dimensions $\Delta_i$.  A priori, this seems justified by the fact that, at late times, the $(d+1)$-dimensional de Sitter Ward identities reduce 
to the $d$-dimensional conformal Ward identities featuring precisely these shadow dimensions. One way to understand why the Ward identities take this form is to consider the Schwinger-Keldysh formulation and its relation to AdS/CFT. Recall that the de Sitter correlators may be computed from the partition function in \eqref{Z_SK}, where $S_{\pm}$ are related by analytic continuation to the AdS action. In addition, the source couplings $J_{\pm} \varphi_{\pm}$ tend to $J_{\pm} \varphi_{(0)}$ as $\tau \to 0$. Effectively, the sources $J_\pm$ behave as the dual operator $\O_i$ and the late-time coupling $J_{\pm} \varphi_{(0)}$ implements a Legendre transform.  It is well known that a Legendre transform in a CFT exchanges fields with shadow fields (see for example \cite{Klebanov:1999tb}), and thus the de Sitter Ward identities for generic dimensions should have the same form as the conformal Ward identities with the shadow dimensions.  It is also known, however, that the connection between fields and shadow fields via the Legendre transform breaks down when the correlators require renormalisation \cite{Bzowski:2015pba}.  One might therefore anticipate that a possible description of dS amplitudes via a shadow CFT will also break down. 

In this section, we show that these expectations are indeed born out by explicit tree-level computations.
After reviewing the de Sitter Ward identities, we re-express all tree-level de Sitter correlators as AdS/CFT correlators of the shadow dimensions rescaled by specific dimension-dependent  factors. These factors are directly linked to the analytic continuations discussed in earlier sections.
Examination of the cases requiring renormalisation shows however that this shadow description breaks down in an apparently irretrievable fashion. 
This suggests that a direct description of de Sitter correlators in terms of a shadow CFT is not in fact  the correct holographic paradigm.

\subsection{de Sitter Ward identities}

Correlators in de Sitter obey Ward identities  stemming from  the bulk  isometries.
Under a diffeomorphism $x^\mu \rightarrow x^\mu-\xi^\mu$ by a  Killing vector $\xi^\mu$, the variation
\[
\delta_\xi\<\vphi(x_1)\ldots\vphi(x_n)\> = \<\delta_\xi\vphi(x_1)\ldots\vphi(x_n)\>+\ldots
+\<\vphi(x_1)\ldots\delta_\xi\vphi(x_n)\>
\]
vanishes giving the $(d+1)$-dimensional de Sitter Ward identity
\[\label{bulkWI}
0 = \sum_{i=1}^n\xi^\mu(x_i)\frac{\partial}{\partial x_i^\mu}\<\vphi(x_1)\ldots\vphi(x_n)\>.
\] 
Since the isometry group  $SO(4,1)$ of four-dimensional de Sitter  coincides with that of the three-dimensional Euclidean conformal group, these Killing vectors include the de Sitter counterparts $\xi^\mu_D$ and $\xi^\mu_{SCT}$ of dilatations and special conformal transformations.
Writing $x^\mu=(\tau,\bs{x})$, 
these are
\begin{align}
\xi_D^\mu\partial_\mu = \tau\partial_\tau+\bs{x}\cdot\bs{\partial},
\qquad
\xi_{SCT}^\mu\partial_\mu = -2(\bs{b}\cdot\bs{x})\tau\partial_\tau + [(-\tau^2+x^2)\,\bs{b}-2(\bs{b}\cdot\bs{x})\,\bs{x}]\cdot\bs{\partial}.
\end{align}
If we evaluate the late-time limit of \eqref{bulkWI} for correlators where all insertions are localised on a fixed-time slice,
\begin{align}
0&=    \lim_{\tau\rightarrow 0^-}\Big[(-\tau)^{n(\Delta-d)}\sum_{i=1}^n\xi^\mu(x_i)\frac{\partial}{\partial x_i^\mu}\<\vphi(\tau,\x_1)\ldots\vphi(\tau,\x_n)\>\Big],
\end{align}
assuming the asymptotic behaviour \eqref{dSasympt} leads to the Ward identities
\begin{align}\label{dSWI1}
    0&=\sum_{i=1}^n\Big(\bar{\Delta}_i+\x_i\cdot\bs{\partial}_i\Big)\<\vphi_{(0)}(\x_1)\ldots\vphi_{(0)}(\x_n)\>,
\\0&=\sum_{i=1}^n\Big(-2\bar{\Delta}_i\bs{b}\cdot\x_i + \big(x_i^2\,\bs{b}-2(\bs{b}\cdot\x_i)\,\x_i\big)\cdot\bs{\partial}_i\Big)\<\vphi_{(0)}(\x_1)\ldots\vphi_{(0)}(\x_n)\>
\label{dSWI2}
\end{align}
where $\bar{\Delta}_i$
is the shadow dimension as defined in \eqref{shadowdef}.
Formally, \eqref{dSWI1} and \eqref{dSWI2} are precisely the dilatation and special conformal Ward identities for CFT correlators of the shadow field $\O_{\bar{\Delta}}$.  
(For their corresponding form in momentum space, see \cite{Bzowski:2013sza, Baumann:2019oyu}.)
It should be emphasised however that these Ward identities generally hold only in cases where the de Sitter correlators are finite: renormalisation leads to anomalous conformal Ward identities containing additional terms, see \cite{Bzowski:2015pba}.

\subsection{de Sitter correlators  as shadow CFT correlators} 
\label{sec:shadowdS}

Given the Ward identities \eqref{dSWI1} and \eqref{dSWI2}, one expects that de Sitter correlators can be expressed as shadow CFT correlators, at least in the dimensionally regulated theory where the asymptotic behaviour \eqref{dSasympt} and hence these Ward identities hold.
In fact, as shown in \cite{Sleight:2020obc, Sleight:2021plv} (see also \cite{DiPietro:2021sjt}), a still stronger statement holds: de Sitter amplitudes can be expressed as {\it AdS amplitudes} of the shadow dimensions, up to multiplication by specific dimension-dependent factors.  For exchange diagrams in de Sitter, the corresponding AdS exchanges consist of a  linear combination of the exchanged field and its shadow.\footnote{While the unitarity bound $\Delta \ge d/2-1$ might seem to preclude having operators of dimension  $\Delta$ {\it and} $\bar{\Delta}$, note that it is not {\it a priori} clear that one should require that the CFT dual to de Sitter is unitary (or, in the Euclidean case, reflection positive).}

Working in dimensional regularisation so that  -- at least for now -- all divergences are absent, and setting  both $\ell_P$ and $L_{(A)dS}$ to unity,  these relations read:
\begin{align}
\label{2ptshadowdS0}
ds_{[\Delta\Delta]}&=\frac{1}{(2\bar{\beta})^2\mathcal{C}_{[\bar{\Delta}\bar{\Delta}]}}\,\ifin_{[\bar{\Delta}\bar{\Delta}]},\\[2ex]
\label{3ptshadowdS0}
ds_{[\Delta_1 \Delta_2\Delta_3]} &=\prod_{j=1}^3\Big(\frac{1}{2\bar{\beta}_j \mathcal{C}_{[\bar{\Delta}_j\bar{\Delta}_j]}}\Big)\,\mathcal{C}_{[\bar{\Delta}_1\bar{\Delta}_2\bar{\Delta}_3]}\ifin_{[\bar{\Delta}_1\bar{\Delta}_2\bar{\Delta}_3]},\\[2ex]
\label{4ptcontactshadow0}
ds_{[\Delta_1\Delta_2\Delta_3\Delta_4]} &=\prod_{j=1}^4\Big(\frac{1}{2\bar{\beta}_j \mathcal{C}_{[\bar{\Delta}_j\bar{\Delta}_j]}}\Big)\,\mathcal{C}_{[\bar{\Delta}_1\bar{\Delta}_2\bar{\Delta}_3\bar{\Delta}_4]}\ifin_{[\bar{\Delta}_1\bar{\Delta}_2\bar{\Delta}_3\bar{\Delta}_4]},\\[2ex]
\label{shadowdSexch0}
 \dsno_{[\Delta_1 \Delta_2; \Delta_3 \Delta_4 x \Delta_x]} &=\prod_{j=1}^4\Big(\frac{1}{2\bar{\beta}_j \mathcal{C}_{[\bar{\Delta}_j\bar{\Delta}_j]}}\Big)\,\Big[
\frac{\mathcal{C}_{[\bar{\Delta}_1\bar{\Delta}_2\bar{\Delta}_x]}\mathcal{C}_{[\bar{\Delta}_3\bar{\Delta}_4\bar{\Delta}_x]}}{\mathcal{C}_{[\bar{\Delta}_x\bar{\Delta}_x]}}\ifin_{[\bar{\Delta}_1 \bar{\Delta}_2; \bar{\Delta}_3 \bar{\Delta}_4 x \bar{\Delta}_x]} 
\nn\\&\qquad\qquad\qquad\qquad \qquad  +
\frac{\mathcal{C}_{[\bar{\Delta}_1\bar{\Delta}_2\Delta_x]}\mathcal{C}_{[\bar{\Delta}_3\bar{\Delta}_4\Delta_x]}}{\mathcal{C}_{[\Delta_x\Delta_x]}}\ifin_{[\bar{\Delta}_1 \bar{\Delta}_2; \bar{\Delta}_3 \bar{\Delta}_4 x \Delta_x]} \Big], 
\end{align}
where 
\[\label{Cdef}
\bar{\beta}_j = \bar{\Delta}_j-\frac{d}{2}  = -\beta_j,\qquad
\mathcal{C}_{[\bar{\Delta}_1,\ldots,\,\bar{\Delta}_n]} = 2\sin\Big[\frac{\pi}{2}\Big(d-\sum_{j=1}^n\bar{\Delta}_j\Big)\Big].
\]
At 2-points, we can equivalently write this as $\mathcal{C}_{[\bar{\Delta}_j\bar{\Delta}_j]}=-2\sin(\pi\bar{\beta}_j)$.
Note that no analytic continuation is involved in these formulae.

By inspection, the 2-point amplitude and all contact amplitudes take a common form, 
with only the exchange amplitude receiving contributions from both $\bar{\Delta}_x$ and $\Delta_x$.
Notice too that, in all relations, the external operators are effectively rescaled by factors of $2\bar{\beta}_j$.  
These factors are related to the different way we normalise the dS amplitudes relative to those in AdS.
From a bulk perspective external legs carry a factor of the bulk-to-bulk propagator, but in AdS/CFT the AdS amplitudes are normalised such that the external legs have bulk-to-boundary propagators. The near-boundary limit of the bulk-to-bulk propagator differs from the bulk-to-boundary one precisely by the factors of $2 \bar{\beta}$, see \eqref{GAdSbdy} or \eqref{GdSbdy}.

Since all finite AdS shadow amplitudes on the right-hand sides  satisfy the shadow CFT Ward identities, the de Sitter Ward identities \eqref{dSWI1} and \eqref{dSWI2} are now manifestly satisfied. 
Actually, one can understand the structure of these equations from general principles. Since 2- and 3-point functions are uniquely fixed by conformal invariance and the AdS amplitudes are solutions of the conformal Ward identities the dS 2- and 3-point amplitudes have to be proportional to the corresponding AdS-amplitudes.  
Similarly in the case of 4-point functions one may use in addition the analytic structure of the tree-level diagrams to argue that the right-hand sides should take the form in \eqref{4ptcontactshadow0}, \eqref{shadowdSexch0}. So we only need to explain the coefficients
that multiply the AdS amplitudes. In this subsection we fix them by direct computation.

The relations \eqref{2ptshadowdS0}-\eqref{shadowdSexch0} are easily derived by noting that, from the standard properties of Bessel functions,  the shadow AdS propagators are 
\begin{align}
\label{Kmb}
\mathcal{K}_{\bar{\beta}}(q,z) &= -4\beta\sin(\pi\beta)\k_\beta q^{-2\beta}\mathcal{K}_{\beta}(q,z),\\
\mathcal{G}_{\bar{\beta}}(q,z_1,z_2) &= 
\mathcal{G}_{\beta}(q,z_1,z_2) +2 \sin(\pi\beta)  \k_\beta q^{-2\beta}\mathcal{K}_\beta(q,z_1)\mathcal{K}_\beta(q,z_2), \label{Gmb}
\end{align}
where the coefficient $\k_\beta$ (originally defined in \eqref{kbeta})
is that appearing in the de Sitter 2-point function, namely 
\[
ds_{[\Delta\Delta]}(q) = \k_\beta q^{-2\beta},
\qquad
\k_{\beta} = \frac{4^{\beta-1} \Gamma^2(\beta)}{\pi}.
\]
For the contact amplitudes, we then use our earlier relations \eqref{3pt_in_dS} and \eqref{4pt_contact} while for the AdS 2-point function we use  the normalisation \eqref{AdS2pt}.
For the exchange amplitude, we write the  analytic continuations \eqref{Gpmcont} and \eqref{Gppcont} of the Schwinger-Keldysh propagators  as
\begin{align}
G_{+-}(q, \I z_1, - \I z_2) & = G_{-+}(q, -\I z_1, \I z_2) =\frac{1}{2\sin\pi\beta}\Big(\mathcal{G}_{-\beta}(q,z_1,z_2)-\mathcal{G}_\beta(q,z_1,z_2)\Big),\\
    G_{\pm\pm}(q,\pm\I z_1,\pm\I z_2) &= \frac{e^{\mp \I \pi d/2}}{2\sin\pi\beta}\Big(e^{\pm \I \pi\beta}\mathcal{G}_{-\beta}(q,z_1,z_2)-e^{\mp\I\pi\beta}\mathcal{G}_\beta(q,z_1,z_2)\Big).
\end{align}
Re-evaluating \eqref{Ippampl} and \eqref{Ipmampl}, the $++$ and $+-$ contributions are then
\begin{align}
I_{++} 
& =  (-\tau_0)^{2d-\beta_T}  e^{\frac{\I \pi}{2} (\beta_T - d)} \Big(\prod_{j=1}^4 \k_{\beta_j} q_j^{-2 \beta_j}\Big)\times \nn\\
&\qquad \times\frac{1}{2\sin(\pi\beta_x)} \left[e^{-\I\pi\beta_x}\ino_{[\Delta_1 \Delta_2; \Delta_3 \Delta_4 x \Delta_x]} -e^{\I\pi\beta_x}\ino_{[\Delta_1 \Delta_2; \Delta_3 \Delta_4 x (d-\Delta_x)]} \right],\\[2ex]
I_{+-} &=  (-\tau_0)^{2d-\beta_T}e^{\frac{\I \pi}{2} (\beta_1+\beta_2-\beta_3-\beta_4)}
\Big(\prod_{j=1}^4 \k_{\beta_j} q_j^{-2 \beta_j}\Big)\times\nn \\
&\qquad \times\frac{1}{2\sin(\pi\beta_x)} \left[-\ino_{[\Delta_1 \Delta_2; \Delta_3 \Delta_4 x \Delta_x]} +\ino_{[\Delta_1 \Delta_2; \Delta_3 \Delta_4 x (d-\Delta_x)]} \right],
\end{align}
so that overall
\begin{align}
& \dsno_{[\Delta_1 \Delta_2; \Delta_3 \Delta_4 x \Delta_x]} = 2 \lim_{\tau_0 \rightarrow 0^{-}} 
(-\tau_0)^{\beta_T-2d} \re \left[ I_{++} + I_{+-} \right]\nn\\
&\qquad  =\Big(\prod_{j=1}^4 \k_{\beta_j} q_j^{-2 \beta_j}\Big)\left[
\mathcal{A}(\beta_x)\ifin_{[\Delta_1 \Delta_2; \Delta_3 \Delta_4 x \Delta_x]} +\mathcal{A}(-\beta_x)\ifin_{[\Delta_1 \Delta_2; \Delta_3 \Delta_4 x \,(d-\Delta_x)]} \right] \label{dsAbeta}
\end{align}
where the coefficients are given by the function
\begin{align}
    \mathcal{A}(\beta_x)&= \frac{1}{\sin(\pi\beta_x)} \mathrm{Re}\Big[e^{\frac{\I\pi}{2}(\beta_T-d-2\beta_x)}-e^{\frac{\I\pi}{2}(\beta_1+\beta_2-\beta_3-\beta_4)}\Big]\nn\\&
    = - \frac{2}{\sin(\pi\beta_x)}  \sin\left(\frac{\pi}{2}(\beta_1+\beta_2-\beta_x-\frac{d}{2})\right)\sin\left(\frac{\pi}{2}(\beta_3+\beta_4-\beta_x-\frac{d}{2})\right). \label{Abetaco}
\end{align}
Using \eqref{Kmb}, 
we can then rewrite the dS exchange diagram in terms of AdS exchanges with shadow operators for the external legs,
\begin{align}\label{shadowdSexch}
& \dsno_{[\Delta_1 \Delta_2; \Delta_3 \Delta_4 x \Delta_x]} =
\mathcal{B}(\beta_x)\ifin_{[\bar{\Delta}_1 \bar{\Delta}_2; \bar{\Delta}_3 \bar{\Delta}_4 x \Delta_x]} +\mathcal{B}(\bar{\beta}_x)\ifin_{[\bar{\Delta}_1 \bar{\Delta}_2; \bar{\Delta}_3 \bar{\Delta}_4 x \bar{\Delta}_x]}, 
\end{align}
where 
\begin{align}\label{shadowdSexchB}
    \mathcal{B}(\bar{\beta}_x)
     &=
     - \frac{2 \sin\left(\frac{\pi}{2}(\bar{\beta}_1+\bar{\beta}_2+\bar{\beta}_x+\frac{d}{2})\right)\sin\left(\frac{\pi}{2}(\bar{\beta}_3+\bar{\beta}_4+\bar{\beta}_x+\frac{d}{2})\right)}{\sin(\pi\bar{\beta}_x)\prod_{j=1}^{4}4\bar{\beta}_j\sin(\pi\bar{\beta}_j)}.
\end{align}
This result is equivalent to \eqref{shadowdSexch0}.

\subsection{Breakdown of the shadow paradigm} 

The shadow formulae \eqref{2ptshadowdS0}-\eqref{shadowdSexch0} for de Sitter correlators are valid in dimensional regularisation, and for generic dimensions where
divergences are absent.  However, as we shall see, 
they fail to hold whenever the de Sitter correlators require renormalisation.  This is because the corresponding divergences in the shadow CFT cannot be removed due to a lack of suitable local counterterms built from sources and operators of the shadow dimensions.  In contrast, for a dual CFT description involving  operators of the canonical AdS/CFT dimensions, the dimensions of sources and operators are switched and local counterterms (of the beta function type) can indeed be constructed.  
A related set of problematic cases arises when the shadow CFT correlators diverge but the corresponding de Sitter correlators do not.  For these, however, the shadow formulae can be considered to hold, albeit only in a limiting sense.  For a dual CFT of the canonical dimensions, these latter cases correspond to those involving conformal anomalies where the anomaly is projected out by the canonical holographic formulae.

As shown in \cite{Bzowski:2015pba}, the divergences of the $n$-point correlator in the shadow CFT correspond to solutions of the singularity condition\footnote{Noting that sources have dimension $d-\bar{\Delta}_j = d/2-\bar{\beta}_j$ while operators have dimension $\bar{\Delta}_j = d/2+\bar{\beta}_j$, this condition is equivalent to the existence of a dimension-$d$ counterterm containing  $k$ boxes.}
\[\label{singcond}
(n-2)\frac{d}{2}+ \sum_{j=1}^n\sigma_j\bar{\beta}_j = -2k, \qquad k=0,1,2,\ldots 
\]
for any independent choice of the signs $\{\sigma_j\in \pm 1\}$ or constant $k\in\mathbb{Z}^+$.  Cases where all the $\{\sigma_j\}$ are minus correspond to conformal anomalies of the shadow theory: such divergences have  an ultralocal momentum dependence and can be removed by the addition of a dimension-$d$ counterterm containing $k$ boxes acting on $n$ sources.  Cases where a single  $\sigma_j$ is plus and the rest are minus correspond to the sources of the shadow operator $\O_{\bar{\Delta}_j}$ acquiring a beta function.  Divergences of this type have a semilocal momentum dependence and are removed by a counterterm containing $k$ boxes acting on $\O_{\bar{\Delta}_j}$ and $n-1$ sources.  (We exclude the case $n=2$ however as here the counterterm is simply the standard source for $\O_{\bar{\Delta}_j}$.)
When $n=3$ and two or more of the $\{\sigma_j\}$ are plus, the leading divergence has a nonlocal momentum dependence.  Such nonlocal divergences cannot be removed by local counterterms in any QFT.  Moreover, no local counterterms for such cases exist.\footnote{Counterterms involving a  product of two or more local operators are excluded since these introduce additional divergences whose renormalisation  then modifies the operator dimensions.  As a result, such multi-operator counterterms are generically no longer of dimension $d$, see \cite{Bzowski:2015pba}.}
 Instead, wherever divergences of this type arise, the dimensionally-regulated correlator must be associated with an overall coefficient that vanishes as the regulator is removed such that the limit is finite.

If we restrict to fields of dimension $\frac{d}{2}<\Delta_j\le d$ such that 
\[\label{shadowrange}
0\le \bar{\Delta}_j<\frac{d}{2},\qquad -\frac{d}{2}\le \bar{\beta}_j<0,
\]
the only potential divergences of the shadow 2-point function are of the type $\{++\}$ with $\bar{\beta}=-k$.  With the holographic normalisation \eqref{AdS2pt}, such cases are however vanishing since
\[
\ino_{[\bar{\Delta}\bar{\Delta}]} = -\frac{\Gamma(1-\bar{\beta})}{2^{2\bar{\beta}-1}\Gamma(\bar{\beta})}q^{2\bar{\beta}}.
\] 
The shadow formula \eqref{2ptshadowdS0} is nevertheless consistent since the factor of $\mathcal{C}_{[\bar{\Delta}\bar{\Delta}]}=-2\sin(\pi\bar{\beta})$ in the denominator cancels the zero in $\ino_{[\bar{\Delta}\bar{\Delta}]}$ such that the de Sitter correlator is a finite power $q^{-2k}$ as expected. 
The corresponding de Sitter Ward identities  \eqref{dSWI1} and \eqref{dSWI2} are then obeyed. 

\begin{table}[t]
\centering
\begin{tabular}{|c|c|c|}
\hline
dS amplitude & Shadow correlator & Shadow singularity type \\
\hline
$\dsno_{[222]}$ & $i_{[111]}$ & $+++$\\[-1ex]
$\dsno_{[322]}$ & $i_{[011]}$ & $++-$\\[-1ex]
$\dsno_{[332]}$ & $i_{[001]}$ & $+++$\\[-1ex]
$\dsno_{[333]}$ & $i_{[000]}$ & $++-$\\
\hline
\end{tabular}
\caption{Singularity type of shadow CFT correlators for de Sitter 3-point functions. \label{shadowtype3pt}} 
\end{table}

At three points, for fields satisfying \eqref{shadowrange}, we encounter only singularities of the types
\begin{align}
&\{+++\}: \quad \bar{\Delta}_t -d =  \bar{\beta}_t+\frac{d}{2} = -2k \nn\\
 \qquad 
 &\{++-\}:\quad \bar{\Delta}_1+\bar{\Delta}_2-\bar{\Delta}_3 = \bar{\beta}_1+\bar{\beta}_2-\bar{\beta}_3 +\frac{d}{2} = -2k
\end{align}
along with permutations. 
For $\Delta=2,3$ in $d=3$  ({\it i.e.,}  $\bar{\Delta}=1,0$ respectively), these cases are shown in table \ref{shadowtype3pt}.  
As no counterterms are available, in all of these cases a finite shadow 3-point function can only be obtained through multiplication by a vanishing coefficient.  The renormalised 3-point function then corresponds to the leading divergence of the associated triple-$K$ integral \cite{Bzowski:2015pba}, namely
\begin{align}\label{shadow3ptfns}
\iren_{[111]} &=\frac{c_{[111]}}{q_1q_2q_3},
&\iren_{[001]} &=c_{[001]}\frac{q_1^2+q_2^2-q_3^2}{q_1^3 q_2^3 q_3},\qquad\nn\\
\iren_{[011]} &=\frac{c_{[011]}}{q_1^3}\Big(\frac{1}{q_2}+\frac{1}{q_3}
\Big), 
&\iren_{[000]} &=c_{[000]}\sum_{i<j}^3\frac{1}{q_i^3q_j^3},
\end{align}
where the $c_{[\bar{\Delta}_1\bar{\Delta}_2\bar{\Delta}_3]}$ are finite constants.
All these 3-point functions are fully nonlocal and satisfy the corresponding homogeneous conformal Ward identities. 

Turning to the shadow formula  \eqref{3ptshadowdS0}, however, we see that the  factor $\mathcal{C}_{[\bar{\Delta}_1\bar{\Delta}_2\bar{\Delta}_3]}$ in the numerator vanishes for all $\{+++\}$ cases by virtue of \eqref{Cdef}, while all $\mathcal{C}_{[\bar{\Delta}_j\bar{\Delta}_j]}$ factors are nonzero. 
For the de Sitter 3-point functions to be nonzero then requires that the shadow 3-point functions ({\it i.e.,} the constants $c_{[111]}$ and $c_{[001]}$ in \eqref{shadow3ptfns}) are divergent.
It could nevertheless be argued that \eqref{3ptshadowdS0} is consistent when viewed as a limiting case within dimensional regularisation.  In the regularisation scheme \eqref{half-beta_scheme},  selecting  $c_{[111]} =-\ep^{-1}$, and using $\mathcal{C}_{[111]}=-\pi\ep$, $\mathcal{C}_{[11]}=2$, we find that  \eqref{3ptshadowdS0} correctly reproduces the de Sitter 3-point function $\dsno_{[222]}$ in  \eqref{ds222}.
Likewise, using $\mathcal{C}_{[001]}=\pi \ep$ and $\mathcal{C}_{[00]}=-2$, and selecting $c_{[001]}=(9/2)\ep^{-1}$ 
we find that \eqref{3ptshadowdS0}  reproduces  $\dsno_{[332]}$ in  \eqref{ds332}.
The homogeneous de Sitter Ward identities \eqref{dSWI1} and \eqref{dSWI2} are moreover obeyed.

However, such a reconciliation cannot be achieved for the remaining $\{++-\}$ cases.  As all coefficients $\mathcal{C}_{[\bar{\Delta}_1\bar{\Delta}_2\bar{\Delta}_3]}$ and $\mathcal{C}_{[\bar{\Delta}_j\bar{\Delta}_j]}$ in \eqref{3ptshadowdS0}  are nonzero,  to obtain a finite de Sitter 3-point function requires the corresponding shadow 3-point function (and hence the constants $c_{[011]}$ and $c_{[000]}$ above) to be finite.
However, the resulting  de Sitter 3-point functions obtained via \eqref{3ptshadowdS0} are then manifestly incorrect:
 for both  $\dsno_{[333]}$ and  $\dsno_{[322]}$, the actual de Sitter 3-point functions \eqref{ds333} and \eqref{ds322}  depend on $\mu$ and contain logarithms absent in \eqref{shadow3ptfns}.
 From the shadow CFT perspective, however, there is no way to introduce such $\mu$-dependent logarithms as there are no local dimension-$d$ countertems.
Likewise,  the actual de Sitter 3-point functions \eqref{ds333} and \eqref{ds322} (as opposed to the output of  \eqref{3ptshadowdS0}) do not satisfy the homogeneous de Sitter Ward identities \eqref{dSWI1} and \eqref{dSWI2}.  From a bulk perspective, this is due to the appearance of logarithms violating the pure power-law asymptotic behaviour \eqref{dSasympt} assumed in their derivation. 

Thus, already at the 3-point level, the notion of a dual description based on the shadow CFT is problematic.  For correlators satisfying the $\{+++\}$ condition, consistency with the de Sitter results can be achieved only at the price of allowing divergent CFT correlators, while for correlators satisfying the $\{++-\}$ condition, we cannot recover the corresponding renormalised de Sitter 3-point functions.  This latter case arises in particular for inflationary correlators of three  non-derivatively coupled massless scalars.   In contrast, no such problems arise for a dual CFT with fields of the  canonical AdS/CFT dimensions.  Here, the signs appearing in the singularity condition \eqref{singcond} are reversed since $\beta_j=-\bar{\beta}_j$ meaning the $\{+++\}$ and $\{++-\}$ cases for the shadow CFT correspond respectively to the $\{---\}$ and $\{--+\}$ cases in the canonical CFT. 
Both of these latter singularity types can be eliminated via counterms giving rise to anomalies and beta functions respectively. Anomalies are then projected out by the canonical holographic formulae due to their ultralocal momentum dependence.  The resulting de Sitter correlators are then independent of  $\mu$ consistent with the finiteness of the corresponding de Sitter correlators.  Beta function contributions are semi-local, however, and survive the holographic formulae reproducing the renormalised de Sitter correlators with logarithms and a nontrivial dependence on the RG scale.

The same pattern extends to  4-point contact diagrams.
For shadow singularities of type $\{+ + + \,+\}$, including those shown in table \ref{shadowtype4pt} (evaluated in \eqref{dsren3222}-\eqref{dsren3333}),  the shadow formulae \eqref{4ptcontactshadow0} once again holds when viewed as a limiting case within dimensional regularisation due to the vanishing of $\mathcal{C}_{[\bar{\Delta}_1\bar{\Delta}_2\bar{\Delta}_3\bar{\Delta}_4]}$.  However, for shadow singularities of type $\{+ + + \,-\}$, the shadow formula \eqref{4ptcontactshadow0} fails to correctly reproduce the renormalised de Sitter correlators which depend on the RG scale and contain logs.

\begin{table}[t]
\centering
\begin{tabular}{|c|c|c|}
\hline
dS amplitude & Shadow correlator & Shadow singularity type \\
\hline
$\dsno_{[3222]}$ & $i_{[0111]}$ & $+++ \,+$\\[-1ex]
$\dsno_{[3322]}$ & $i_{[0011]}$ & $+++\,-$\\[-1ex]
$\dsno_{[3332]}$ & $i_{[0001]}$ & $+++\,+$\\[-1ex]
$\dsno_{[3333]}$ & $i_{[0000]}$ & $+++\,-$\\
\hline
\end{tabular}
\caption{Singularity type of shadow CFT correlators for dS 4-point contact diagrams. \label{shadowtype4pt}} 
\end{table}

For exchange diagrams, with the exception of $\dsno_{[22,22x2]}$, all the de Sitter amplitudes we constructed required renormalisation as per table \ref{fig:deg4}.  The shadow formula \eqref{shadowdSexch0} then fails to reproduce these de Sitter amplitudes since the latter contain logarithms and depend on the RG scale whereas the  shadow CFT correlators do not since no local counterterms are available.
As we saw in previous sections, however, all these cases can be correctly handled using a dual CFT of the canonical AdS/CFT dimensions and the holographic formulae \eqref{amp2}-\eqref{amp4x}.

\section{Discussion} 

We had two objectives in writing this paper. The first was to set up a renormalisation procedure for IR divergences in de Sitter, and the second was to use this information to refine our understanding of 
the possible duality between dS and CFT. 

We have succeeded in setting up 
a renormalisation procedure.
This result is independent of holography and the connection to AdS, and should be useful 
more generally 
beyond the 
context of this paper. We have shown that one can renormalise the IR divergences by adding local counterterms at future infinity in dS. The renormalised correlators have an associated (finite) scheme-dependence, and it would be interesting to understand what physics ({\it i.e.,} normalisation conditions) fixes this dependence. 

Here, we discussed how to renormalise IR divergences at tree-level. IR divergences in dS at loop level have a long history \cite{Starobinksky:1984, Ford:1984hs, Antoniadis:1985pj, Starobinsky:1986fx, Salopek, Starobinsky:1994bd, Tsamis:2005hd, Weinberg:2005vy, Riotto:2008mv, Burgess:2009bs, Senatore:2009cf, Giddings:2010nc, Burgess:2010dd, Marolf:2010zp, Rajaraman:2010xd,Senatore:2012nq, Pimentel:2012tw, Polyakov:2012uc, Senatore:2012ya, Serreau:2013psa, Akhmedov:2017ooy, Gorbenko:2019rza, Mirbabayi:2019qtx, Baumgart:2019clc,  Green:2020txs, Cohen:2020php, Baumgart:2020oby}, with recent works closer in spirit to ours including \cite{Heckelbacher:2020nue, Cespedes:2023aal, Beneke:2023wmt, Chowdhury:2023arc}; see also the reviews \cite{Seery:2010kh, Hu:2018nxy}. 
It would be interesting to revisit the issue of loops using the methodology developed in this paper. In this respect, 
we expect an interesting interplay between bulk UV and IR issues similar to that observed for AdS in \cite{Banados:2022nhj}.

The renormalisation of bulk IR divergences is consistent with the holographic duality computing the wavefunction of the universe in terms of the partition function of the dual CFT, upon a specific analytic continuation. The IR renormalisation then corresponds to the UV renormalisation of the CFT. This provides  structural support for the duality that goes beyond symmetry considerations.
To further emphasise  this point, note that while symmetry considerations imply the regulated dS amplitudes can be expressed in terms of CFT correlators of shadow operators without any analytic continuation, the dual CFT cannot be a local CFT involving shadow operators:  the UV structure of such a theory does not match the IR structure of the bulk, and one cannot promote the regulated relations to renormalised ones.

The bulk IR singularities imply that the dS Ward identities are modified. Note that the dS in-in amplitudes do not suffer from conventional conformal anomalies. Instead, the renormalisation is associated with renormalising the late-time bulk fields. On the CFT side, this maps to renormalisation of the sources that couple to the dual operators. One may work out by standard methods the effect of this renormalisation on the conformal Ward identities \cite{Bzowski:2015pba}, and then use the holographic map to work out the modified dS Ward identity. It would interesting to work this out in full generality; see \cite{Wang:2022eop} for a recent work in this direction. It would also be interesting to derive the modified dS Ward identity directly in the bulk. It is these modified Ward identities that should be the starting point for cosmological bootstrap considerations.

In this paper we studied scalar fields on a fixed dS background. It would be interesting to extend our analysis to spinning correlators on dS, and more generally, to gauge-invariant cosmological perturbations about accelerating FRLW spacetimes, extending our earlier analysis \cite{McFadden:2011kk} beyond
the level of 3-point functions.

\section*{Acknowledgments}

We thank Paolo Benincasa, Arthur Lipstein, 
Scott Melville, Enrico Pajer, Guilherme Pimentel,  Charlotte Sleight, Massimo Taronna, Ayngaran	Thavanesan, Aron Wall, Dong-Gang Wang, and the participants of the Correlators in Cortona workshop for discussions.
AB is supported by the NCN POLS grant No.~2020/37/K/ST2/02768 financed from the Norwegian Financial Mechanism 2014-2021 \includegraphics[width=12pt]{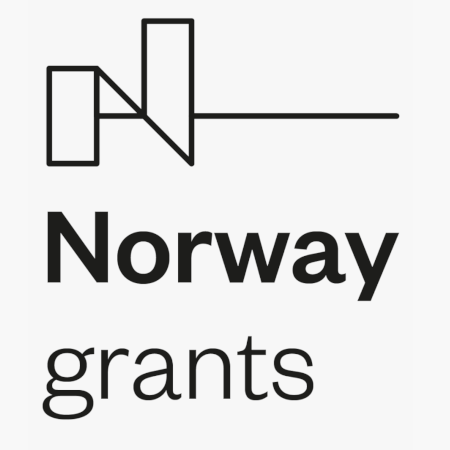} \includegraphics[width=12pt]{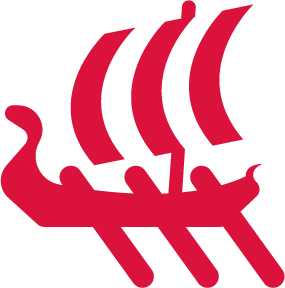}.
PM is supported in part by the UKRI consolidated grant ST/T000708/1. KS is supported in part by UKRI consolidated grants ST/P000711/1 and ST/T000775/1.

\appendix

\section{Definitions for momenta} \label{sec:notation}

\begin{itemize}
\item External momenta are denoted $\bs{q}_j$, with lengths or magnitudes $q_j = |\bs{q}_j|$, where $j=1,2,\ldots$. The Mandelstam variables are
\begin{align} \label{def:stu}
& s = | \bs{q}_1 + \bs{q}_2 |, && t = | \bs{q}_1 + \bs{q}_3 |, && u = | \bs{q}_2 + \bs{q}_3 |
\end{align}
without squares. For convenience, we also adopt the convention $q_s = s$.
\item The $3$- and $4$-point total magnitudes are denoted
\begin{align} \label{total}
& q_t = q_1 + q_2 + q_3, && q_T = q_1 + q_2 + q_3 + q_4.
\end{align}
\item A contact diagram with $n$ identical particles has symmetry group $S_n$ corresponding to permutations of the external momenta: $\bs{q}_j \mapsto \bs{q}_{\sigma(j)}$ for any $\sigma \in S_4$.  

\item We use $\s{m}{J}$ to denote  the corresponding  $m$-th symmetric polynomial on the set of indices $J$. 
 To be precise, let $J$ be an ordered set of indices and let $m$ be an integer such that $1 \leq m \leq |J|$. Then,
\begin{align} \label{def:sigma}
\s{m}{J} = \sum_{\substack{L \subseteq J\\|L|=m}} q_{L_1} \ldots q_{L_m},
\end{align}
where the sum is taken over all ordered subsets $L \subseteq J$ of cardinality $m$. In particular
\begin{align}
\s{1}{12} & = q_1 + q_2, & \s{1}{123} & = q_1 + q_2 + q_3, \\
\s{2}{12} & = q_1 q_2, & \s{2}{123} & = q_1 q_2 + q_1 q_3 + q_2 q_3, \\
&& \s{3}{123} & = q_1 q_2 q_3,
\end{align}
\begin{align}
\s{1}{1234} & = q_1 + q_2 + q_3 + q_4, \\
\s{2}{1234} & = q_1 q_2 + q_1 q_3 + q_1 q_4 + q_2 q_3 + q_2 q_4 + q_3 q_4, \label{def:sigma2} \\
\s{3}{1234} & = q_1 q_2 q_3 + q_1 q_2 q_4 + q_1 q_3 q_4 + q_2 q_3 q_4, \\
\s{4}{1234} & = q_1 q_2 q_3 q_4.
\end{align}
We also allow for the indices to take the value $s$, so that, for example, $\sigma_{(1)12s} = q_1 + q_2 + s$ and so on.

\item For 4-point exchange diagrams we define the following variables,
\begin{align}\label{mij}
& \m{ij} = q_i + q_j - | \bs{q}_i + \bs{q}_j|, && \p{ij} = q_i + q_j + | \bs{q}_i + \bs{q}_j|.
\end{align} 
In particular all 4-point exchange diagrams will contain the following combinations,
\begin{align}
& \m{12} = q_1 + q_2 - s, && \p{12} = q_1 + q_2 + s, \\
& \m{34} = q_3 + q_4 - s, && \p{34} = q_3 + q_4 + s,
\end{align}
which we will use in addition to the standard momentum variables. 

Note that all $\m{ij}$ and $\p{ij}$ are non-negative. Furthermore, $\p{ij} = 0$ corresponds to $\bs{q}_i = \bs{q}_j = 0$, while $\m{ij} = 0$ indicates the collinear limit, $\bs{q}_i \parallel \bs{q}_j$, when the two momenta are parallel.

\item The highest possible symmetry group of an exchange diagram $12 \mapsto 34$, arising when all external particles are identical, is the dihedral group $D_4$.  This contains the eight permutations generated by swapping the numbers within each pair, $12;34 \mapsto 21;34$ and $12;34 \mapsto 12;43$, as well as exchanging the pairs, $12;34 \mapsto 34;12$. 

The following dilogarithmic quantities then arise in exchange diagrams:
\begin{align}
\PLp & = \Li_2 \left( \frac{\m{34}}{q_T} \right) + \Li_2 \left( \frac{\m{12}}{q_T} \right) + \log \left( \frac{\p{12}}{q_T} \right) \log \left( \frac{\p{34}}{q_T} \right) - \frac{\pi^2}{6}, \label{defD+} \\
\PLm & = \Li_2 \left( \frac{\m{34}}{q_T} \right) - \Li_2 \left( \frac{\m{12}}{q_T} \right) + \frac{1}{2} \log^2 \left( \frac{\p{12}}{q_T} \right) - \frac{1}{2} \log^2 \left( \frac{\p{34}}{q_T} \right). \label{defD-}
\end{align}
$\PLp$ is invariant under the group $D_4$,
\begin{align}
\PLp(\bs{q}_1, \bs{q}_2; \bs{q}_3, \bs{q}_4) = \PLp(\bs{q}_2, \bs{q}_1; \bs{q}_3, \bs{q}_4) = \PLp(\bs{q}_3, \bs{q}_4; \bs{q}_1, \bs{q}_2),
\end{align}
while $\PLm$ acquires a sign when any two pairs of indices are exchanged,
\begin{align}
\PLm(\bs{q}_1, \bs{q}_2; \bs{q}_3, \bs{q}_4) = \PLm(\bs{q}_2, \bs{q}_1; \bs{q}_3, \bs{q}_4) = - \PLm(\bs{q}_3, \bs{q}_4; \bs{q}_1, \bs{q}_2).
\end{align}

\end{itemize}

\section{Holographic 1-point functions in the presence of sources}\label{1pt_app}

In this appendix, we outline the derivation of the standard holographic formula \eqref{1ptfn} for 1-point functions in the presence of sources.  A similar calculation applies for computing wavefunction coefficients leading to \eqref{dS1ptfn}.

\subsection{In AdS} \label{app:1pt_AdS}

We follow the conventions of \cite{Skenderis:2002wp} and define $W$, the generating functional of the connected diagrams as $Z_{QFT}[\varphi_{(0)}] = e^{W[\varphi_{(0)}]}$, where $Z_{QFT}[\varphi_{(0)}]$ is the partition function of the dual QFT. Accoring to the AdS/CFT correspondence, $Z_{QFT} = Z_{AdS}$, which in the saddle point approximation becomes $W[\varphi_{(0)}] = - S_{AdS}[\varphi_{(0)}]$. On the right-hand side here, the AdS action \eqref{EAdSaction} is evaluated on the unique classical bulk solution $\varphi$ which near the boundary approaches $\varphi_{(0)}$, according to \eqref{AdSasympt}. In general this on-shell action $S_{AdS}[\varphi_{0}]$ is divergent due to the infinite volume of AdS. Indeed, by substituting the near-boundary expansion \eqref{AdSasympt} into the free field part of \eqref{EAdSaction} one finds that the action diverges near $z = 0$ for any values of the dimensions $d$ and $\Delta$ \cite{deHaro:2000vlm}. 

It is more  convenient to analyse this issue from the point of view of the 1-point function, $\<\O(\bs{x})\>_s$, where the subscript $s$ indicates that this is an expectation value  in the presence of sources. To do this, using the chain rule we  write,
\begin{align} \label{1ptfn_expanded}
\<\O(\bs{x})\>_s =-\frac{\delta\ln Z_{QFT}}{\delta\vphi_{(0)}(\x)}= \int \D z\, \int\D^d \bs{x}' \, \frac{\delta S_{AdS}}{\delta \varphi(z, \bs{x}')} \frac{\delta \varphi(z, \bs{x}')}{\delta \varphi_{(0)}(\bs{x})}.
\end{align}
The second functional derivative follows from the near-boundary expansion \eqref{AdSasympt}. From now on, let us assume that the conformal dimension $\Delta$ satisfies $\frac{d}{2} < \Delta < d$ and the interaction potential $V_{int}$ in \eqref{EAdSaction} is a polynomial in $\varphi$ or at least has a regular Taylor expansion around $\varphi = 0$. This guarantees that the source term of order $z^{d - \Delta}$ is the most leading term in the near-boundary expansion of the bulk field. In particular,
\begin{align} \label{second_f_deriv}
\frac{\delta \varphi(z, \bs{x}')}{\delta \varphi_{(0)}(\bs{x})} = z^{d - \Delta} \delta(\bs{x} - \bs{x}') + o(z^{d-\Delta}).
\end{align}

The first functional derivative in \eqref{1ptfn_expanded} produces equations of motion for the bulk field plus a boundary term. As we work on-shell, we can drop the bulk term proportional to the equations of motion. The boundary term occurs due to the integration by parts of the kinetic term in \eqref{EAdSaction} and is supported on the boundary of the spacetime. It is equal to the radial canonical momentum $\Pi_z$ flowing through the boundary \cite{Papadimitriou:2004ap, Papadimitriou:2004rz},
\begin{align} \label{defPi} 
\frac{\delta S_{AdS}}{\delta \varphi(z, \bs{x})} = \sqrt{\gamma^z} \Pi_z(z, \bs{x}) \delta(z).
\end{align}
For the action \eqref{EAdSaction}, the radial canonical momentum reads
\begin{align} \label{Pican}
\Pi_z = - \frac{(\ell^{(AdS)}_P)^{1-d}}{L_{AdS}} z \partial_z \varphi(z, \bs{x}),
\end{align}
where $\gamma^z_{ij}$  denotes the induced metric on  constant $z$-slices such that $\sqrt{\gamma^z} = (z/L_{AdS})^{-d}$.

When put together, the 1-point function reads
\begin{align}
\<\O(\bs{x})\>_s & = L_{AdS}^d \lim_{z \rightarrow 0} z^{-\Delta} \Pi_z(z, \bs{x}).
\end{align}
This is the general formula relating the 1-point function of the dual operator to the canonical momentum flowing through the boundary. However, for the canonical momentum in \eqref{Pican} the limit diverges,
\begin{align}
\<\O(\bs{x})\>_s & = - \left( \frac{L_{AdS}}{\ell^{(AdS)}_P} \right)^{d-1} \lim_{z \rightarrow 0} z^{-\Delta} \times z \partial_{z} \varphi \nn\\
& = - \left( \frac{L_{AdS}}{\ell^{(AdS)}_P} \right)^{d-1} \lim_{z \rightarrow 0} \left[ (d - \Delta) z^{d - 2 \Delta} \varphi_{(0)} + \ldots + \Delta \varphi_{(\Delta)} + \ldots \right].
\end{align}
The underlying philosophy of holography is that the 1-point function should be proportional to the vev coefficient $\varphi_{\Delta}$. Unfortunately, the source term $z^{d - 2 \Delta} \varphi_{(0)}$ in the above expansion is always more leading than the vev term. On the other hand, under the assumptions on dimensions and the form of the potential stated above, there exists only a finite number of terms more leading than the vev term. Furthermore, all such terms are local in the source $\varphi_{(0)}$. Thus, they can be removed by the addition of finite and local counterterm action localised at the boundary. 

Consider the counterterm action \cite{deHaro:2000vlm}
\begin{align} \label{AdS_Sct}
S_{\rm ct} = \frac{(\ell^{(AdS)}_P)^{1-d}}{2 L_{AdS}} (d - \Delta) \int \D^d \bs{x} \sqrt{\gamma^z} \, \varphi^2.
\end{align}
This action is located on the boundary and does not alter the bulk equations of motion. On the other hand it does contribute to the canonical momentum defined in \eqref{defPi}. With $S = S_{AdS} + S_{\rm ct}$, the radial canonical momentum reads
\begin{align}
\Pi_z = \frac{(\ell^{(AdS)}_P)^{1-d}}{L_{AdS}} \left[ - z \partial_z \varphi(z, \bs{x}) + (d - \Delta) \varphi \right].
\end{align}
With this modification the 1-point function becomes
\begin{align} \label{towards_1pt}
\<\O(\bs{x})\>_s & = - \left( \frac{L_{AdS}}{\ell^{(AdS)}_P} \right)^{d-1} \lim_{z \rightarrow 0} \left[ 0 \times z^{d - 2 \Delta} \varphi_{(0)} + \ldots + (2 \Delta - d) \varphi_{(\Delta)} + \ldots \right].
\end{align}
The holographic formula \eqref{1ptfn} follows if the limit exists, \textit{i.e.}, if the omitted terms more leading than the vev term vanish.

In order to argue that the troublesome terms in \eqref{towards_1pt} vanish or are irrelevant, two paths can be taken. In the standard holographic renormalisation \cite{deHaro:2000vlm,Bianchi:2001kw,Skenderis:2002wp}, a finite number of counterterms are added in order to remove all such divergences. To do so, the theory is regulated by a near-boundary cut-off at $z = \epsilon$, all counterterms are placed on the cut-off surface and ultimately the $\epsilon \rightarrow 0$ limit is taken. It is worth noting that in some special cases, \textit{i.e.}, for special values of the dimensions $d$ and $\Delta$, the near-boundary expansion \eqref{AdSasympt} may acquire secular terms involving logarithms of the radial variable $z$. Such terms may lead to logarithmic divergences in the canonical momentum. These cases are related to conformal anomalies \cite{Henningson:1998gx, deHaro:2000vlm}, and their removal of logarithmic divergences has associated scheme-dependence parametrised by finite  local counterterms. Such counterterms, in turn, may also modify the relation \eqref{1ptfn} by an ultralocal functional of the source $\varphi_{(0)}$. An alternative (and equivalent) procedure is to expand the canonical momentum in terms of eigenfunctions of the dilatation operator $\delta_D$ \cite{Papadimitriou:2004ap, Papadimitriou:2004rz},
\begin{equation} \label{radialM_exp}
    \Pi_z = \Pi_{(d-\Delta)} + \cdots + \Pi_{(\Delta)} + \cdots, 
\end{equation}
where $\delta_D \Pi_{(n)}=-n \Pi_{(n)}$.
Such expansion can be obtained via a covariant form of the near-boundary analysis,  and the renormalised 1-point function is the eigenfunction with eigenvalue equal to $\Delta$,
\begin{equation} \label{1pt_radialM}
 \<\O(\bs{x})\>_s = \Pi_{\Delta}.
\end{equation}
Comparing with \eqref{towards_1pt} we find
\begin{equation} \label{radialM}
  \Pi_{\Delta} = - \left( \frac{L_{AdS}}{\ell^{(AdS)}_P} \right)^{d-1} (2 \beta) \varphi_{(\Delta)}.  
\end{equation}
This procedure avoids the need to explicitly construct the counterterms. 

In the spirit of this paper, however, we may apply dimensional renormalisation instead \cite{Bzowski:2016kni,Bzowski:2019xri}. This method is particularly convenient in our context as the conformal dimensions $\Delta$ are treated as parameters and are not \textit{a priori} fixed to any special values. The method is applicable since there exists a non-empty open set of the parameters for which the limit in \eqref{towards_1pt} is well-defined. If the potential $V_{int}$ in \eqref{EAdSaction} is a polynomial or at least exhibits a regular Taylor expansion at $\varphi = 0$, the vev term in the expansion in \eqref{towards_1pt} becomes the leading term if
\begin{align} \label{dim_cond}
\frac{d}{2} < \Delta < \min \left( \frac{d}{2} + 1, \frac{2 d}{3} \right).
\end{align}
For such values of $d$ and $\Delta$ the limit exists and \eqref{towards_1pt} becomes \eqref{1ptfn}. Furthermore, using the explicit form of the propagators, one can show that all correlation functions are analytic functions of the dimensions in this range. Using analytic continuation one obtains unique expressions for the correlation functions except at a number of poles, corresponding to the special cases. Only when such  special cases are encountered are counterterms needed. The counterterms turn out to be in a one-to-one correspondence with the counterterms of the dimensionally regulated dual theory.

\subsection{In dS}

In de Sitter, the calculation runs along similar lines.  We define
\begin{align} 
\psi_s(\x) =-\frac{\delta\ln \Psi_{dS}}{\delta\vphi_{(0)}(\x)}= -i\int \D \tau\, \int\D^d \bs{x}' \, \frac{\delta S_{dS}}{\delta \varphi(\tau, \bs{x}')} \frac{\delta \varphi(\tau, \bs{x}')}{\delta \varphi_{(0)}(\bs{x})}.
\end{align}
where
\begin{align} 
\frac{\delta \varphi(\tau, \bs{x}')}{\delta \varphi_{(0)}(\bs{x})} = (-\tau)^{d - \Delta} \delta(\bs{x} - \bs{x}') + o\big((-\tau)^{d-\Delta}\big).
\end{align}
The variation of the action is 
\begin{align} 
\frac{\delta S_{dS}}{\delta \varphi(\tau, \bs{x})} = \sqrt{\gamma^\tau}\, \Pi_\tau(\tau, \bs{x}) \delta(\tau).
\end{align}
where $\Pi_\tau(\tau, \bs{x})$ is the (standard) canonical momentum, and from the action \eqref{dSaction2}, 
\begin{align} 
\Pi_\tau = - \frac{(\ell_P^{(dS)})^{1-d}}{L_{dS}} \tau \partial_\tau \varphi(z, \bs{x}),
\end{align}
where $\gamma^\tau_{ij}$  denotes the induced metric on  constant $\tau$-slices such that $\sqrt{\gamma^\tau} = (-\tau/\ell_{dS})^{-d}$.
The 1-point function then reads
\begin{align}
\psi_s(\x) & = -i\,L_{dS}^d \lim_{\tau \rightarrow 0} (-\tau)^{-\Delta} \Pi_\tau(\tau, \bs{x}).
\end{align}
Including the boundary counterterm
\begin{align}
S_{\rm ct} = \frac{(\ell_P^{(dS)})^{1-d}}{2 L_{dS}} (d - \Delta) \int \D^d \bs{x} \sqrt{\gamma^\tau} \, \varphi^2,
\end{align}
the canonical momentum becomes
\begin{align}
\Pi_\tau = \frac{(\ell_P^{(dS)})^{1-d}}{L_{dS}} \left[ - \tau \partial_\tau \varphi(\tau, \bs{x}) + (d - \Delta) \varphi \right].
\end{align}
With this modification the 1-point function becomes
\begin{align} 
\psi_s(\x) & = - i \left( \frac{L_{dS}}{\ell_P^{(dS)}} \right)^{d-1} \lim_{\tau \rightarrow 0} \left[ 0 \times (-\tau)^{d - 2 \Delta} \varphi_0 + \ldots + (d-2 \Delta) \varphi_{\Delta} + \ldots \right].
\end{align}
leading to the holographic formula \eqref{dS1ptfn}, where the subleading divergences have either been removed by additional counterterms or dimensional regularisation and analytic continuation is used, as in the discussion above of the AdS case.

\section{Singularity conditions for individual amplitudes}
\label{app:sings}

In this appendix we summarise the singularities of both AdS and dS contact and exchange amplitudes for general values of the operator and spacetime dimensions, working in the dimensionally regulated theory throughout.

\subsection{Contact diagrams}

Contact diagrams have singularities whenever the singularity condition 
\[\label{contsings}
(n-2)\frac{d}{2}+\sum_{i=1}^n \sigma_i \beta_i = - 2k,\qquad k\in\mathbb{Z}^+
\]
is satisfied, 
where $n$ is the number of points, $k\in\mathbb{Z}^+$ is a non-negative integer ($k=0,1,2\ldots$), and the signs $\sigma_i\in\pm 1$ are chosen independently \cite{Bzowski:2015pba, Bzowski:2019kwd}.  In AdS, any choice of signs is permitted, but in dS the case where all signs are negative is excluded.  This all-minus  case is non-singular in dS because the corresponding all-minus divergence in AdS has an ultralocal momentum dependence ({\it i.e.,} is analytic in all the squared momenta), and hence is projected out when passing to dS via the holographic formulae, see \eqref{3pt} and \eqref{4pt_contact}.  This projection can  equivalently be seen from the zero of the sine factor in, {\it e.g.,} \eqref{sin_to_im_i3}.
This singularity condition is sufficient to account for all the contact diagram singularities in tables \ref{fig:deg3} and \ref{fig:deg4}.

\subsection{AdS exchanges}\label{sec:AdSxsings}

To identify the analogous singularity conditions for AdS exchange diagrams, we use the identity
\[\label{KtoI}
K_{\beta}(z) = \frac{\pi}{2\sin\pi\beta}(I_{-\beta}(z)-I_{\beta}(z))
\]
to re-write the bulk-bulk propagator \eqref{AdSGdef} as
\begin{align}
\mathcal{G}_{\beta}(q, z_1, z_2) = \frac{\pi (z_1 z_2)^{d/2}}{2\sin\pi\beta}\Big[-I_\beta(qz_1)I_\beta(q z_2)+\Big(I_{-\beta}(qz_1)I_\beta(qz_2)\Theta(z_1-z_2)+(z_1\leftrightarrow z_2)\Big)\Big].
\end{align}
The $s$-channel exchange diagram then decomposes into a sum of  factorised and nested integrals,
\begin{align}
&\ireg_{[\Delta_1 \Delta_2; \Delta_3 \Delta_4 x \Delta_x]}(q_i,s) \nn\\
&\quad
= 
\frac{\pi}{2\sin\pi\beta_x}\Big[-\int_0^\infty\frac{\D z_1}{z_1^{d/2+1}}\mathcal{K}_{\beta_1}(q_1,z_1)\mathcal{K}_{\beta_2}(q_2, z_1)I_{\beta_x}(s z_1)\nn\\ &\quad\qquad\qquad \qquad\qquad \times\int_0^\infty\frac{\D z_2}{z_2^{d/2+1}}\mathcal{K}_{\beta_3}(q_3, z_2)\mathcal{K}_{\beta_4}(q_4,z_2)I_{\beta_x}(s z_2)\nn\\
&\quad
+\Big(\int_0^\infty\frac{\D z_1}{z_1^{d/2+1}}\mathcal{K}_{\beta_1}(q_1, z_1)\mathcal{K}_{\beta_2}(q_2,z_1)I_{-\beta_x}(s z_1)\int_0^{z_1}\frac{\D z_2}{z_2^{d/2+1}}\mathcal{K}_{\beta_3}(q_3, z_2)\mathcal{K}_{\beta_4}(q_4, z_2)I_{\beta_x}(s z_2)\nn\\&\quad\qquad\qquad \qquad\qquad+ (z_1\leftrightarrow z_2)\Big)\Big].
\label{nestedAdSintegrals}
\end{align}
The singularities of the factorised integrals are found by expanding their integrands about the lower limits and looking for the appearance of $z^{-1}$ terms \cite{Bzowski:2015pba}.  As $I_{\beta}(z)=z^\beta(1+O(z^2))$ while $K_\beta(z)=z^{-\beta}(1+O(z^2))+z^\beta(1+O(z^2))$, this gives the two singularity conditions
\begin{align}\label{vertexconds}
\frac{d}{2}\pm \beta_1\pm \beta_2 + \beta_x = -2k_L,\qquad
\frac{d}{2}\pm \beta_3\pm \beta_4 + \beta_x = -2k_R,\qquad
k_L,\,k_R\in\mathbb{Z}^+
\end{align}
where the $\beta_x$ term appears only with a plus sign (in contrast to the analogous singularity condition for a 3-point contact diagram).
For the remaining nested integrals, the singularities can again be found by expanding around the lower limit of the outer integral.   For the first nested integral in \eqref{nestedAdSintegrals}, keeping track only of powers of $z_1$ and $z_2$, we obtain a sum of terms of the form
\begin{align}\label{gennestedterm}
&\int_0^\infty\D z_1\, z_1^{d/2-1+\sigma_1\beta_1+\sigma_2\beta_2-\beta_x+2k_1}\int_0^{z_1}\D z_2\, z_2^{d/2-1+\sigma_3\beta_3+\sigma_4\beta_4+\beta_x+2k_2}\nn\\ &\qquad\qquad\qquad \sim
\int_0^\infty\D z_1\, z_1^{d-1+2k_T+\sum_{i=1}^4\sigma_i\beta_i}
\end{align}
where the $z_1^{2k_1}$ and $z_2^{2k_2}$ contributions arise from subleading terms in the series expansions of the Bessel functions and hence $k_1,k_2\in\mathbb{Z}^+$.  To obtain the right-hand side, we evaluated the inner integral assuming singularity conditions
\eqref{vertexconds} (which we have already identified\footnote{Note there can be no cancellation of singularities between the nested and factorised integrals when either (or both) of the conditions \eqref{vertexconds} are satisfied: the singularity of the factorised integral is always non-analytic in $s^2$ (of the form $\sim s^{2\beta_x+2k}$ for some $k\in\mathbb{Z}^+$) 
whereas that of the nested integral is analytic ($\sim s^{2k}$) due to the opposite indices on the two Bessel $I$.
}) 
are not satisfied. A singularity then arises whenever we obtain a $z_1^{-1}$ term, namely when
\[\label{totalcond}
d+\sum_{i=1}^4\sigma_i\beta_i = -2k_T, \qquad k_T\in\mathbb{Z}^+.
\]
Note in particular that the $\beta_x$ contributions in \eqref{gennestedterm} have cancelled since the Bessel $I$ functions from which they originated have opposite indices. 
As both nested integrals in \eqref{nestedAdSintegrals} are related by exchanging $(\beta_1,\beta_2,-\beta_x)\leftrightarrow(\beta_3,\beta_4,\beta_x)$, the singularity condition we obtain from the second nested integral is again \eqref{totalcond}, and no cancellation can occur as both terms contribute with the same sign.

The singularities of AdS exchange diagrams are thus predicted by the combination of the `vertex' and `total' singularity conditions \eqref{vertexconds} and \eqref{totalcond}.  These conditions are sufficient to account for the degrees of divergence (obtained by direct integration) in table \ref{fig:deg4}.  In particular, the quadratic divergences arise when multiple singularity conditions are satisfied simultaneously
(see also \cite{Bzowski:2015pba}).

\subsection{dS exchanges}

The singularity conditions for dS exchanges can be obtained using the holographic formula  \eqref{4pt_exchange}.  
Once again, taking the imaginary parts in this formula acts to project out all singularities of AdS correlators that have a purely ultralocal momentum dependence, namely those arising from any all-minus condition.  This projection can equivalently be seen from the zeros of the sine factors in \eqref{4ptindSsin}.  
The holographic formula relates the dS exchange diagram  to both the AdS exchange diagram, whose singularities are given by \eqref{vertexconds} and \eqref{totalcond}, and a product of AdS 3-point functions, whose singularities are given by \eqref{contsings} with $n=3$.
From the AdS exchange contribution, the dS exchange inherits the `total' singularity condition
\[\label{dStotalcond}
d+\sum_{i=1}^4\sigma_i\beta_i = -2k_T, \qquad k_T\in\mathbb{Z}^+
\]
where any independent choice of signs $\sigma_i\in\pm 1$ is permitted {\it except} for the all-minus case.
The `vertex' singularity conditions arise from both the AdS exchange contribution and the product of AdS 3-point functions, and read 
\begin{align}\label{dSvertexconds}
\frac{d}{2}+\sigma_1 \beta_1+\sigma_2 \beta_2 + \sigma_x^L\beta_x = -2k_L,\quad
\frac{d}{2}+\sigma_3 \beta_3+\sigma_4 \beta_4 + \sigma_x^R\beta_x = -2k_R,\quad
k_L,\,k_R\in\mathbb{Z}^+.
\end{align}
Note that both signs $\sigma_x^L,\sigma_x^R\in \pm 1$ can appear here, since while only the plus sign appears in the vertex condition for AdS exchanges \eqref{vertexconds}, both signs appear in the 3-point conditions \eqref{contsings}.  (This explains why
the dS amplitude $\dsno_{[22,22x3]}$ is divergent while the corresponding AdS amplitude $\ino_{[22,22x3]}$ is finite, see table \ref{fig:deg4}.)
However, the all-minus cases  $(\sigma_1,\sigma_2,\sigma_x^L)=(-,-,-)$ and $(\sigma_3,\sigma_4,\sigma_x^R)=(-,-,-)$ are excluded since the corresponding AdS singularities are ultralocal and hence projected out when we pass to dS.
Moreover, an analysis using the shadow holographic formulae in the following section 
shows that the $(\sigma_1,\sigma_2,\sigma_x)=(-,-,+)$ and $(\sigma_3,\sigma_4,\sigma_x)=(-,-,+)$ cases of \eqref{dSvertexconds} also excluded, {\it i.e,} do not give rise to singularities of the dS exchange diagram.  This is due to a cancellation between the singularities of the AdS exchange and 3-point contributions in the holographic formula \eqref{4pt_exchange} as we will see shortly.

The conditions above, \eqref{dStotalcond} and \eqref{dSvertexconds} (where all cases for which $(\sigma_1,\sigma_2,\sigma_3,\sigma_4)=(-,-,-,-)$ are excluded), are sufficient to account for the degrees of divergence for dS exchange diagrams in table \ref{fig:deg4}.  As above, quadratic divergences arise when more than one condition is satisfied.

\subsection{Compatibility with the shadow formula}\label{sec:shadowsingcompatibility}

The singularity conditions for dS exchanges can also be analysed via the shadow formula \eqref{shadowdSexch0}, or equivalently \eqref{dsAbeta} and \eqref{Abetaco}. 
These formulae relate the dS exchange to a linear combination of the AdS exchange and shadow exchange diagrams, where the latter is obtained by replacing $\beta_x\rightarrow -\beta_x$.
Starting from the exchange singularity conditions in AdS, \eqref{vertexconds} and \eqref{totalcond}, we then recover the dS singularity conditions \eqref{dSvertexconds} and \eqref{dStotalcond} after resolving the following two subtleties.

Firstly, we need to establish that the all-minus case of the `total' singularity condition   \eqref{dStotalcond} is absent.  This occurs due to  a cancellation of singularities between the AdS exchange and shadow exchange diagrams.  The `total' singularity for both these diagrams is the same, as can be seen from the analysis in section \ref{sec:AdSxsings}, with the cancellation then following from the identity
\begin{align}\label{trigid}
&\sin\Big[\frac{\pi}{2}(\frac{d}{2}-\beta_1-\beta_2+\beta_x)\Big]\sin\Big[\frac{\pi}{2}(\frac{d}{2}-\beta_3-\beta_4+\beta_x)\Big] \nn\\& \qquad
-\sin\Big[\frac{\pi}{2}(\frac{d}{2}-\beta_1-\beta_2-\beta_x)\Big]\sin\Big[\frac{\pi}{2}(\frac{d}{2}-\beta_3-\beta_4-\beta_x)\Big] \nn\\[1ex]
&= \sin(\pi\beta_x)\sin\Big[\frac{\pi}{2}(d-\beta_T)\Big].
\end{align}
The relative sign between the two terms on the left-hand side arises since the sign of $\beta_x$ is flipped between the exchange and the shadow exchange, and the denominator in \eqref{Abetaco} contains a factor of $\sin(\pi\beta_x)$.

The second point to be understood is the  absence 
of `vertex'-type dS singularities corresponding to \eqref{dSvertexconds} for the cases $(\sigma_1,\sigma_2,\sigma_x)=(-,-,+)$ and $(\sigma_3,\sigma_4,\sigma_x)=(-,-,+)$.
For these cases, the sine factors appearing in the shadow formula \eqref{Abetaco} have zeros cancelling the corresponding singularities, however an equivalent cancellation is not immediately apparent in our earlier approach based on the holographic formula \eqref{4pt_exchange}.

Closer inspection shows, however,  that a cancellation of singularities between the AdS exchange and 3-point contributions to the holographic formula \eqref{4pt_exchange} indeed occurs.   
When either $d/2-\beta_1-\beta_2+\beta_x = -2k_L$ and/or $d/2-\beta_3-\beta_4+\beta_x=-2k_R$, the first term in the trigonometric identity \eqref{trigid} vanishes.
The numerator of the holographic formula \eqref{4ptindSsin} then becomes
\begin{align}
&\sin \left( \frac{\pi}{2} (\beta_T - d) \right)\ireg_{[\Delta_1 \Delta_2; \Delta_3 \Delta_4 x \Delta_x]}  \label{Icancel}
\\& \qquad
+ 2 \k_{\beta_x} s^{-2 \beta_x} \ireg_{[\Delta_1 \Delta_2 \Delta_x]}(q_1, q_2, s) \ireg_{[\Delta_x \Delta_3 \Delta_4]}(s, q_3, q_4)  \nn\\
& \qquad\qquad \times
\sin \left( \frac{\pi}{2} (\frac{d}{2}-\beta_1 - \beta_2 - \beta_x ) \right) \sin \left( \frac{\pi}{2} (\frac{d}{2}- \beta_3 - \beta_4 -\beta_x ) \right)\nn\\[1ex]
&=
\sin \left( \frac{\pi}{2} (\beta_T - d) \right)
\Big[
\ireg_{[\Delta_1 \Delta_2; \Delta_3 \Delta_4 x \Delta_x]} \nn\\&\qquad\qquad 
+ 2\sin(\pi\beta_x) \k_{\beta_x} s^{-2 \beta_x} \ireg_{[\Delta_1 \Delta_2 \Delta_x]}(q_1, q_2, s) \ireg_{[\Delta_x \Delta_3 \Delta_4]}(s, q_3, q_4) \Big].\nn
\end{align}
The singularities of the integrals on the right-hand side now cancel.  
This can most readily be seen by using the relation \eqref{Gmb} between the AdS bulk-bulk and shadow bulk-bulk propagator, which yields
\begin{align}
&   \ireg_{[\Delta_1 \Delta_2; \Delta_3 \Delta_4 x \Delta_x]}
+ 2\sin(\pi\beta_x) \k_{\beta_x} s^{-2 \beta_x} \ireg_{[\Delta_1 \Delta_2 \Delta_x]}(q_1, q_2, s) \ireg_{[\Delta_x \Delta_3 \Delta_4]}(s, q_3, q_4) 
\nn\\[1ex]&\qquad = \ireg_{[\Delta_1 \Delta_2; \Delta_3 \Delta_4 x\, d-\Delta_x]}. 
\end{align}
This shadow AdS exchange diagram is manifestly finite as 
the conditions $d/2-\beta_1-\beta_2+\beta_x = -2k_L$ and $d/2-\beta_3-\beta_4+\beta_x=-2k_R$ correspond to $d/2-\beta_1-\beta_2-\bar{\beta}_x = -2k_L$ and $d/2-\beta_3-\beta_4-\bar{\beta}_x=-2k_R$ where $\bar{\beta}_x=\bar{\Delta}_x-d/2 = - \beta_x$.  From the analysis of the AdS exchange diagram in section \ref{sec:AdSxsings}, however, there are no all-minus `vertex' singularities of this type, see \eqref{vertexconds}.

In summary, dS `vertex'-type singularities corresponding to \eqref{dSvertexconds} for $(\sigma_1,\sigma_2,\sigma_x)=(-,-,+)$ and/or $(\sigma_3,\sigma_4,\sigma_x)=(-,-,+)$ are indeed absent.  This is manifest in the shadow holographic formula due to the vanishing of the sine factors, but arises through a non-trivial cancellation in the conventional holographic formula.
Conversely, the absence of the all-minus `total' singularity in dS is readily visible in the conventional holographic formula, but arises from a non-trivial cancellation in the shadow holographic formula.

\section{Shadow relations in AdS \label{app: shadow}}

In the absence of divergences, amplitudes in AdS satisfy the shadow relations
\begin{align}
\ifin_{[\bar{\Delta}\bar{\Delta}]}(q)  &= -\frac{4\beta^2}{\ifin_{[\Delta \Delta]}(q)}, \label{app: sh1} \\
\ifin_{[\bar{\Delta}_1,\bar{\Delta}_2,\bar{\Delta}_3]}(q_i)&=\left(\prod_{j=1}^3\frac{2\beta_j}{
\ifin_{[\Delta_j\Delta_j]}(q_j)}\right)
\ifin_{[\Delta_1,\Delta_2,\Delta_3]}(q_i),\\
\ifin_{[\bar{\Delta}_1,\bar{\Delta}_2, \bar{\Delta}_3, \bar{\Delta}_4]}(q_i)&=\left(\prod_{j=1}^4\frac{2\beta_j}{
\ifin_{[\Delta_j\Delta_j]}(q_j)}\right)
\ifin_{[\Delta_1 \Delta_2, \Delta_3 \Delta_4]}(q_i),\\
\ifin_{[\bar{\Delta}_1,\bar{\Delta}_2, \bar{\Delta}_3, \bar{\Delta}_4 x\bar{\Delta}_x]}(q_i,s)&=\left(\prod_{j=1}^4\frac{2\beta_j}{
\ifin_{[\Delta_j\Delta_j]}(q_j)}\right)\Bigg[
\ifin_{[\Delta_1 \Delta_2, \Delta_3 \Delta_4 x\Delta_x]}(q_i,s)\nn\\[1ex]&\qquad -\frac{\ifin_{[\Delta_1\Delta_2\Delta_x]}(q_1,q_2,s)\ifin_{[\Delta_x\Delta_3\Delta_4]}(s,q_3,q_4)}{\ifin_{[\Delta_x\Delta_x]}(s)}\Bigg] \label{app: sh4}
\end{align}
where $\bar{\Delta}_j=d-\Delta_j$ are the shadow dimensions.  

In deriving these formulae, we used the standard holographic normalisations (see \eqref{AdS2pt} for the 2-point function, with higher-point functions as defined in \cite{Bzowski:2022rlz}), and applied the same normalisations to shadow fields replacing $\beta_j\rightarrow \bar{\beta}_j = \bar{\Delta}_j-d/2=-\beta_j$.  In addition, we used the identities \eqref{Kmb} and \eqref{Gmb} connecting the AdS propagators to their shadows.

Alternatively, these formulae can be understood as arising from the Legendre transform: for each field, we add to the partition function,
$Z_{AdS}[\varphi_{(0)}]$ in \eqref{ZAdSdef}, the product of the source $\varphi_{(0)}$ with its conjugate variable $\Pi_{(\Delta)}$ (see appendix  \ref{app:1pt_AdS}), namely 
$\int d^d x \sqrt{\gamma} \Pi_{(\Delta)} \varphi_{(0)}$,
and integrate over $\varphi_{(0)}$. Completing squares, one may then integrate out $\varphi_{(0)}$ giving
\begin{equation}
    \varphi_{(0)} = \frac{2 \beta}{\ifin_{[\Delta\Delta]}} \varphi_{(\Delta)}.
\end{equation}
Now $\varphi_{(\Delta)}$ acts as a source for the shadow operator of dimension $\bar{\Delta}=d-\Delta$, and functionally differentiating with respect to it,  one may obtain \eqref{app: sh1}-\eqref{app: sh4}.

\bibliographystyle{JHEP}
\bibliography{dS}
\end{document}